\documentclass[a4paper,fleqn,usenatbib]{mnras}

\renewcommand{\footnoterule}{%
  \kern -19pt
  \hrule width 2in
  \kern 2.6pt
}
\usepackage{mathptmx}

\usepackage[T1]{fontenc}
\usepackage{ae,aecompl}

\usepackage{tikz,xcolor,hyperref}

\definecolor{lime}{HTML}{A6CE39}
\DeclareRobustCommand{\orcidicon}{%
	\begin{tikzpicture}
	\draw[lime, fill=lime] (0,0) 
	circle [radius=0.16] 
	node[white] {{\fontfamily{qag}\selectfont \tiny ID}};
	\draw[white, fill=white] (-0.0625,0.095) 
	circle [radius=0.007];
	\end{tikzpicture}
	\hspace{-4mm}
}

\foreach \x in {A, ..., Z}{%
	\expandafter\xdef\csname orcid\x\endcsname{\noexpand\href{https://orcid.org/\csname orcidauthor\x\endcsname}{\noexpand\orcidicon}}
}

\usepackage{graphicx}	
\usepackage{amsmath}	
\usepackage{amssymb}	
\usepackage{scrextend}  
\usepackage{xcolor}     
\usepackage{subfig}     
\usepackage{longtable}

\title[SF in perturbed galaxies: kinematic distributions]{Star formation in CALIFA survey perturbed galaxies.\\
III. Stellar \& ionized-gas kinematic distributions}

\author[A. Morales-Vargas et al.]{
A. Morales-Vargas\orcidA{}$^{1}$\thanks{E-mail: abdmoralesv@gmail.com (AMV).},
J. P. Torres-Papaqui\orcidB{}$^{2}$,
F. F. Rosales-Ortega\orcidC{}$^{3}$,
\
\newauthor
M. Chow-Mart\'{i}nez\orcidE{}$^{4}$,
R. A. Ortega-Minakata\orcidG{}$^{5}$,
A. C. Robleto-Or\'{u}s\orcidH{}$^{6}$
\
\newauthor
\& the CALIFA survey Collaboration
\\
$^{1}$CITEVA, Centro de Astronom\'\i{}a, Universidad de Antofagasta, Avenida U. de Antofagasta 02800, Antofagasta, Chile\\
$^{2}$Departamento de Astronom\'\i{}a, Universidad de Guanajuato, Apartado Postal 144, Guanajuato 36000, Mexico\\
$^{3}$Instituto Nacional de Astrof\'\i{}sica, \'{O}ptica y Electr\'{o}nica, Luis Enrique Erro 1, Tonantzintla 72840, Mexico\\
$^{4}$Instituto de Geolog\'{i}a y Geof\'{i}sica, Universidad Nacional Aut\'{o}noma de Nicaragua, Rotonda Universitaria\\
$^{}$Rigoberto L\'{o}pez P\'{e}rez 150 metros al Este, Managua 663, Nicaragua\\
$^{5}$Instituto de Radioastronom\'{i}a y Astrof\'{i}sica (IRyA), UNAM, Apartado Postal 72-3, Morelia, Michoac\'{a}n 58089, Mexico\\
$^{6}$Instituto de Astronom\'\i{}a, Universidad Nacional Aut\'{o}noma de M\'exico (UNAM), Apartado Postal 70-264, M\'exico D. F. 04510, Mexico\\
}

\date{Accepted XXX. Received YYY; in original form ZZZ}

\pubyear{2023}

\begin{document}
\label{firstpage}
\pagerange{\pageref{firstpage}--\pageref{lastpage}}
\maketitle

\begin{abstract}

We obtain the kinematic distributions of stars (synthetic model line absorption) and ionized gas (H$\alpha$ line emission) for star-forming regions residing 
in CALIFA survey tidally perturbed (perturbed) and non-tidally perturbed (control) galaxies. We set the uncertainties of the velocity dispersion by measuring 
the statistical variability of the datasets themselves. Using these adopted uncertainties and considering the sensitivity of the grating device, we establish 
thresholds of reliability that allow us to select reliable velocity dispersions. From this selection, we pair the star-forming spaxels between control and 
perturbed galaxies at the closest shifts in velocity (de-redshifting). We compare their respective distributions of velocity dispersion. In perturbed
galaxies, median velocity dispersions for the stellar and gaseous components are minimally higher and equal, respectively, than those in control galaxies.
The spread in velocity dispersion and the velocity shift\,-\,velocity dispersion space agree with this result. Unlike the well-known trend in strongly
interacting systems, the stellar and ionized-gas motions are not disturbed by the influence of close companions. For the gaseous component, this result
is due to the poor statistical variability of its data, a consequence of the tightness in velocity dispersion derived from high spectral line intensities.
This analysis concludes the series, which previously showed star-forming regions in galaxies with close companions undergoing more prominent gas inflows,
resulting in differences in their star formation and consequent metal content.

\end{abstract}

\begin{keywords}
galaxies: evolution -- galaxies: interactions -- galaxies: kinematics and dynamics -- galaxies: star formation -- galaxies: statistics.
\end{keywords}



\section{Introduction}
\label{sec:int}

As the star formation (SF) and its consequent metal renewal, the dynamics of the stars and ionized gas use to vary from isolated to interacting galaxies. Due to 
the realization of kiloparsec (kpc)-scale studies, our knowledge of galaxy kinematics has improved significantly in detail. The Calar Alto Legacy Integral Field 
Area \citep[CALIFA,][]{San12,Hus13,Wal14,GarB15,San16} survey is well suited to this subject. It is custom-made to study large-scale kinematic deviations or long-axis 
rotation in galaxies \citep{Fac17}. Prove is that it has delivered results in good agreement with those of pioneering integral field spectroscopic surveys such 
as DiskMass \citep{Ber10} and ATLAS$^{3\mathrm{D}}$ \citep{Cap11}. That is one of the many legacies the CALIFA survey has left for us: a comprehensive picture 
of galaxy kinematics. For instance, \citet{BaBa14,BaBa15a} obtain, for isolated and merging galaxies, respectively, the line-of-sight (LOS) velocity distributions 
of the stellar and ionized-gas components and compare their orientations with that one from photometry. For practically all their galaxies in isolation, \citet{BaBa14} 
find differences in the kinematic position angles that are $<$\,22\,$\deg$ for both the stars and gas. Between both, the differences are $<$\,16\,$\deg$. 
Contrarily, \citet{BaBa15a} find these little discrepancies absent in around half of their sample tracing all merger stages (mostly absent with the presence of morphological 
signatures of interaction). From an averaged major kinematic axis, the radial deviations are considerable in the stars. In the gas, for a significant fraction of their 
sample, the deviations are larger than the values for isolated galaxies. In agreement, \citet{GarL15} find that the ionized-gas kinematics is rarely consistent with 
simple co-planar circular motions. Most of their objects with kinematic lopsidedness between receding and approaching sides of the ionized-gas velocity fields are 
interacting galaxies. All this confirms that gas kinematics is more easily affected than stellar one.

From the SAMI Galaxy Survey \citep[\textit{e.g.}][]{Bry15} Collaboration, \citet{Oh16} use deeper imaging to best reveal the interacting stage. They find prominent 
misalignment between the photometric and the stellar kinematic major axes of morphologically distorted galaxies. They suggest that mergers (and perhaps flybys) contribute 
to determining the angular momentum characteristics in galaxies. From the same collaboration, \citet{Bl17} used the ratio of H$\alpha$ half-light radius to \textit{r}-band 
continuum half-light as an approximation of centrally-concentrated SF to confirm that the latter correlates with the kinematic asymmetry of the H$\alpha$ line 
emission.\footnote{\citet{Ell13,BaBa15b,Mor15}, among others, have already reported this correlation.} Their relationship gains strength with stellar mass 
(M$_{*}$). Later on, \citet{Bl18} confirm the influence of M$_{*}$ by finding the distance to the nearest neighbour inversely correlated with asymmetry of the H$\alpha$ 
line emission for galaxies with log$_{10}$\,M$_{*}\,>$\,10\,M$_{\odot}$. However, an absent correlation distinguishes lower M$_{*}$. For given asymmetry values, they 
find low-M$_{*}$ galaxies as more H$\alpha$-velocity-dispersion supported than high-M$_{*}$ ones. Contrastingly, for very bright group galaxies (log$_{10}$\,M$_{*}\,\sim$\,10.6\,M$_{\odot}$, 
typically early-type), \citet{Rao21} find the gas-star misalignment more strongly correlated with S\'{e}rsic index than with M$_{*}$. They propose that the dynamical 
state of galaxy groups characterizes the internal kinematics of stars and ionized gas in the brightest objects. Lastly, \citet{Ris22} find external processes such as 
gas accretion as the dominant one within the most probable causes of kinematic decoupling between stars and gas.

Proving the frequency of gaseous-stellar kinematic misalignment, or even counter-rotation\footnote{Counter-rotation implies two components (either both stellar, both 
gaseous or one of each) that oppositely rotate concerning each other.}, due to either secular or external processes comes into place. From the SDSS-IV MaNGA survey 
\citep[\textit{e.g.}][]{Bun15} Collaboration, \citet{Chen16} find evidence that suggests collisions between accreted and pre-existing gas, so its velocity dispersion 
ends slightly increased. These collisions make the stellar and gas components counter-rotate, so the central regions of blue star-forming galaxies rejuvenate their 
stellar populations (SPs) and intensify their SF. Including more types of misaligned objects, \citet{Xu22} find ordered-to-random motion ratios of both stars and gas 
that are lower than those of control galaxies. Their misaligned active galaxies show higher central SF than the control ones. Also, in mostly all misaligned objects, 
\citet{Xu22} find lower nebular metallicities than in control galaxies. However, their misaligned galaxies are not merging or irregular, and they do not mention the 
presence of close companions. Favourably, from a cross-matching between the MaNGA MPL-8 data release and the DESI Legacy Imaging surveys \citep{De19}, \citet{Li21} 
corroborate that the frequency of merging or strongly interacting galaxies is higher if stellar-to-ionized gas misalignments are present. Moreover, \citet{Bev22} study 
galaxies that co-spatially show two counter-rotating stellar disks. Accretion of external gas in retrograde orbits may cause counter-rotation (the new gas in retrograde 
motion and the pre-existing prograde component collide, \textit{e.g.} \citealt{Cocc13}, \citealt{Cor14}, \citealt{Chen16}, \citealt{Bev22}).

Using the EAGLE cosmological hydrodynamical simulation \citep{Sch15}, \citet{Cas22} find that galaxy mergers and gas accretion occur in a similar frequency in aligned 
and misaligned galaxies at fixed M$_{*}$. They suggest morphology itself can quickly torque the gas and align it with the stellar disk. For spheroids, which host most 
of the misalignment events, occur the same, though in a much longer timescale. Though they suggest that external processes are not likely the first source of misalignment, 
they emphasize that broader simulations are essential to study the evolution of the tidal force parameter over time.

To describe how tidal perturbations can affect SF at kpc scales in CALIFA survey galaxies, \citet{Mor20} (hereafter Paper I) confirm that properties or phenomena other 
than kpc-scale stellar mass surface density ($\Sigma_{*}$) can also modulate SF. From statistically-significant linear regression models, the star formation rate (SFR) 
properties of \textit{tidally-perturbed} (perturbed) galaxies increase with increments in their \textit{tidal perturbation parameters} (measurements of the tidal force 
exerted on galaxies by close companions, see Paper I, section 3.1). Such models are motivated after a weaker dependence of SF on $\Sigma_{*}$ is found for perturbed 
than for \textit{non-tidally-perturbed} (control) galaxies (on the resolved plane of the Star Formation Main Sequence, SFMS). Later, \citet{Mor21} (hereafter Paper II) 
inferred that star-forming regions in perturbed galaxies underwent a more intense reactivation of SF in the last 1 Gyr. This conclusion results from statistically significant 
differences in the stellar mass assembly histories and look-back-time annular profiles of SF (star-forming spaxels of perturbed galaxies show higher values than those 
of control ones). 

To explore possible deviations from the kinematics of new-born SPs and the gas they ionize in isolated rotational systems, this last paper of the series analyses the 
stellar (synthetic model line absorption) and ionized-gas (H$\alpha$ line emission) velocity dispersion distributions between star-forming spaxels of control and 
perturbed galaxies. Similarly, as in Papers I and II, we compare our subsamples or galaxy types by only treating star-forming spaxels. For this third analysis, 
however, the comparisons among such spaxels are based on their rest-framed velocity values. We remark that we do not enter into the details of the kinematic orientations 
by mapping each sampled galaxy as a whole. For that, we recommend \citet{BaBa14,BaBa15a}. We expect to find more disordered or disturbed kinematics (stellar, gaseous, or 
both) in star-forming regions of perturbed galaxies. For a good end, we use:

\begin{enumerate}
 \item CALIFA survey Integral Field Spectroscopy (IFS). 
 \item Detailed spectral synthesis techniques that fit stellar components modelled by Simple Stellar Populations\footnote{The models of SSPs that we use are those of 
 \citet{BrCh03}. The libraries of base spectra are those of MILES \citep{SaB06,Fac11}. Finally, we use the Initial Mass Function (IMF) of \citet{Chab03}.} \citep[SSPs, 
 \textit{e.g.}][]{Cid05,Cid07,Asa07}. 
 
\end{enumerate}

We order this paper as follows. The reliability of our kinematic measurements and how we determine their uncertainties are both described in Section~\ref{sec:met}. 
We present the stellar and ionized-gas kinematic distributions in Section~\ref{sec:res}. We discuss the comparisons of our results in Section~\ref{sec:dis}. We state 
our summary and conclusions in Section \ref{sec:conc}.

In this paper, we use a cosmological set of $\mathrm{H_{0}}\,=72\,\mathrm{h_{72}^{-1}}\,\mathrm{km\,s^{-1}\,Mpc^{-1}}$, $\Omega_\mathrm{M}\,=\,0.3$ and $\Omega_\mathrm{\Lambda}\,=\,0.7$. 
In some cases, we present our data distributions as boxplots.\footnote{A box-and-whisker plot (boxplot) shows the median trend (box midline) and general variability 
(spread of the box, \textit{i.e.}, the interquartile range) of a data distribution by depicting its quartiles.} Finally, we use a level of significance of 0.05 for the 
statistics.

\section{Methods}
\label{sec:met}

\begin{figure*}\centering
   \mbox{\includegraphics[width=.6925\columnwidth]{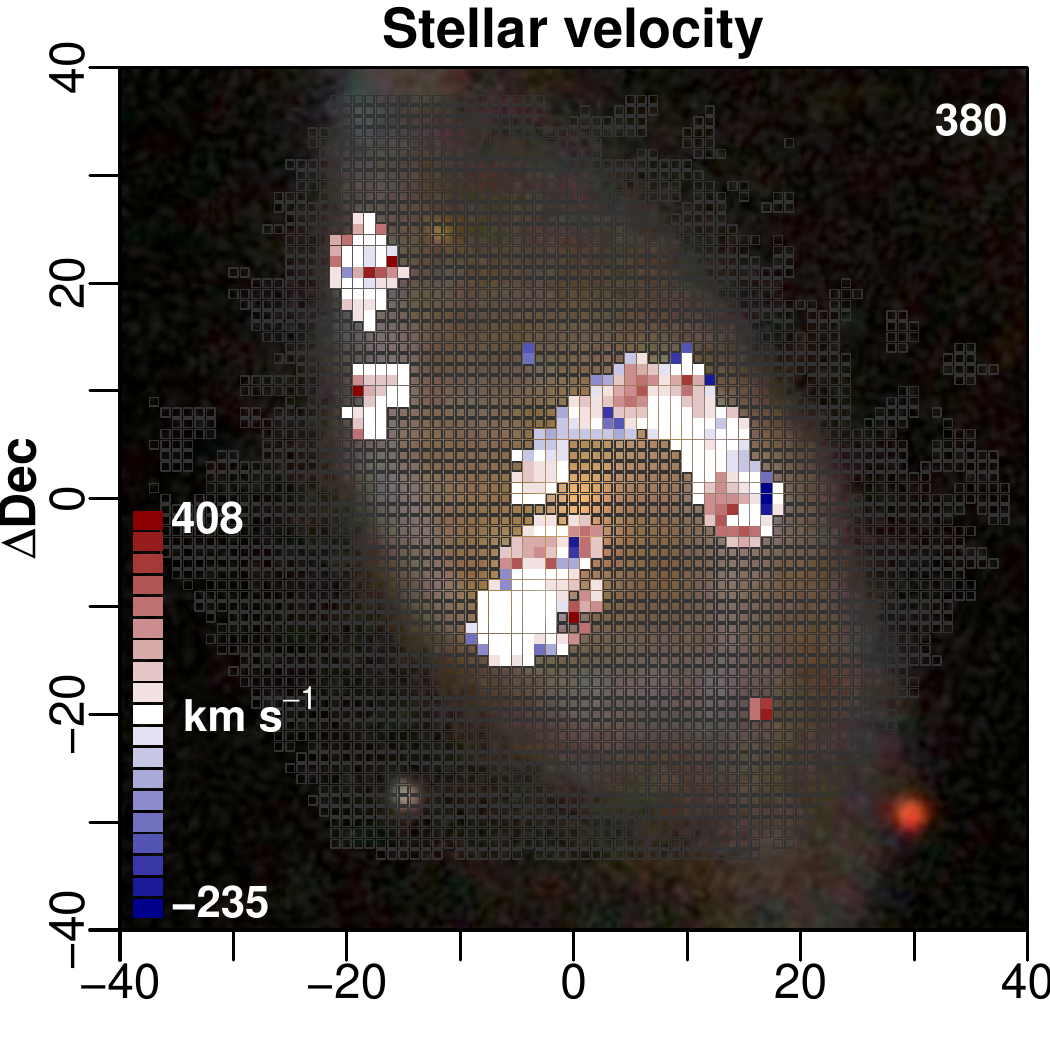}}
   \mbox{\includegraphics[width=.6925\columnwidth]{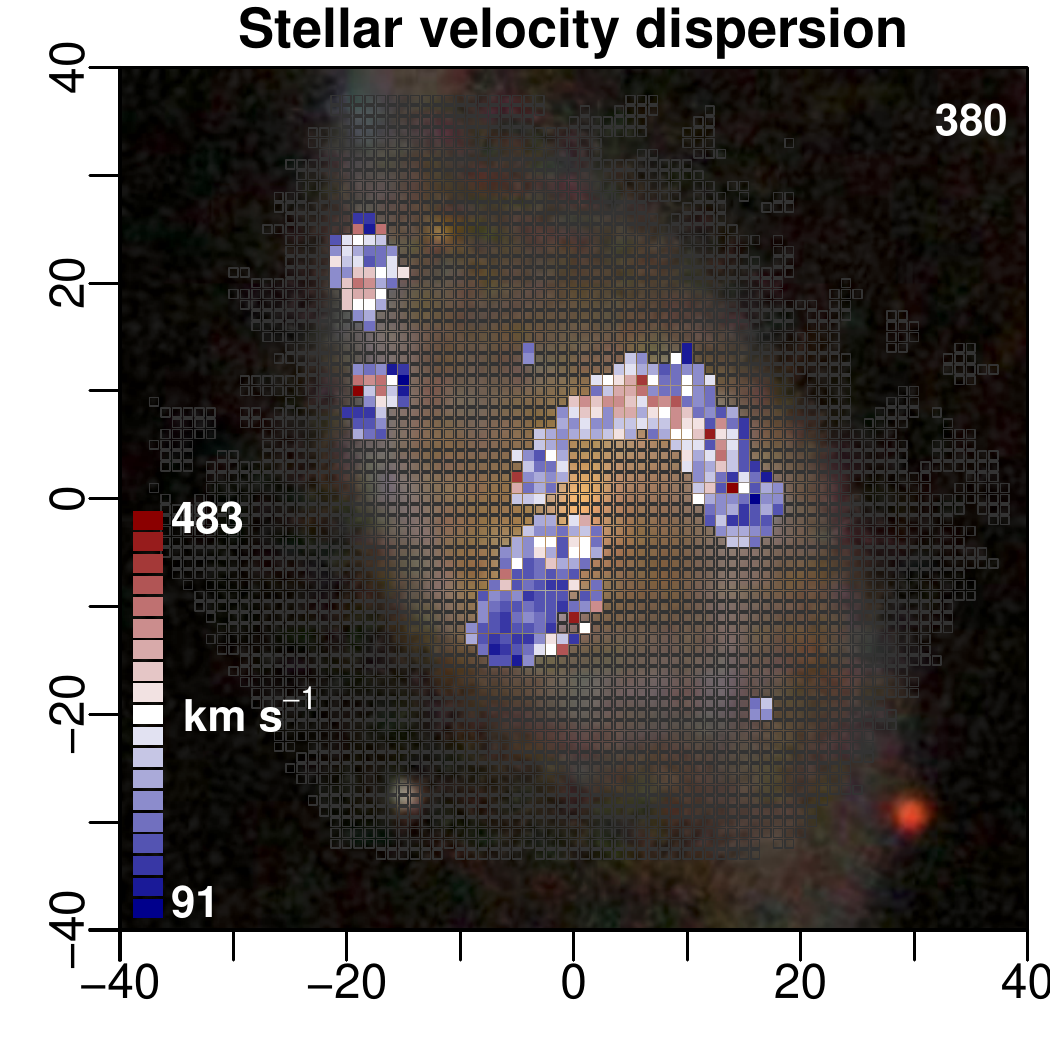}}
   \mbox{\includegraphics[width=.6925\columnwidth]{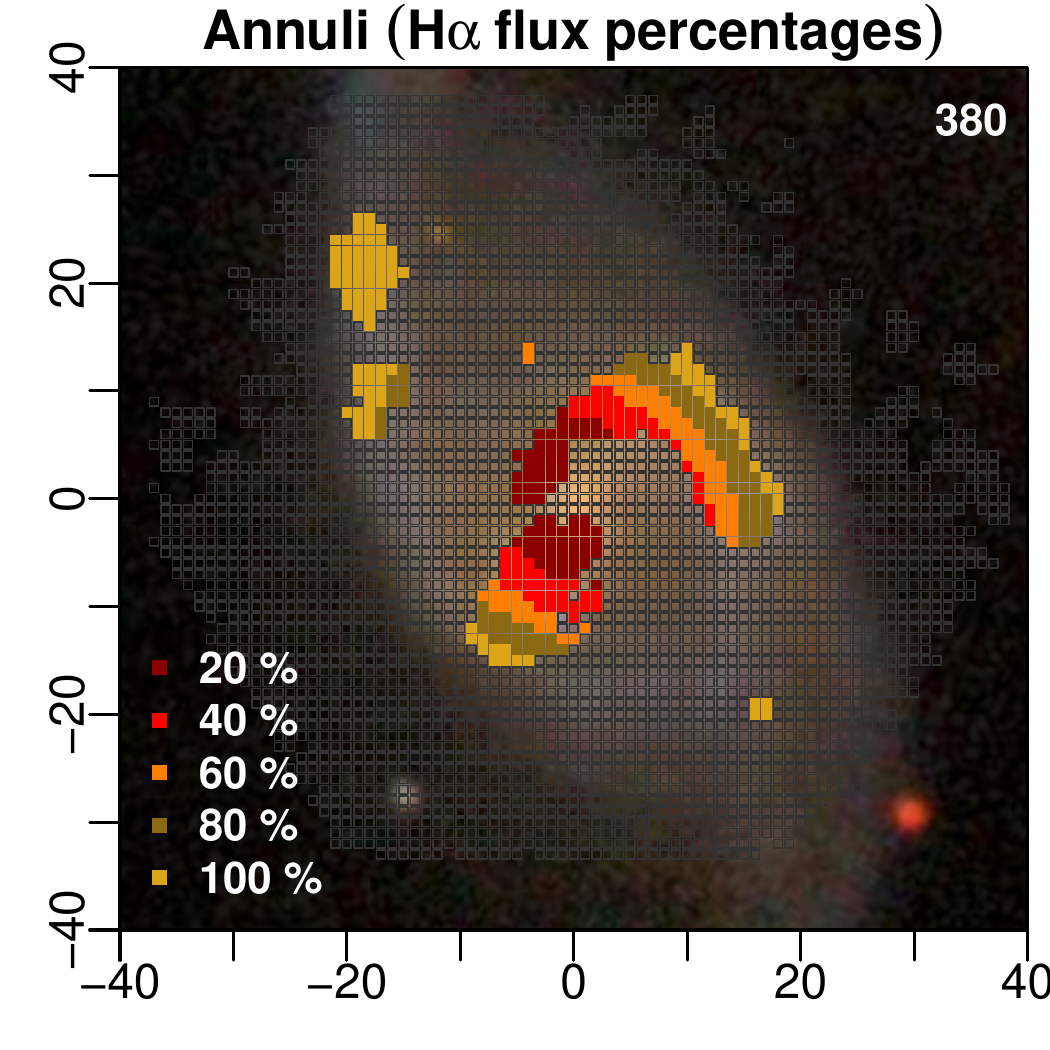}}\\
   \mbox{\includegraphics[width=.6925\columnwidth]{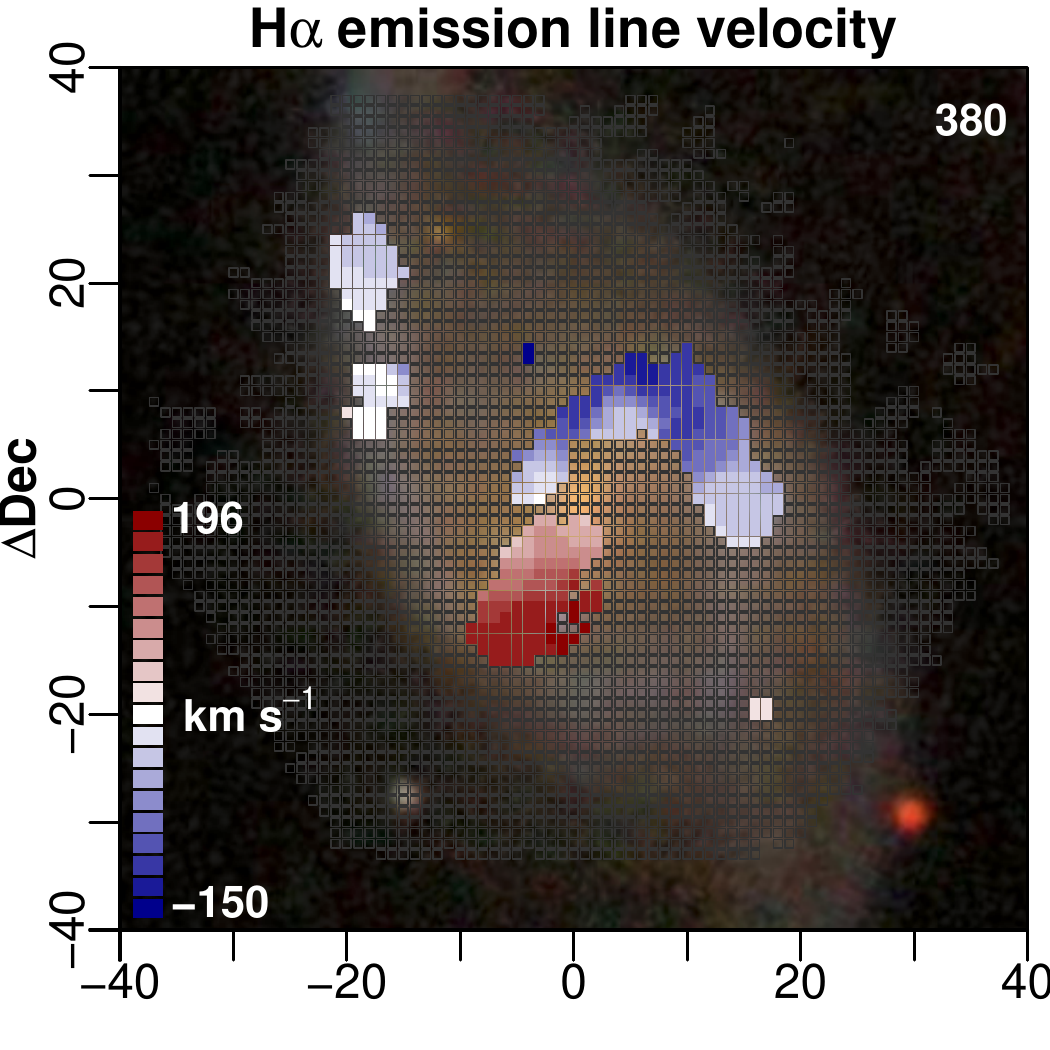}}
   \mbox{\includegraphics[width=.6925\columnwidth]{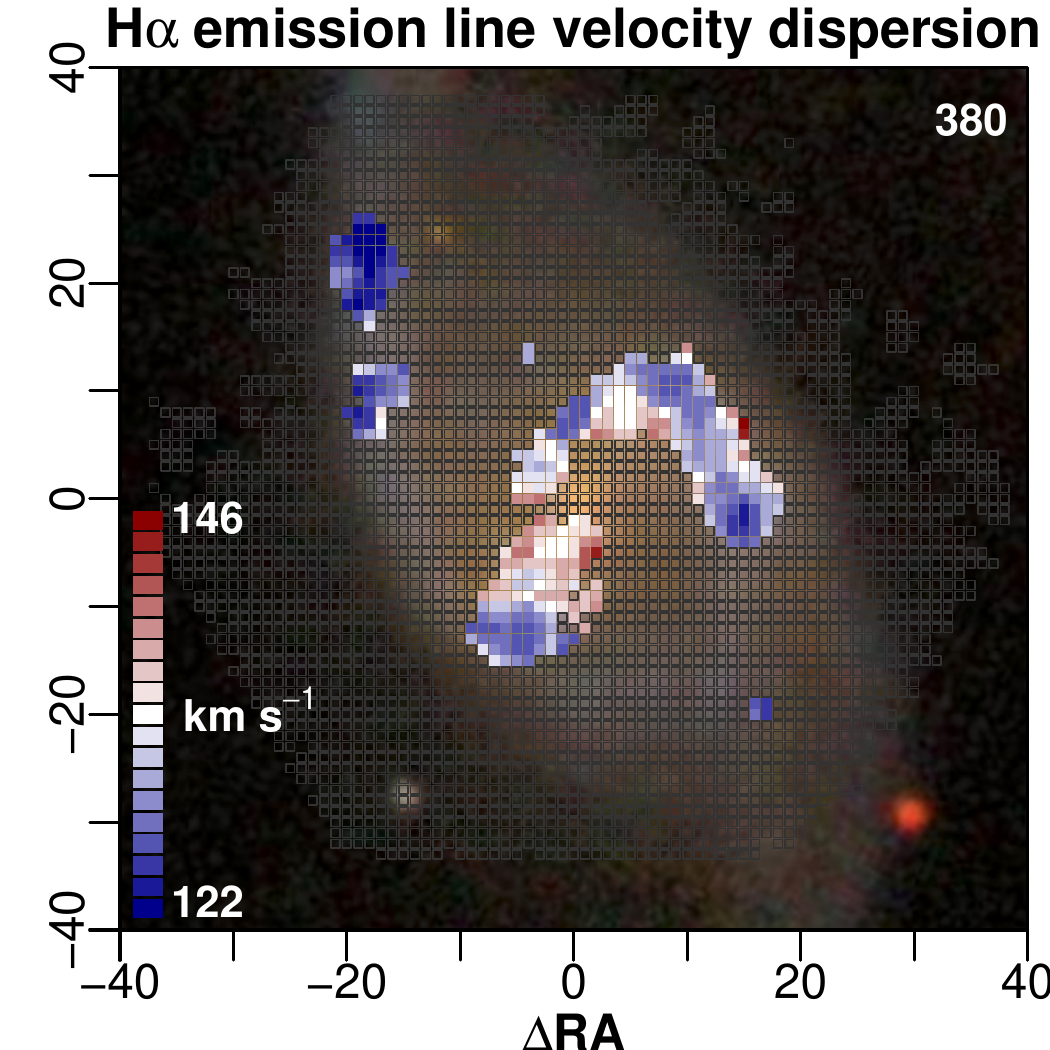}}
   \mbox{\includegraphics[width=.6925\columnwidth]{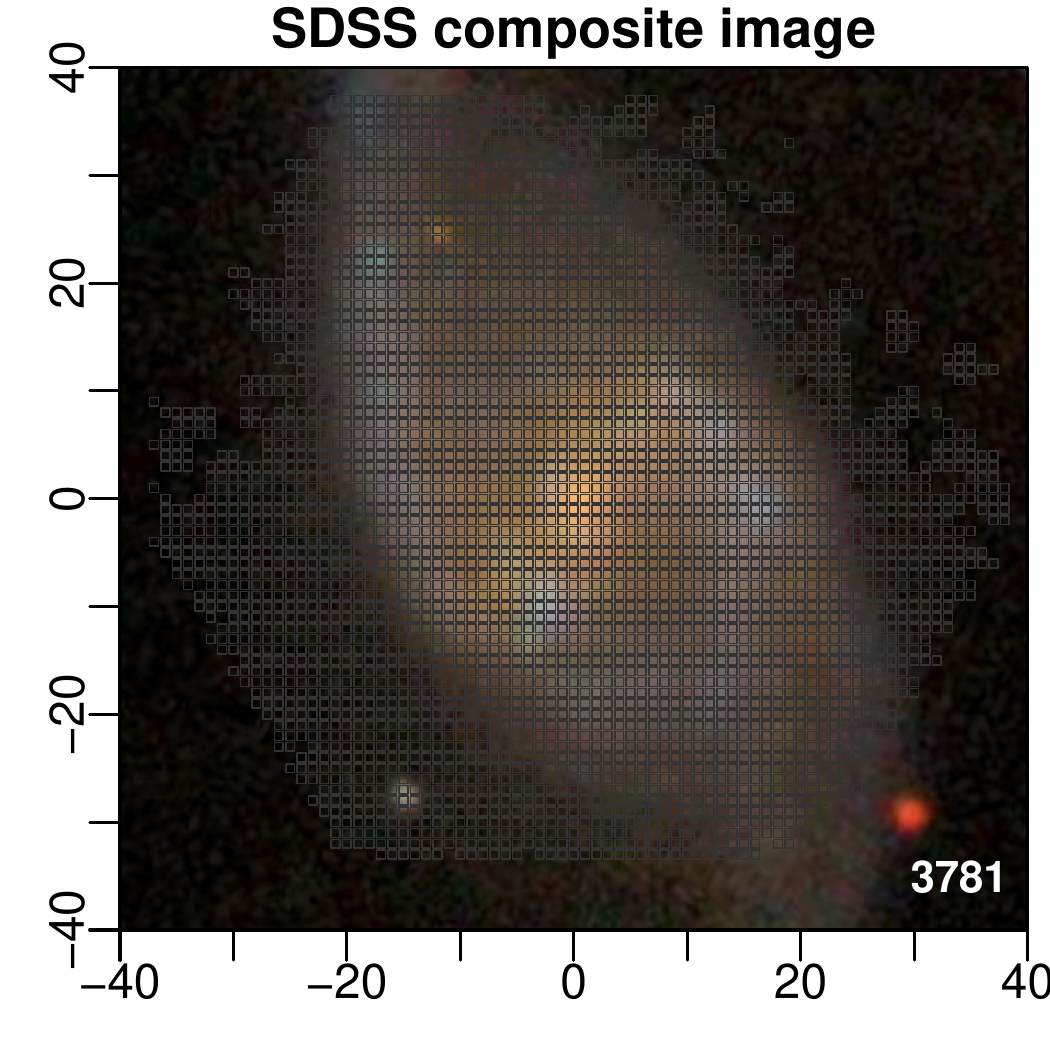}}
\caption{\scriptsize{Kinematics and annular sectioning for the star-forming spaxels selected as summarized in Section~\ref{subsec:SFregions}. An example is the perturbed 
galaxy NGC\,7549 (CALIFA survey id 901), which we classify as SFG, SFG LTS and SFG Red. Following Sections treat the kinematics (velocity dispersion). We deproject 
the spaxel positions, so five consecutive-outward annuli (20, 40, 60, 80 and 100\,\%, top-right), related to the H$\alpha$ emission line flux of each galaxy, indicate 
the radial extension. The totals of star-forming spaxels are at the top. An SDSS composite image shows the spaxels (spectra) solved by the \textsc{starlight} synthesis code of 
SSPs (see Section~\ref{subsec:SSPstellar}).}}
   \label{f0} 
\end{figure*}

\subsection{The star-forming regions}
\label{subsec:SFregions}

Our control and perturbed galaxies are biased to spiral morphologies (only 2\,\%, per sample and overall, are classified as ellipticals or lenticulars, see Table \ref{tab:A1}). 
That is a consequence of using only emission-line galaxies with star-forming regions, \textit{i.e.}, spaxels where the gas is prevailingly excited by massive young stars 
(typical excitation in star-forming galaxies, SFGs). Firstly, these star-forming spaxels are chosen based on the detectability (a line S/N\,$\geq$\,3) and intensity (an 
EW (H$\alpha$) cut-off of $\geq$\,6\,\AA{}\footnote{This cut is crucial to ensure the presence of prevailingly young-massive-ionizing stars. Lower EWs (H$\alpha$) very 
likely characterize regions where SF has started to cease \citep[\textit{e.g.}][]{San14}.}) of the H$\alpha$ emission line. Secondly, we select the spaxels according to 
their position in diagrams that classify the origin of the dominant ionization of the gas. Only spaxels with SFG excitation are considered star-forming. We remark that the 
recombination lines H$\alpha$ (the SFR tracer) and H$\beta$ (for extinction corrections) are subject to strict line criteria, whereas only flux limits restrict forbidden 
lines. The association with harder excitation of the latter, different from SFG, use to reduce the numbers of star-forming spaxels in some galaxy types (see Paper I, 
section 2.3.1). Figure~\ref{f0} shows the resulting star-forming spaxels for an example galaxy.

To avoid observational constraints likely present in IFS surveys, we get rid of star-forming spaxels with log$_{10}\,\Sigma_{*}\,<\,$7.5\,M$_{\odot}\,$kpc$^{-2}$ (see 
section 4.1, Paper I). This cut makes the spaxels in this analysis not to flatten the resolved SFMS (see \citealt{Can19} and references therein).

\subsection{The galaxy subsamples \& annular sectioning}
\label{subsec:samp-prof}

These are general notes on the conduction of the present analysis and the presentation of its results. If needed, Paper I provides much proper complete descriptions. For 
instance, sections 3.3 and 3.4 treat the definition of the samples and the comparisons of some of their properties.

A division on our samples of control and perturbed galaxies defines our subsamples or galaxy types. The base of such division is the gas excitation source that generally 
characterizes each galaxy, whether SFG or due to an active galactic nucleus (AGN-like). We have considered how misleading the diagnostics of gas excitation sources may be 
on spatially-resolved data (for example, shocks may resemble hard excitation from an AGN). The general dominant type of activity in our sampled galaxies is SFG excitation 
(SFGs). To make detailed comparisons among SFGs, we divide them by morphological group, either early-type (ETS) or late-type (LTS) spirals. Apart, we divide SFGs according 
to their photometric position on a colour-magnitude diagram, either along the red sequence (SFG Red), within the green valley (SFG Green) or blue cloud (SFG Blue). For these 
subsamples, we remark on the following. First, we have quantified the presence of galaxies and their contributions of star-forming spaxels. Based on that, we determined that 
these subsamples are not redundant. Secondly, these subsamples have a certain degree of overlapped galaxies, \textit{i.e.}, each one is found in three subsamples at a time 
(either by gas excitation source, morphological group or colour). Appendix \ref{sec:app1} gives further details on these remarks. 

Instead of using photometric metrics to define our annular sectioning, we show the radial extension using five concentric annuli where the positions of the spaxels 
have been deprojected.\footnote{We assume a disk-like component for deprojected galaxies (see Table A1, Paper I). For them, we obtain inclination angles from their 
semiminor-to-semimajor axis ratios. Comparing each angle with 0 deg (parallel to the plane of the sky), we deproject the positions of their spaxels in the direction 
of their semiminor axes by assuming they should be as extended as their semimajor ones.} Each annulus encircles a percentage (20, 40, 60, 80 and 100\,\%, see Fig. 
\ref{f0} for an example) of the H$\alpha$ emission line flux per galaxy.

In line with tidal parameter estimations in the CALIFA survey, the frequency of perturbed galaxies per each control one is $>$\,3. We hence include as many perturbed 
galaxies as possible by showing, troughout this paper series, the trend of ten samples of perturbed galaxies combined (see Paper I, section 3.3). We do that combination 
by merging, \textit{i.e.}, placing all those star-forming spaxels in a pool and then comparing them with those of the control sample.

Previously in the series, we analysed properties related to  M$_{*}$ growth. This time, we treat properties inherent to resolved kinematics so we compare star-forming
spaxels of control and perturbed galaxies at their most alike LOS rest-framed velocities.

\begin{figure*}\centering
   \mbox{\includegraphics[width=.925\textwidth]{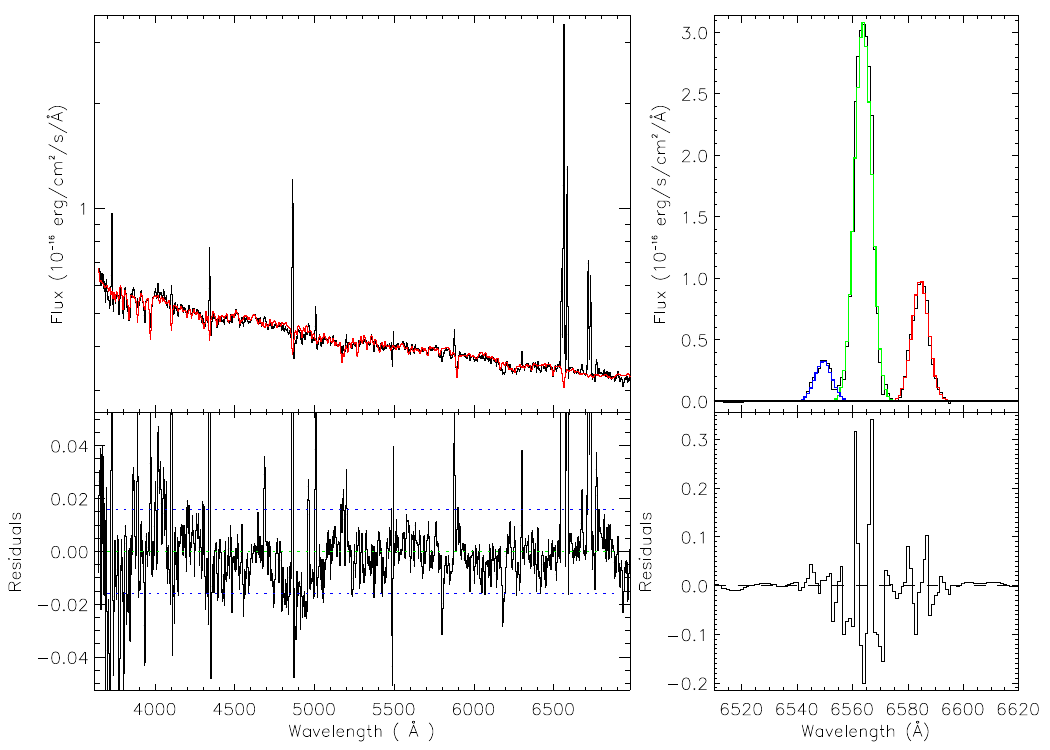}}
\caption{\scriptsize{Stellar component subtraction and line fitting procedures for a spectrum with SFG gas excitation (randomly selected from NGC\,7549). Left: top 
(y-axis in log$_{10}$ scale), observed spectrum (black) and \textsc{starlight} synthesis (red); bottom, the residual emission-line spectrum (blue-dashed lines at $\pm$1\,$\sigma\,=\,$0.16, 
based on the S/N of all the continuum window, 5075-5125\,\AA{}). Right: top, fitting performance for blended lines (\textit{e.g.} H$\alpha$-[\ion{N}{ii}]$\lambda\lambda$6548, 6584) 
and bottom, a close-up of the residual at the position of those lines.}}
   \label{f1} 
\end{figure*}

\subsection{Spectral setups}
\label{subsec:setups}

The CALIFA survey observations used two different and complementary instrumental setups: 1) V500, with a measured spectral resolution of $\sigma_{\mathrm{inst}}$ 
$\sim$ 139\,km\,s$^{-1}$; and 2) V1200, similarly, with a measurement of $\sigma_{\mathrm{inst}}$ $\sim$ 72\,km\,s$^{-1}$. Notice a difference in instrumental 
resolution between the two setups. We use spectra corresponding to the V500 one throughout this analysis (as in Papers I and II). Ideally, the V1200 setup is the 
proper one to obtain more accurate stellar and gaseous kinematics \citep[see][]{San12}. However, \citet{Fac17} compared their LOS stellar velocity dispersions 
($\sigma_{*}$) between both setups (see their figure 5) and reported systematic differences at dispersion values $\lesssim$\,100\,km\,s$^{-1}$. They also compared, 
between both setups, their LOS stellar velocities (v$_{*}$) and found they are well within the uncertainties.

\subsection{Synthesis of SSPs: the derived kinematics}
\label{subsec:SSPstellar}

Our stellar and gaseous kinematics are along the LOS and rest-frame-like (see Fig.~\ref{f0} for an example). According to the population synthesis method used by the 
\textsc{starlight} code \citep{Cid05}, a velocity \textit{shift}, $\mathrm{v_{*}-v_{0}}$ (km\,s$^{-1}$) centred in a Gaussian distribution, models the stellar kinematics. 
$\mathrm{v_{*}}$ represents any velocity, whereas $\mathrm{v_{0}}$ is strictly the nuclear value. De-redshifting is done once the difference is computed. 
This shift\footnote{The velocity shift is a fine tunning for slight errors in our de-redshifting process. The best de-redshifted spectra, \textit{i.e.} those with the best 
S/N ratios, will result in small velocity shifts ($\pm$5-20\,km\,s$^{-1}$).} in velocity should be applied to the model to match the observed spectrum best. It should be 
close to zero because the spectra are rest-framed. We also introduce v$_{\mathrm{d}}$, which is the width (in km\,s$^{-1}$) of the Gaussian filter \textsc{starlight} 
applied to the model to make it better match the observations. It is important to remark that our observed and base spectra (MILES, see Section~\ref{sec:int}) have both 
the same resolutions ($\sim$ 139\,km\,s$^{-1}$) so that v$_{\mathrm{d}}$ well approximates the velocity dispersion ($\sigma_{*}$).\footnote{To approximate v$_{\mathrm{d}}$ 
to a proper velocity dispersion, we must fulfil $\sigma_{*}^{2}\,=\,\mathrm{v}_{\mathrm{d}}^{2}-\sigma_{\mathrm{inst}}^{2}+\sigma_{\mathrm{base}}^{2}$ (where 
$\sigma_{\mathrm{base}}$ is the resolution of the template spectra). This equation defines all variables as\,$>$\,0 while indicating that $\sigma_{\mathrm{inst}}$ should be 
kept low (reliable estimates are not expected when v$_{\mathrm{d}}\,<\,\sigma_{\mathrm{inst}}$ or v$_{\mathrm{d}}\,<\sigma_{\mathrm{base}}$, \textit{i.e.} when the true 
velocity dispersion is smaller). Since our cases of spectra with v$_{\mathrm{d}}\,\sim\,$0 are very few, \textsc{starlight} practically broadens each model spectrum to 
improve each fit.}

We also use \textsc{starlight} to subtract synthesized stellar components (\textit{i.e.}, the best-fit continuum models) and obtain nebular ones from the observed 
spectra. Besides, \textsc{starlight} offsets the effects of the underlying stellar absorption on the nebular component. We estimate emission-line features afterwards by 
fitting Gaussian profiles. These are widely used to approximate spectral features of emission-line gaseous components along the LOS. Figure \ref{f1} exemplifies the 
subtraction of the stellar component and the process of fitting Gaussian profiles. We apply this process to fit each nebular line to extract fluxes and apparent (observed) 
central wavelengths.\footnote{Central wavelength, amplitude, and associated dispersion are variable parameters along the fitting iterations. These stop once they find the 
minimum $\chi^{2}$ residual between the observed line and the profile.} Interested in the H$\alpha$ emission line, we adapt several profiles since the line is 
generally blended, both physically and spectroscopically (see Fig. \ref{f1}) along the LOS, with other gaseous components. The line-fitting routine calculates the line 
offset or wavelength displacement with respect to the rest-frame central wavelength (6562.8 \AA{} in this study). Then the gaseous velocity shift, $\mathrm{v_{ion}-v_{0}}$ 
(km\,s$^{-1}$), is computed based on the Doppler Shift (any displacement with the rest-frame central wavelength is proportional to the shift in velocity). Moreover, the 
routine estimates the corresponding dispersion for each profile (one standard deviation from a vertical line passing through the central wavelength). Such a dispersion is 
turned into velocity by the Doppler Shift equation.

\subsubsection{Uncertainties of the velocity dispersion}
\label{subsubsec:Uncer}

\textsc{starlight} neither attempts to obtain the probability distribution function of the parameters it calculates nor to estimate their errors. \textsc{starlight} 
seeks the single best solutions, and when it achieves the desired convergence (a $\chi^{2}$ minimization that uses the Metropolis algorithm), it simply stops. Therefore, 
we do not possess uncertainty measurements from the population synthesis. We set such measurements regarding the dispersion of velocities as follows. For the stellar 
component, we compute the interquartile range (IQR) of the distribution of $\sigma_{*}$ in each galaxy so that each of its star-forming spaxels adopts the same dispersion 
the IQR represents as their common error. 

\begin{figure}\centering
   \mbox{\includegraphics[width=.7\columnwidth]{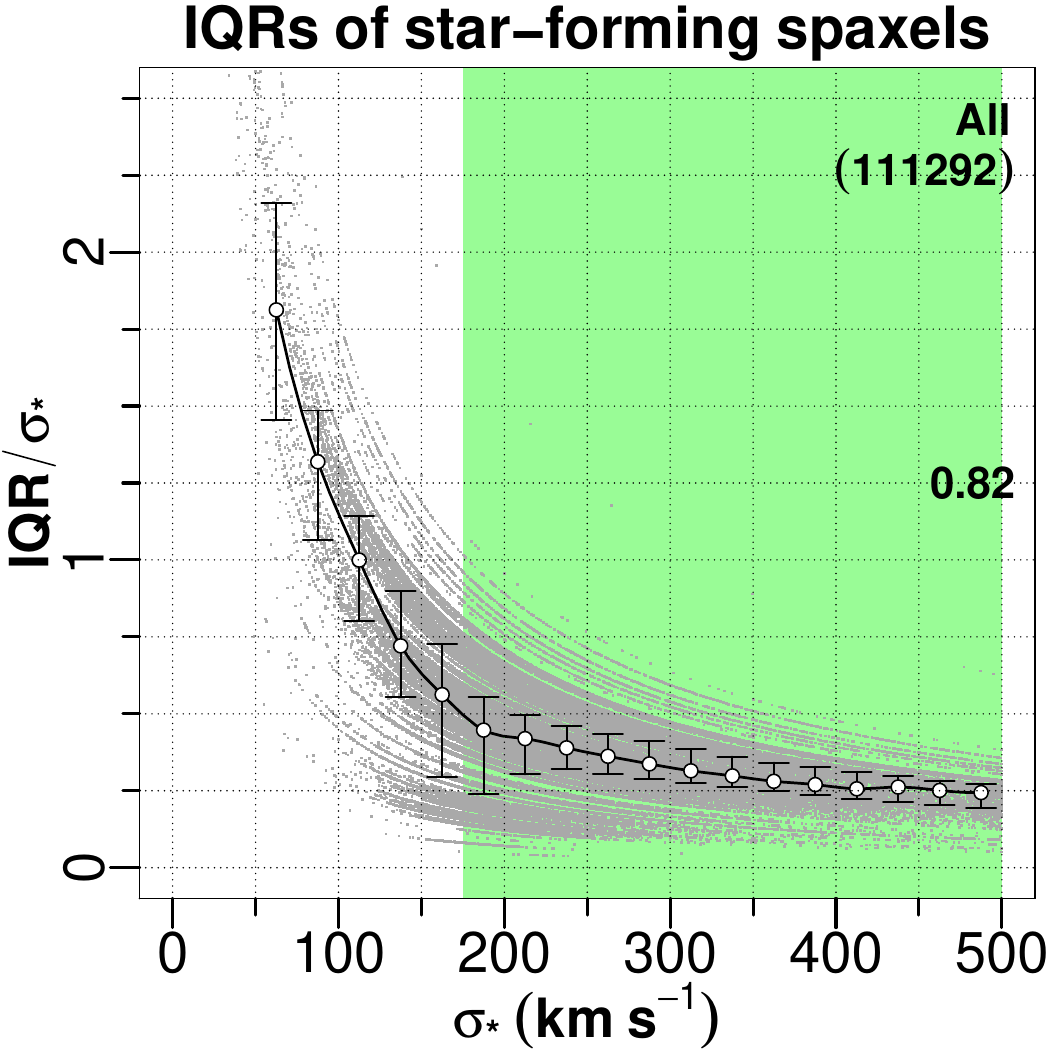}}
   \mbox{\includegraphics[width=.7\columnwidth]{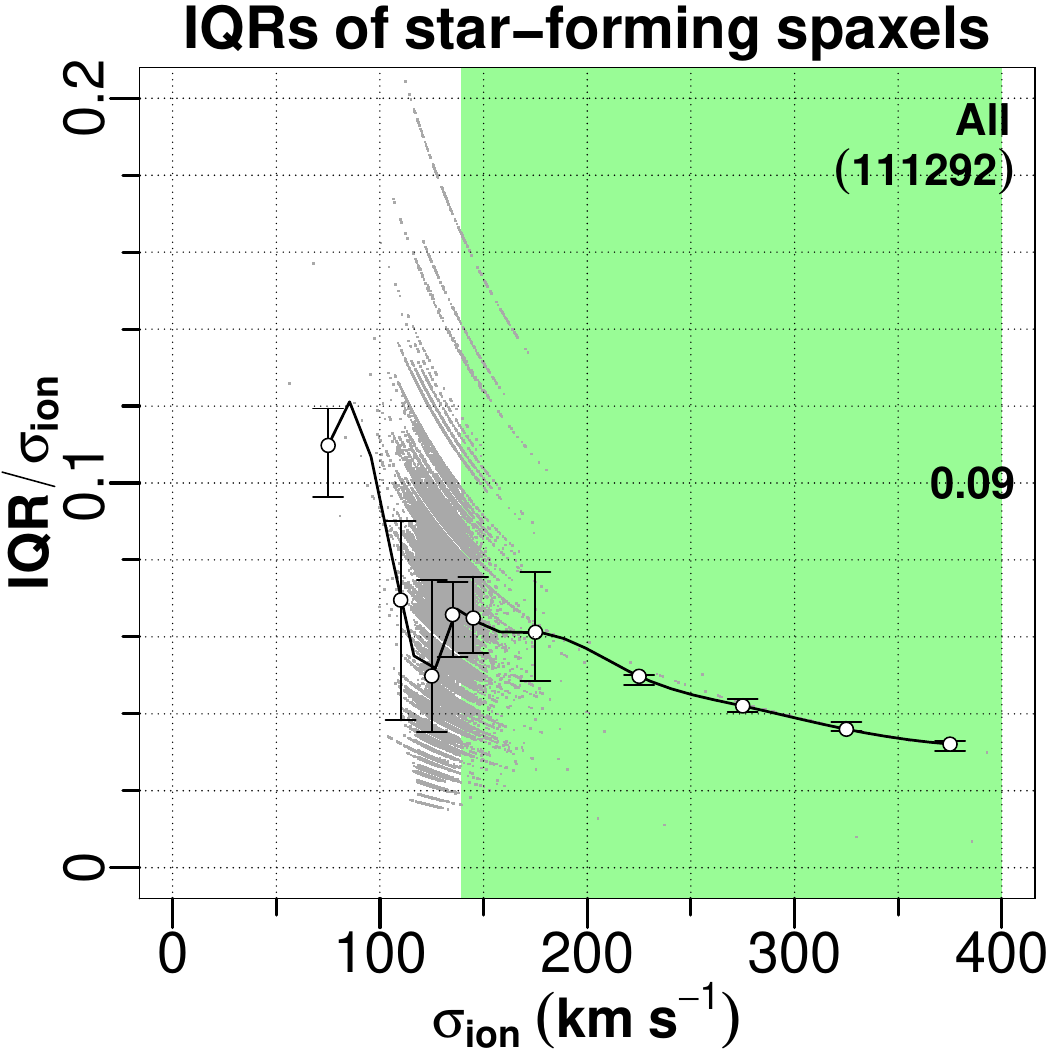}}
\caption{\scriptsize{Adopted uncertainties for our velocity dispersions. We compare the interquartile ranges (IQRs) of the distributions of the sampled galaxies with the 
velocity dispersions of each of their star-forming spaxels. \textit{Top}: we bin the $\sigma_{*}$ values of the spaxels every 25\,km\,s$^{-1}$. 
\textit{Bottom}: the $\sigma_{\mathrm{ion}}$ values of the spaxels are binned every 50\,km\,s$^{-1}$ except for the 100-150\,km\,s$^{-1}$ range, which is 
first binned in 20 and then in 10\,km\,s$^{-1}$ thrice. The open dots and error bars indicate the medians and IQRs of the uncertainties compared with the 
measurements (IQR/$\sigma$) in each bin. Spline interpolation gives the lines connecting each median. The green areas indicate the selected thresholds of reliability for 
the stars and ionized gas ($\geq$\,175 and $\geq$\,139\,km\,s$^{-1}$, respectively). Numbers at the mid-right are the fractions of star-forming spaxels 
to the total (111\,292) that satisfy the thresholds. That one for $\sigma_{\mathrm{ion}}$ reduces the frequency of present galaxies from 224 (see Paper I, 
table A1) to 208, \textit{i.e.}, 5 and 11 are missing for the control and perturbed galaxies, respectively (Table~\ref{tab:1} lists the net frequencies of present galaxies).}}
   \label{f2} 
\end{figure}

Figure~\ref{f2} (top) shows the ratio of the IQRs to the $\sigma_{*}$ of their respective star-forming spaxels against the $\sigma_{*}$ values themselves. We divide the 
$\sigma_{*}$ scale into bins (25\,km\,s$^{-1}$ each), and the open dots and error bars indicate the median and IQR of the ratio within each bin. For $\sigma_{*}\,\leq\,$100\,km\,s$^{-1}$, 
the adopted uncertainties of the spaxels exceed their measurements. To reduce this issue, we select a reliability threshold of $\sigma_{*}\,\geq\,$175\,km\,s$^{-1}$, which 
allows median uncertainties of up to 50\% (green area). Such a threshold permits us to consider as much data as possible (a fraction of $\sim$0.82 of the total of sampled 
star-forming spaxels).

Figure~\ref{f2} (bottom) shows a similar plot now for the velocity dispersion of the ionized gas (H$\alpha$ line emission). For spaxels with $\sigma_{\mathrm{ion}}\,>\,$100\,km\,s$^{-1}$,
the adopted uncertainties are lower than 8\,\% (median values). We remark that 99\,\% of the star-forming spaxels have values in the 100$\,\leq\,\sigma_{\mathrm{ion}}\,<$\,150\,km\,s$^{-1}$ 
range (that is why we bin this range into four, see Fig.~\ref{f2}). These low $\sigma_{\mathrm{ion}}$ values might be due to a relatively narrow range of velocities 
consequence of closeness in redshift.\footnote{More than 60\,\% of our sampled galaxies have redshifts in the 0.0085$\,\lesssim\,z\,\lesssim\,$0.0185 range.} Besides, 
H\,\textsc{ii}/star-forming regions show low-velocity dispersion since narrow emission lines are typical in their spectra \citep[\textit{e.g.}][]{Koo95}. Selecting thresholds 
of $\geq\,$139\,km\,s$^{-1}$, \textit{i.e.}, equal to our instrumental resolution or, for instance, below it, $\geq\,$130\,km\,s$^{-1}$ 
or $\geq\,$120\,km\,s$^{-1}$, implies fractions of star-forming spaxels to the total (see caption of Fig.~\ref{f2}) of 0.09, 0.45 and 0.95, respectively. Preferring data 
quality over data quantity, we select the first threshold ($\geq\,$139\,km\,s$^{-1}$, see the green area, Fig.~\ref{f2} bottom) no matter that the fraction of spaxels used 
gets significantly reduced.

\begin{figure*}\centering
   \mbox{\includegraphics[width=.6925\columnwidth]{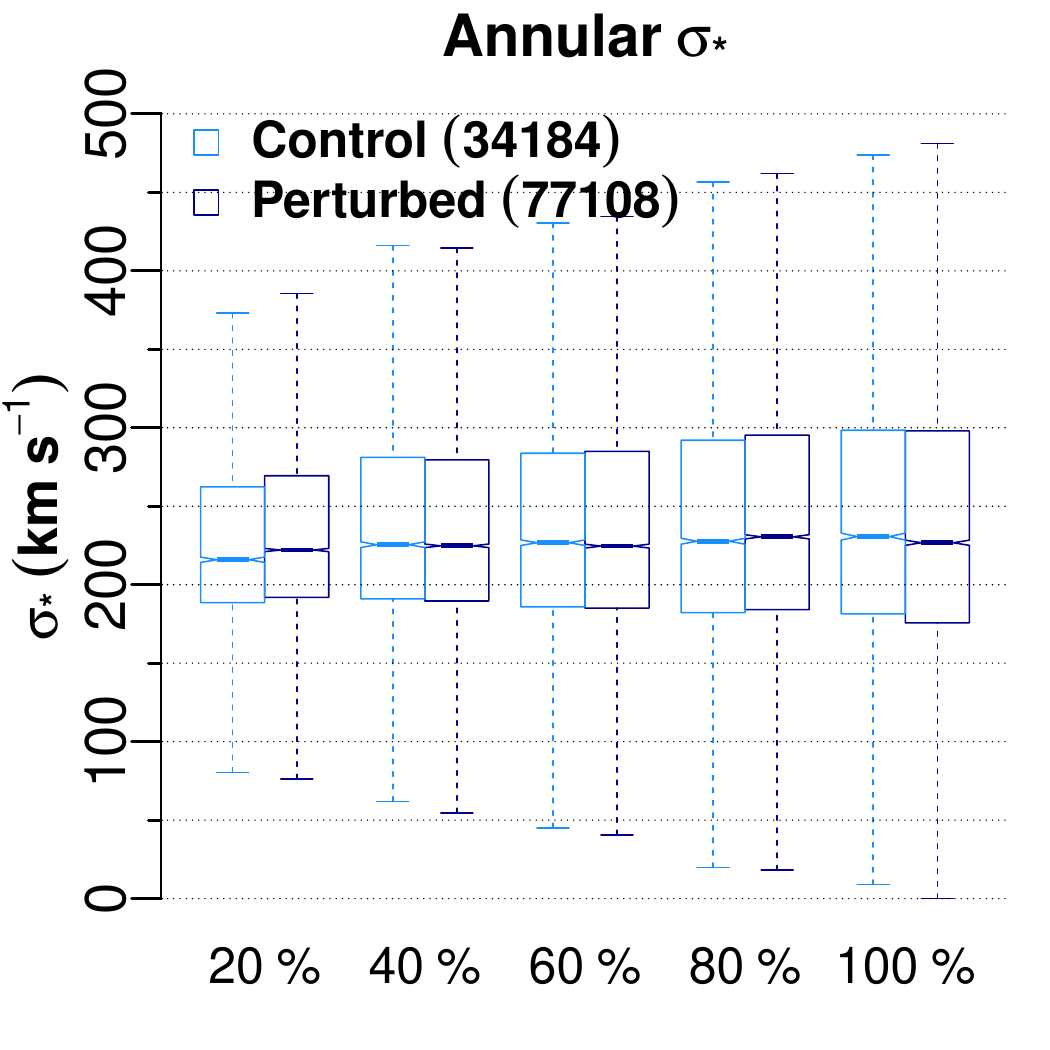}}
   \mbox{\includegraphics[width=.6925\columnwidth]{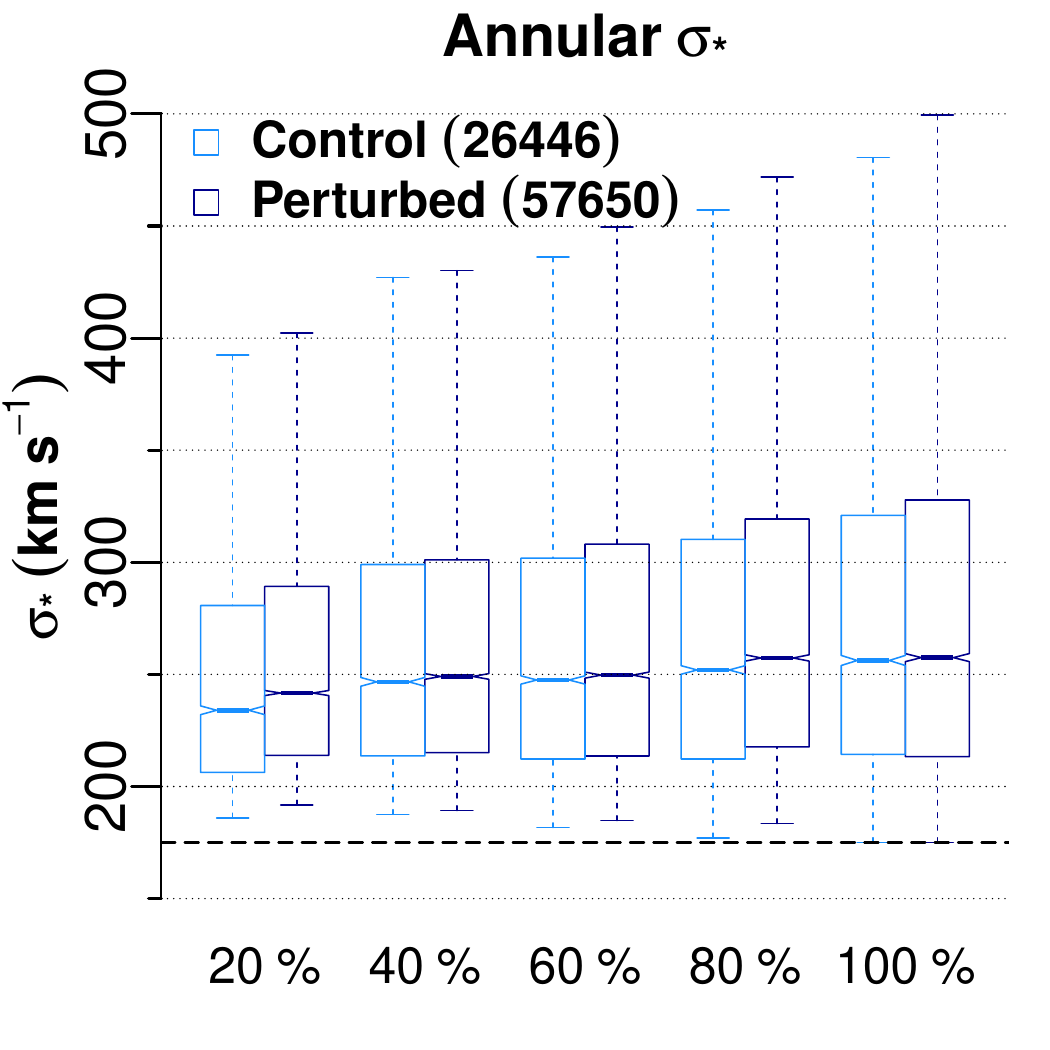}}
   \mbox{\includegraphics[width=.6925\columnwidth]{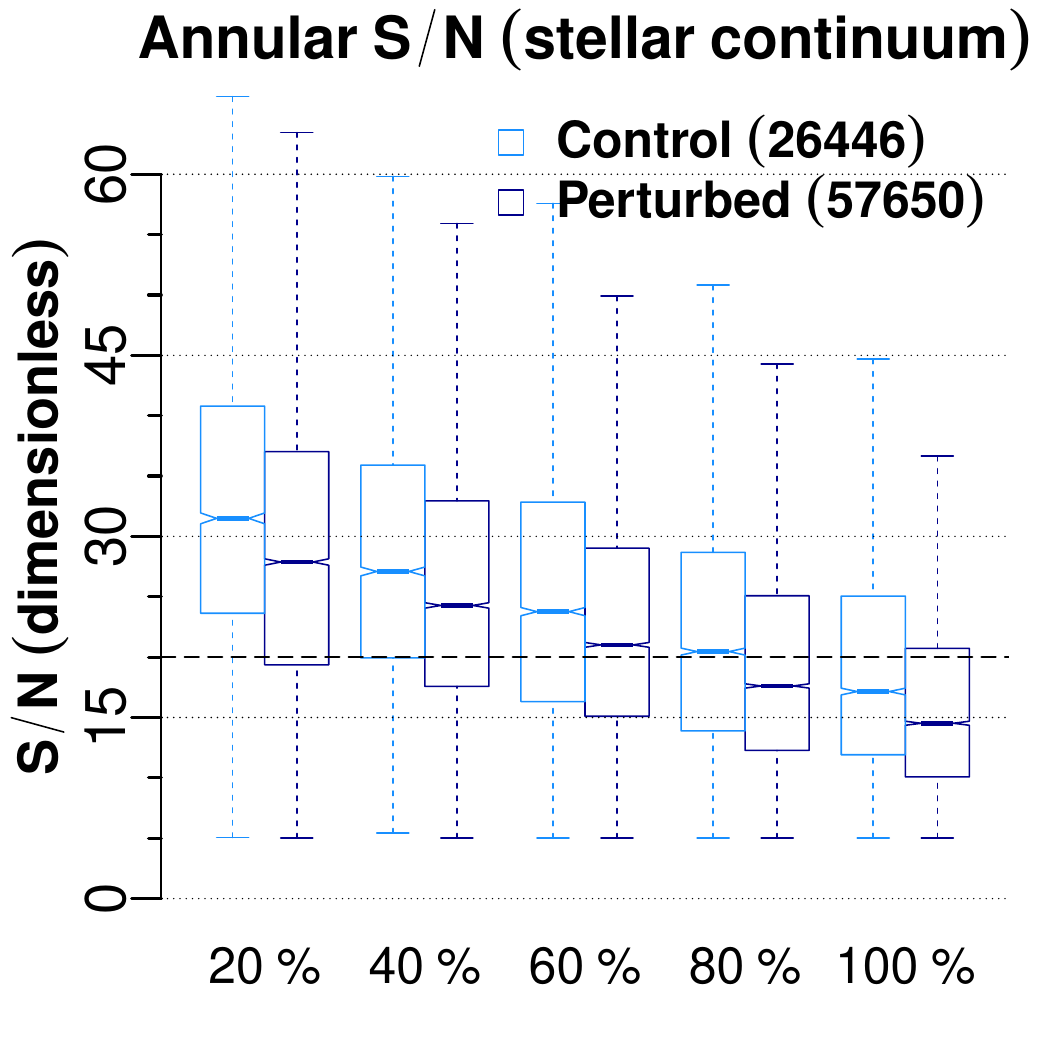}}\\
   \mbox{\includegraphics[width=.6925\columnwidth]{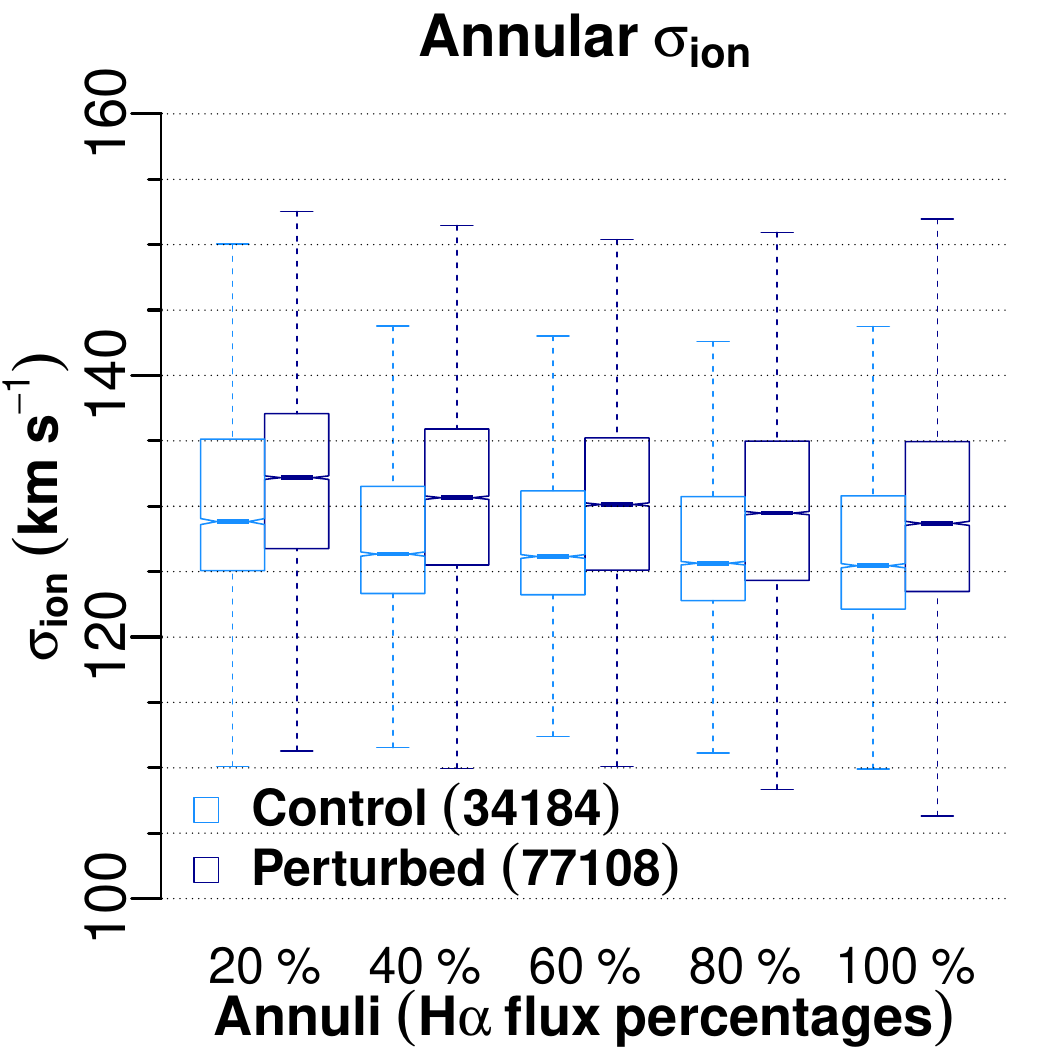}}
   \mbox{\includegraphics[width=.6925\columnwidth]{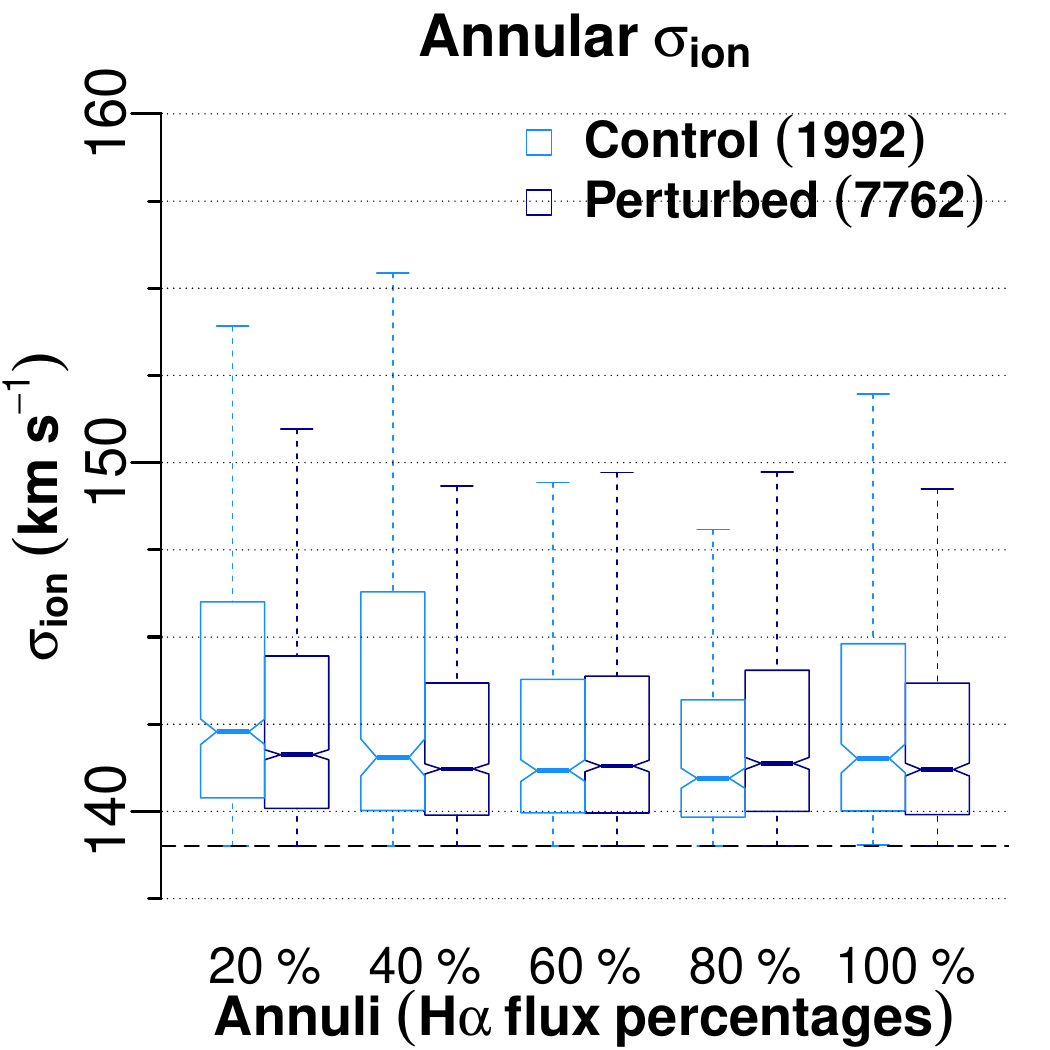}}
   \mbox{\includegraphics[width=.6925\columnwidth]{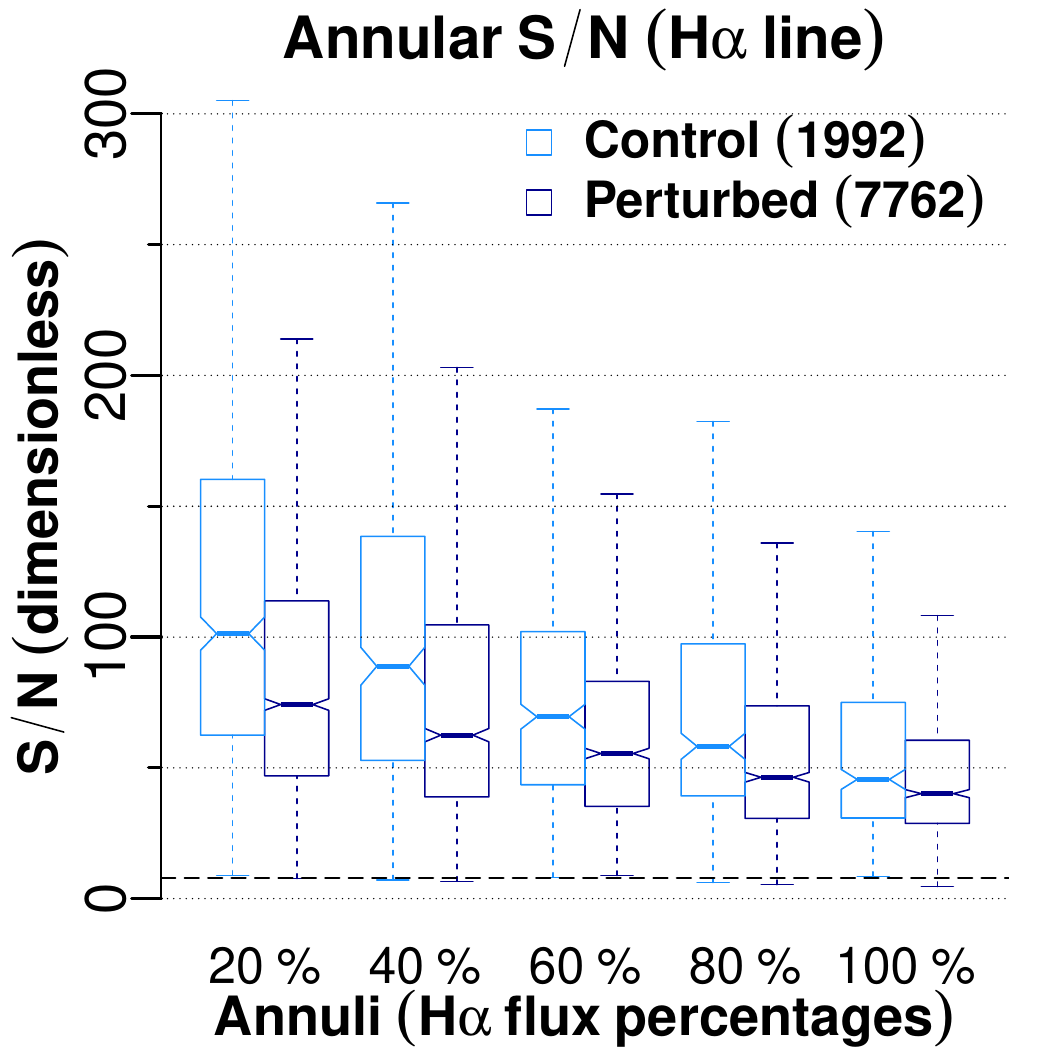}}
\caption{\scriptsize{Boxplots of our annular distributions of velocity dispersion and S/N ratio (stars and ionized gas, \textit{top} and \textit{bottom}, respectively). 
Five consecutive-outward annuli related to the H$\alpha$ flux of each galaxy represent the radial extension (see Section~\ref{subsec:samp-prof}). \textit{Left}: star-forming 
spaxels selected as summarized in Section~\ref{subsec:SFregions}. \textit{Middle}: remaining spaxels from trimming according to the reliability thresholds 
(175 and 139\,km\,s$^{-1}$, see Section~\ref{subsubsec:Uncer}) and as described in the text. \textit{Right}: S/N ratios of the same remaining spaxels (values by \textsc{starlight} 
within a stellar continuum window of 5075-5125\,\AA{}; and by our line fitting process described in Section~\ref{subsec:SSPstellar}). The parentheses give 
the frequencies of present spaxels. The dashed lines indicate our thresholds of velocity dispersion (\textit{middle}) and the S/N ratio thresholds used by \citet{Fac17} 
(S/N $\geq$\,20) and \citet{BaBa14,GarL15} (S/N\,$>$\,8) (\textit{right}).}}
   \label{f3} 
\end{figure*}

\subsection{Reliability of our kinematic measurements}
\label{subsec:relia}

\begin{figure*}\centering
   \mbox{\includegraphics[width=.693\columnwidth]{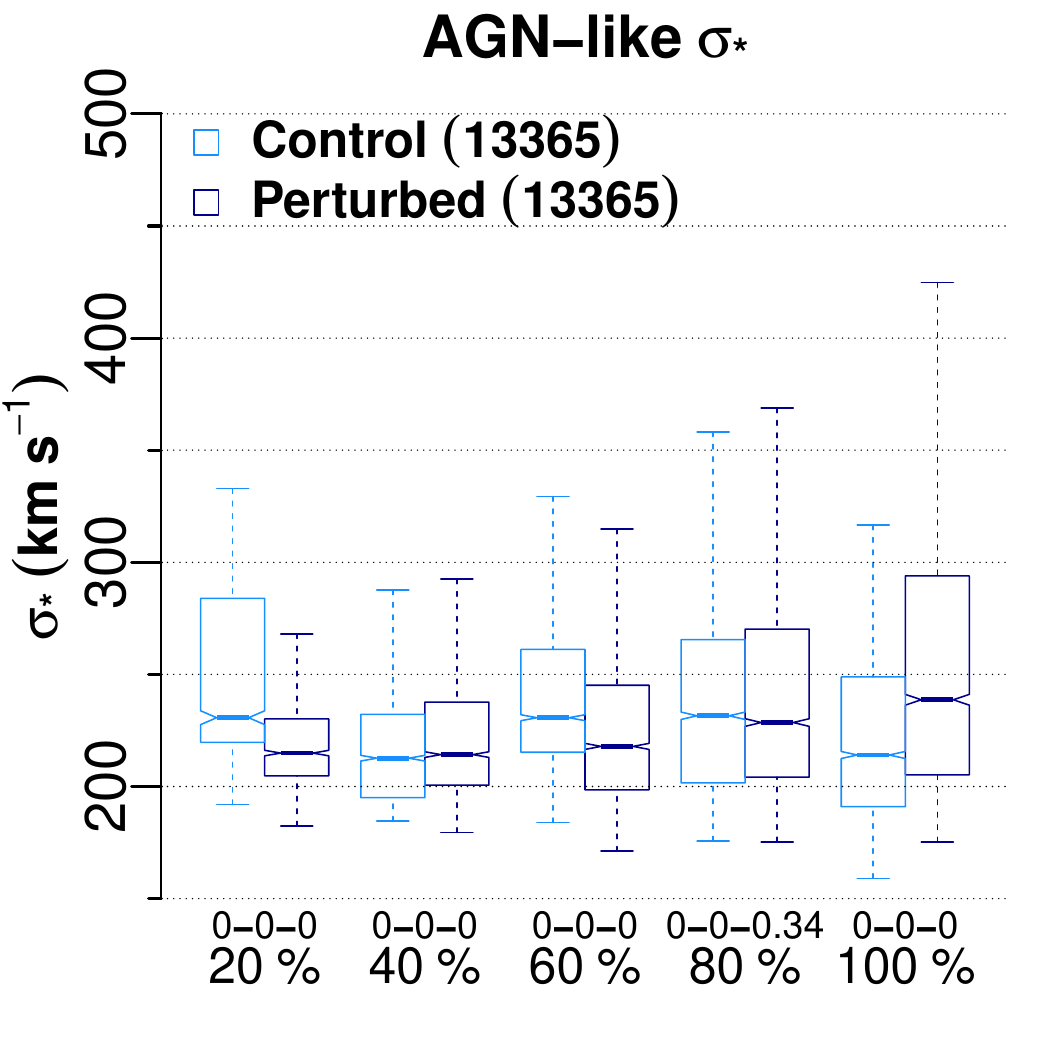}}
   \mbox{\includegraphics[width=.693\columnwidth]{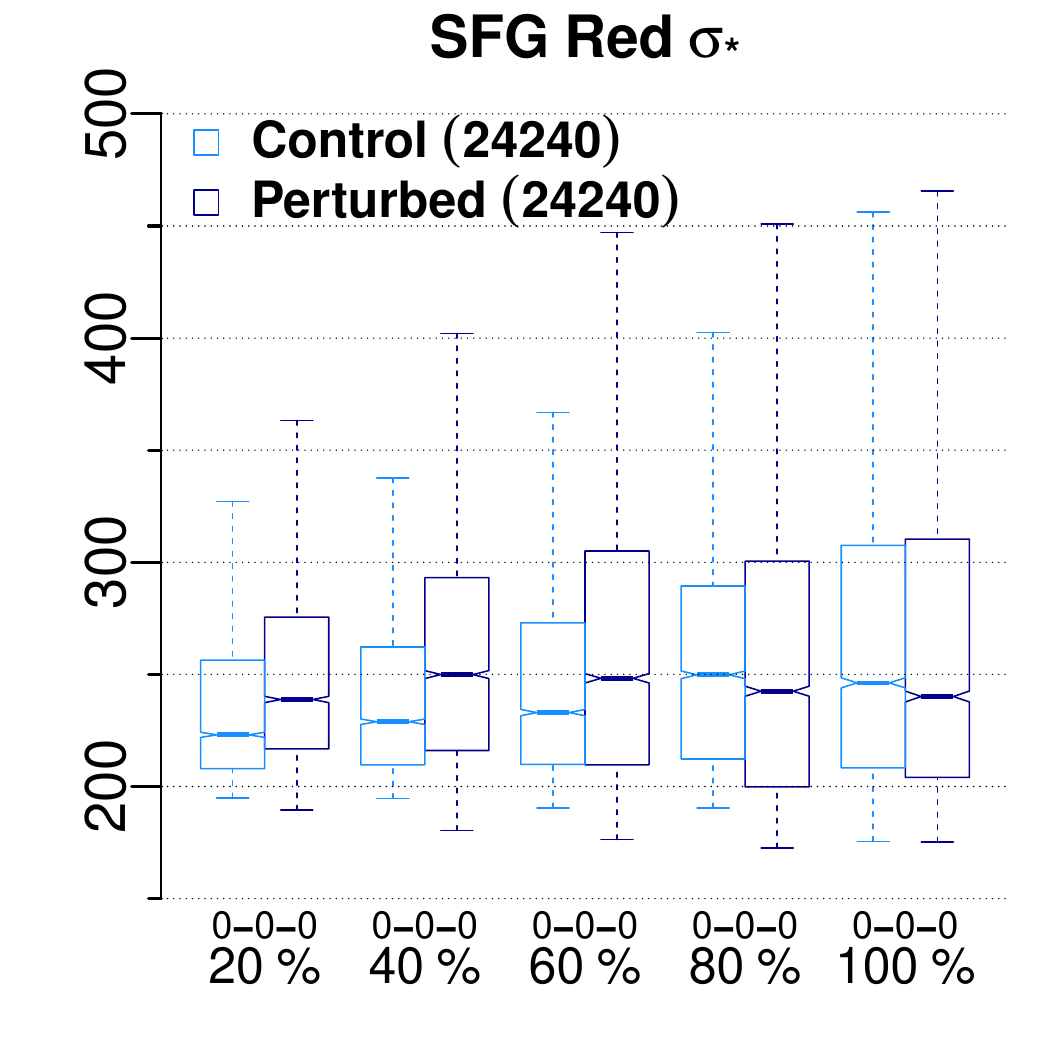}}
   \mbox{\includegraphics[width=.693\columnwidth]{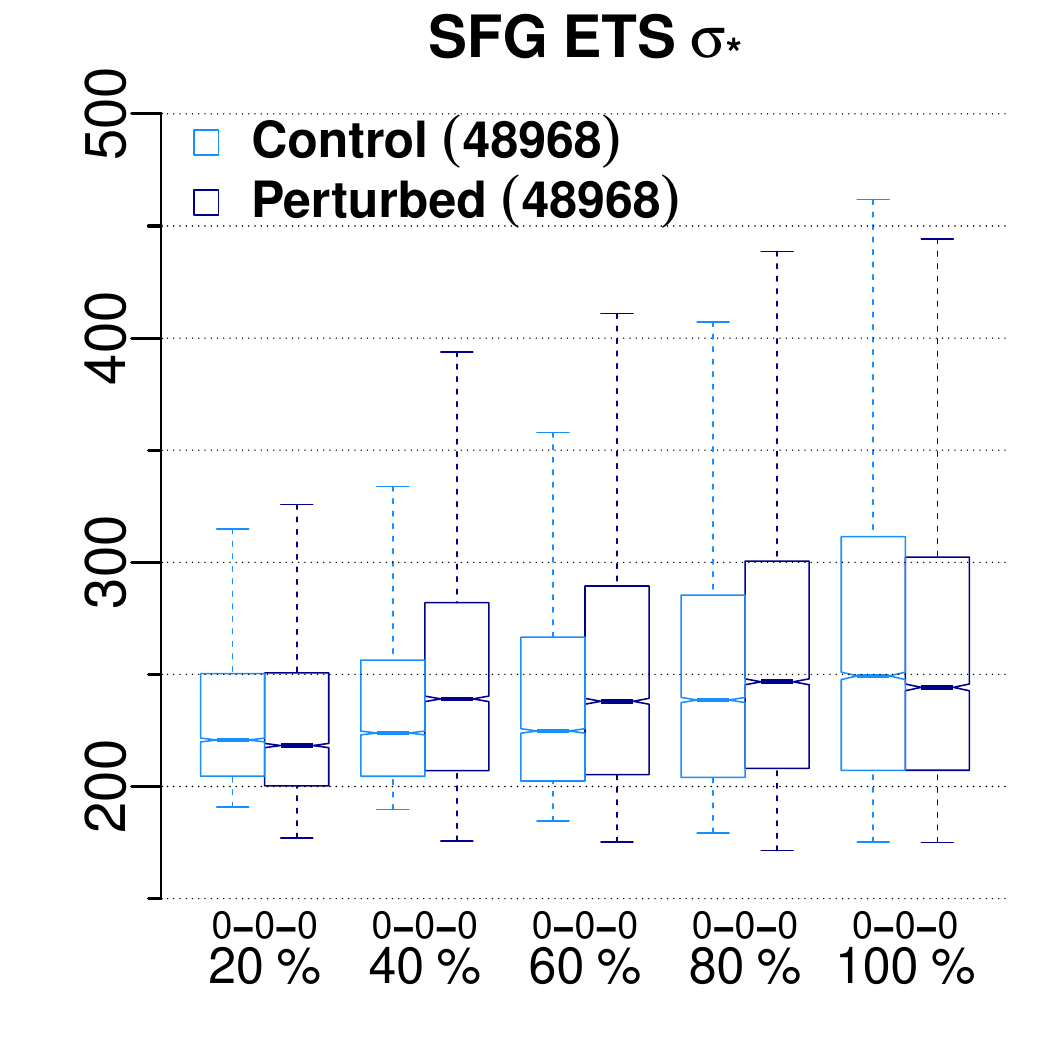}}\\
   \mbox{\includegraphics[width=.693\columnwidth]{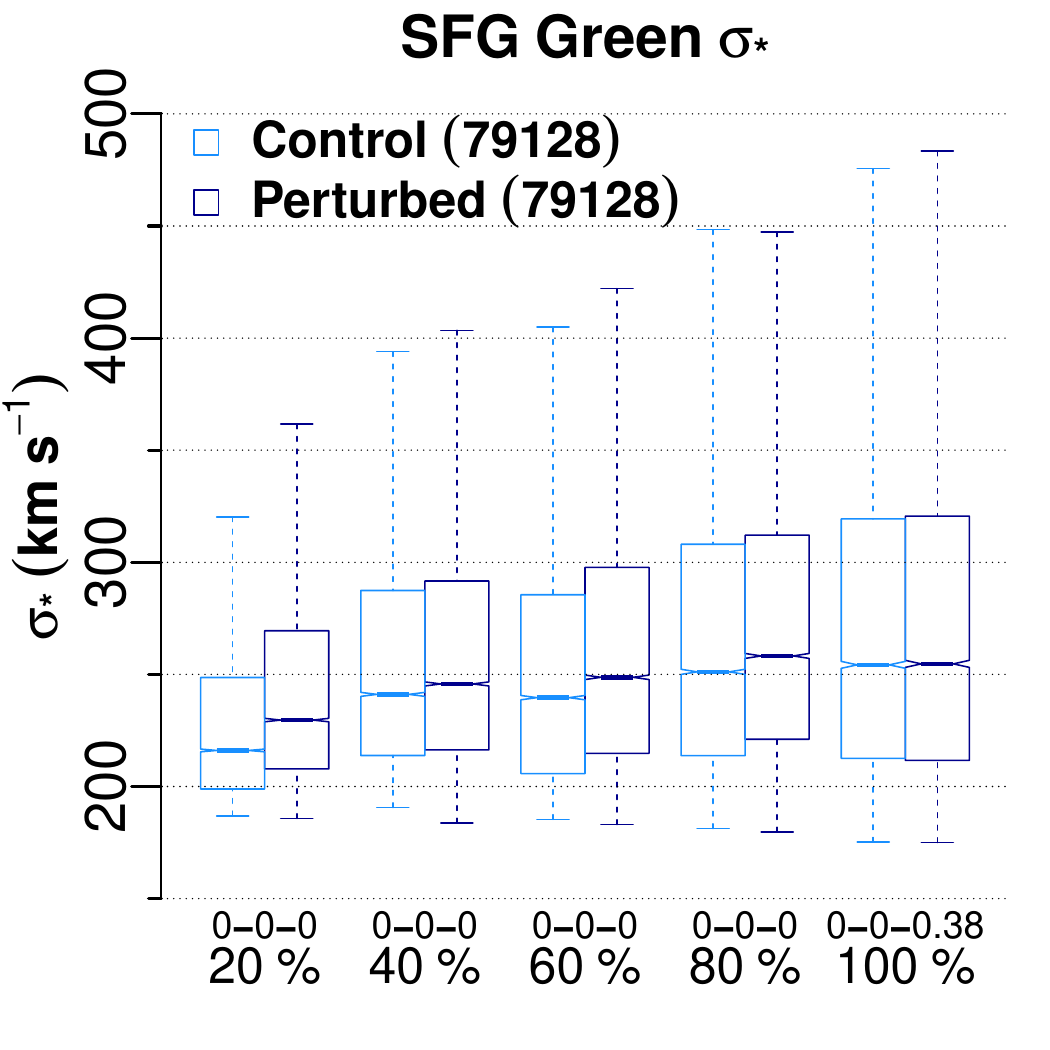}}
   \mbox{\includegraphics[width=.693\columnwidth]{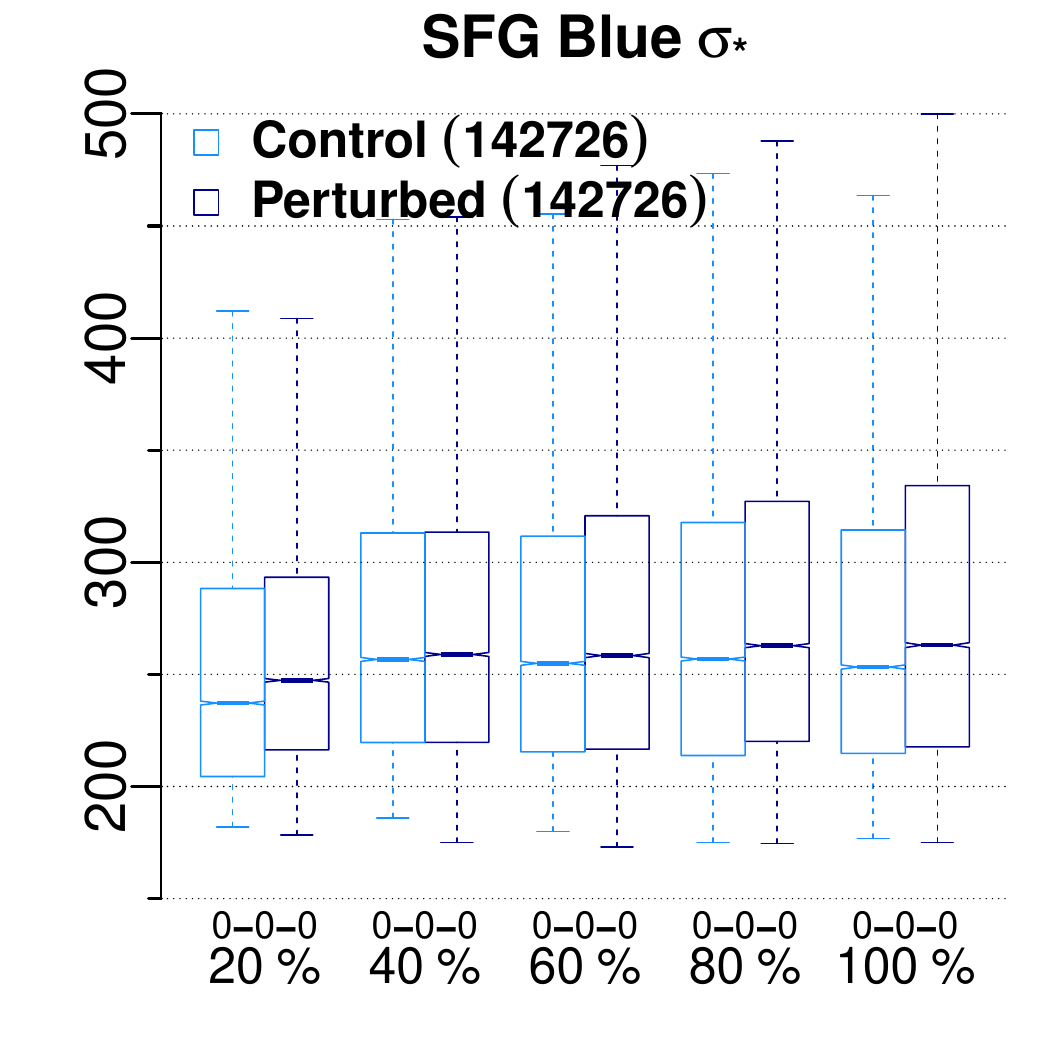}}
   \mbox{\includegraphics[width=.693\columnwidth]{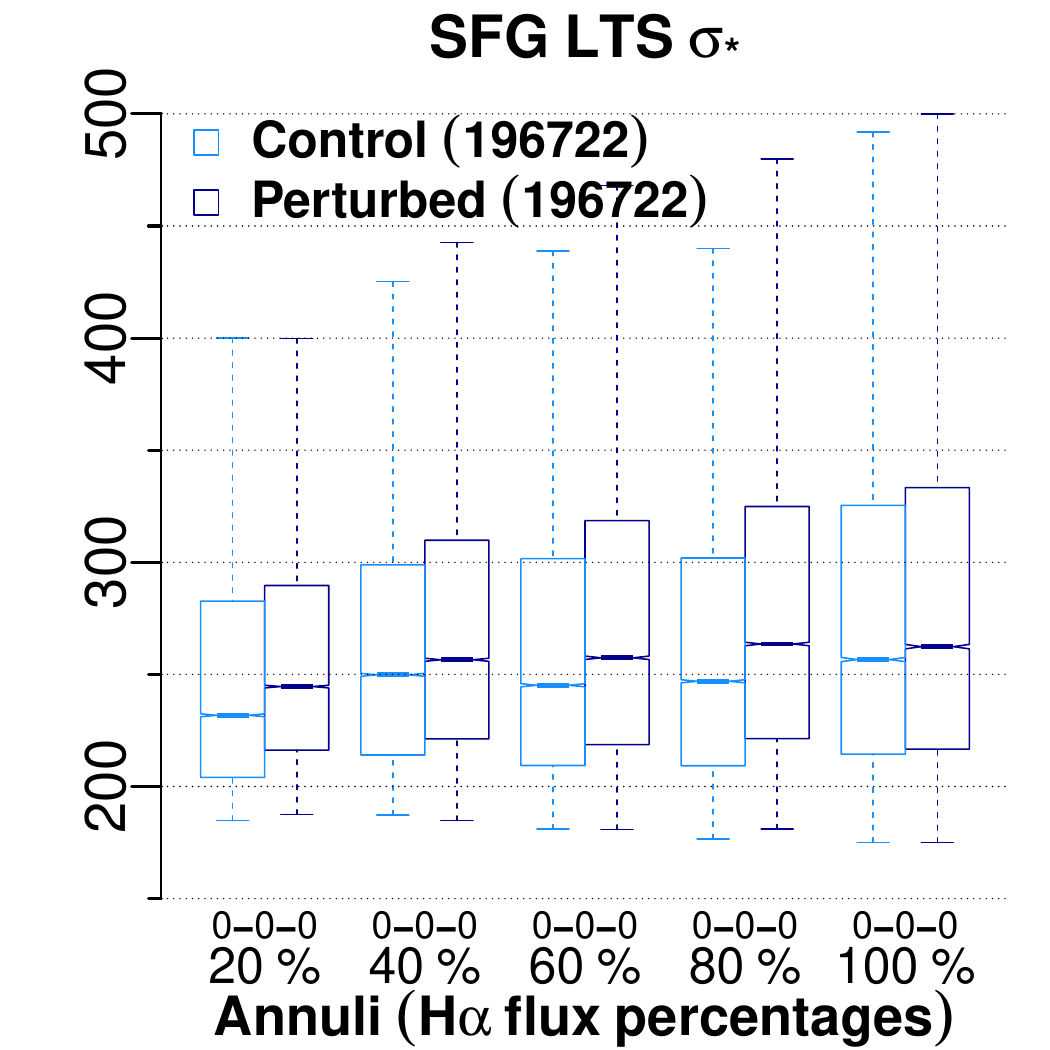}}\\
   \mbox{\includegraphics[width=.693\columnwidth]{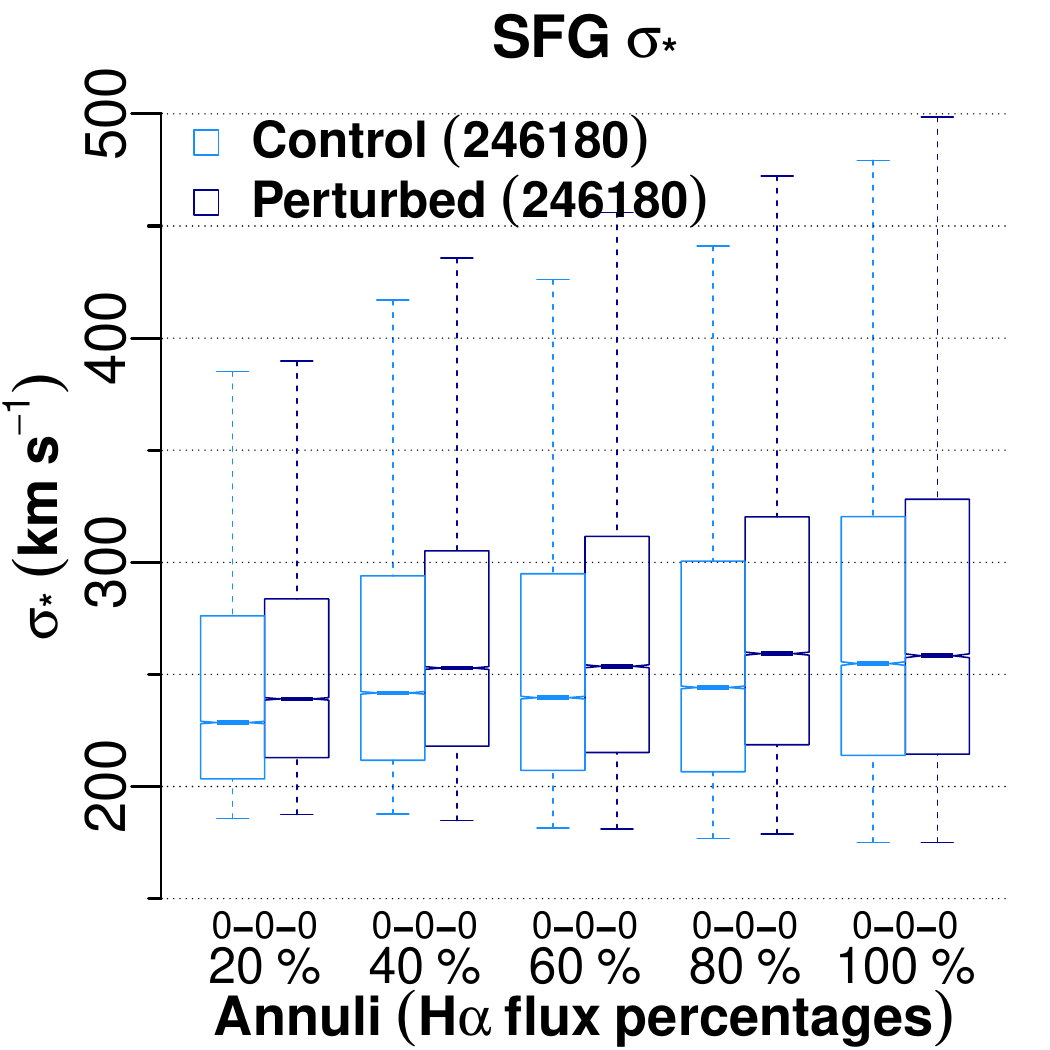}}
   \mbox{\includegraphics[width=.693\columnwidth]{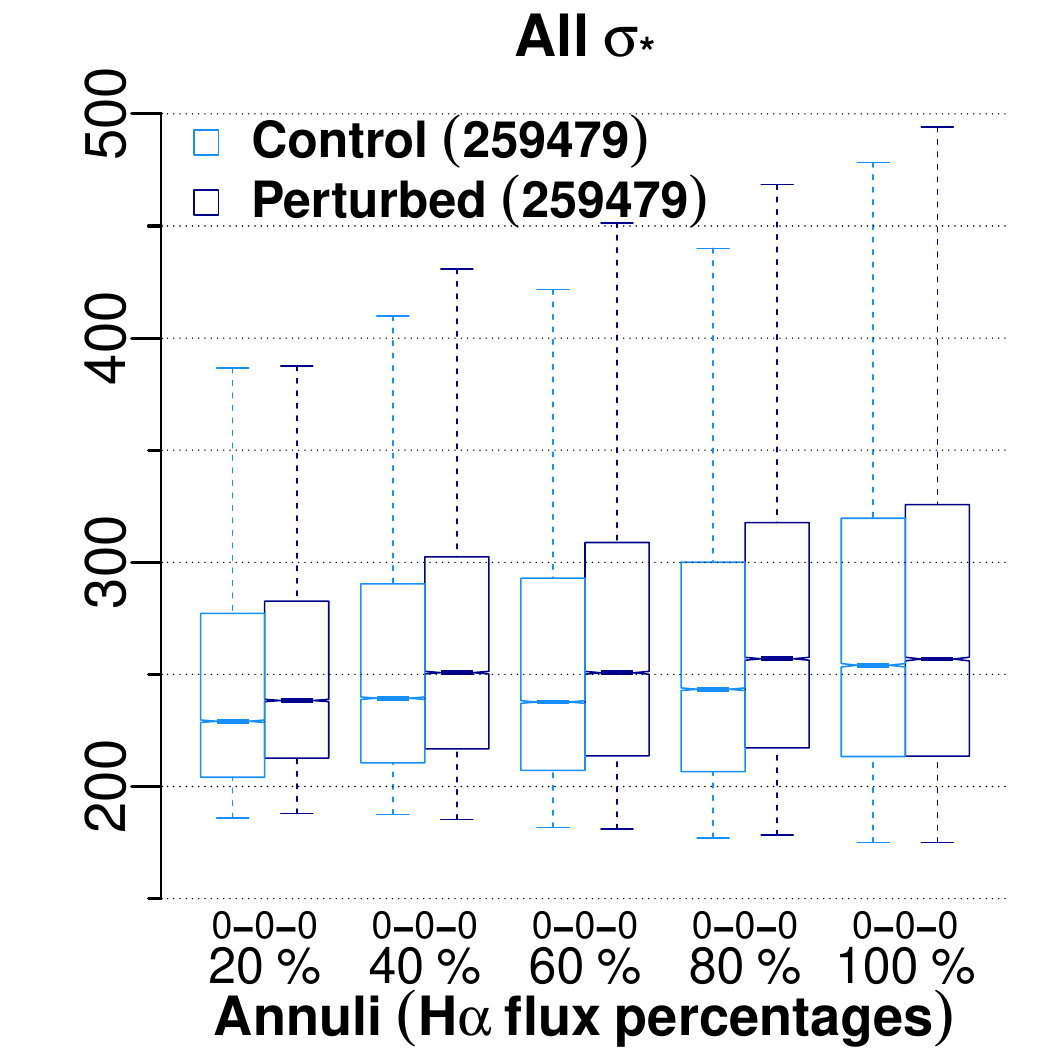}}
   \mbox{\includegraphics[width=.6825\columnwidth]{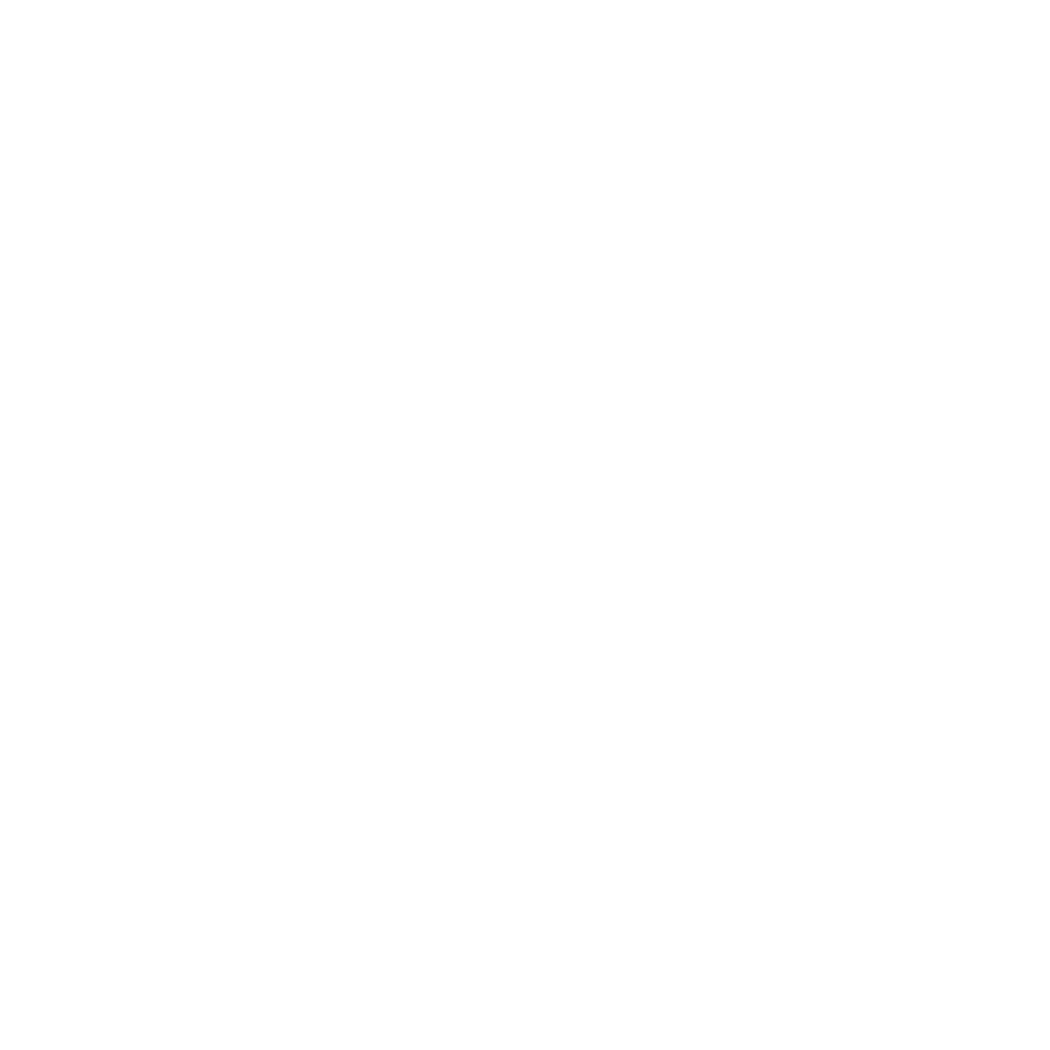}}
\caption{\scriptsize{Boxplots of the annular distributions of stellar velocity dispersion ($\sigma_{*}$) for star-forming spaxels of control and perturbed galaxies. 
Five consecutive-outward annuli, related to the H$\alpha$ flux of each galaxy, represent the radial extension. We keep the subsample division of Papers I and II, 
\textit{i.e.} AGN-like, SFG Red, SFG ETS, SFG Green, SFG Blue, SFG LTS, SFG and all subsamples. We trim the annular distributions according to Section~\ref{subsec:relia}. 
Additionally, all annular distributions result from pairing star-forming spaxels in the perturbed samples to those in the control sample by minimizing their differences 
in LOS rest-framed velocity ($\mathrm{v_{*}-v_{0}}$). The parentheses give the frequencies of present-paired spaxels. Anderson-Darling (AD), permutation of equal densities, 
and Mann-Whitney (MW) test results (AD-permutation-MW) for all pairs of annular distributions are shown right below the boxplots (also see text).}}
   \label{f4} 
\end{figure*}

\begin{table*}
   \setlength{\tabcolsep}{0.5\tabcolsep}
 \begin{minipage}{\textwidth}
\caption{\scriptsize{Fractions ($\frac{\mathrm{med_{P}}}{\mathrm{med_{C}}}$ and $\frac{\mathrm{IQR_{P}}}{\mathrm{IQR_{C}}}$) and uncertainties (see text) for the velocity 
dispersion that compare the medians of the annular distributions (midlines of boxes in Figs.~\ref{f4} and \ref{f5}) and the interquartile ranges (IQRs) of the velocity 
dispersion distributions per galaxy dataset between control (C) and perturbed (P) galaxies. Fractions $>$\,1 are in bold font, whereas those $<$\,1 are underlined. The 
corresponding differences ($\mathrm{med_{P}}-\mathrm{med_{c}}$ and $\mathrm{IQR_{P}}-\mathrm{IQR_{c}}$) and their uncertainties (in km\,s$^{-1}$) are also listed right 
next to each fraction. The ``freq'' columns list the frequencies of present galaxies.}
 \label{tab:1}}
  \centering
 \begin{scriptsize}
 \begin{tabular}{@{\hspace{0.5\tabcolsep}}lcccccccc}
 \hline
         &\multicolumn{5}{c}{Annuli (H$\alpha$ flux percentages)}                                                                                                                       &                     Per                   &    &    \\
         &20\%                              &40\%                              &60\%                              &80\%                              &100\%                             &                   sample                  &freq&freq\\
subsample&\multicolumn{5}{c}{$\frac{\mathrm{med_{P}}}{\mathrm{med_{C}}}$\phantom{0}$\mathrm{med_{P}}-\mathrm{med_{c}}$}                                                                                                               &$\frac{\mathrm{IQR_{P}}}{\mathrm{IQR_{C}}}$\phantom{0}$\mathrm{IQR_{P}}-\mathrm{IQR_{c}}$&C   &P   \\
\cline{1-9}                                                                                                                                                                                                                                   \\
         &\multicolumn{6}{c}{$\sigma_{*}$ (see Fig.~\ref{f4})}                                                                                                                                                                      &    &    \\[4pt]
AGN-like &\underline{0.93}$^{+0.26}_{-0.23}$\phantom{0}$-$15.78$^{+63.65}_{-56.13}$&   \textbf{1.01}$^{+0.23}_{-0.17}$\phantom{0}   1.60$^{+48.11}_{-35.56}$&\underline{0.94}$^{+0.22}_{-0.12}$\phantom{0}$-$12.79$^{+52.69}_{-28.40}$&\underline{0.99}$^{+0.43}_{-0.22}$\phantom{0}$-$3.02$^{+100.85}_{-50.73}$&   \textbf{1.11}$^{+0.41}_{-0.23}$\phantom{0}   24.41$^{+83.81}_{-46.40}$&            \textbf{1.08}$^{+1.07}_{-0.50}$\phantom{0}            3.41$^{+48.03}_{-22.10}$&10  &36  \\[3.5pt]
SFG Red  &   \textbf{1.07}$^{+0.31}_{-0.26}$\phantom{0}   15.76$^{+67.24}_{-56.00}$&   \textbf{1.09}$^{+0.62}_{-0.40}$\phantom{0}   21.03$^{+134.93}_{-88.38}$&   \textbf{1.07}$^{+0.40}_{-0.26}$\phantom{0}   15.29$^{+92.35}_{-58.76}$&\underline{0.97}$^{+0.33}_{-0.26}$\phantom{0}$-$7.32$^{+82.31}_{-66.74}$&\underline{0.98}$^{+0.54}_{-0.39}$\phantom{0}$-$6.08$^{+135.30}_{-96.51}$&            \textbf{1.17}$^{+0.76}_{-0.38}$\phantom{0}            13.36$^{+51.61}_{-25.80}$&12  &18  \\[3.5pt]
SFG ETS  &\underline{0.99}$^{+0.37}_{-0.31}$\phantom{0}$-$2.48$^{+81.70}_{-69.04}$&   \textbf{1.07}$^{+0.54}_{-0.44}$\phantom{0}   15.22$^{+115.38}_{-93.66}$&   \textbf{1.06}$^{+0.39}_{-0.34}$\phantom{0}   13.23$^{+84.56}_{-74.96}$&   \textbf{1.03}$^{+0.43}_{-0.34}$\phantom{0}   8.12$^{+101.41}_{-78.81}$&\underline{0.98}$^{+0.32}_{-0.23}$\phantom{0}$-$5.12$^{+80.75}_{-57.00}$&            \textbf{1.05}$^{+0.19}_{-0.50}$\phantom{0}            4.35$^{+15.61}_{-42.47}$&12  &33  \\[3.5pt]
SFG Green&   \textbf{1.06}$^{+0.53}_{-0.45}$\phantom{0}   13.55$^{+111.28}_{-94.14}$&   \textbf{1.02}$^{+0.53}_{-0.48}$\phantom{0}   4.66$^{+126.90}_{-114.15}$&   \textbf{1.04}$^{+0.41}_{-0.27}$\phantom{0}   9.07$^{+95.86}_{-62.67}$&   \textbf{1.03}$^{+0.43}_{-0.31}$\phantom{0}   7.10$^{+107.55}_{-77.23}$&            1.00$^{+0.36}_{-0.25}$\phantom{0}            0.44$^{+91.55}_{-63.75}$&         \underline{0.95}$^{+0.20}_{-0.18}$\phantom{0}         $-$4.64$^{+18.38}_{-16.35}$&17  &55  \\[3.5pt]
SFG Blue &   \textbf{1.04}$^{+0.54}_{-0.34}$\phantom{0}   10.13$^{+124.47}_{-79.00}$&   \textbf{1.01}$^{+0.46}_{-0.37}$\phantom{0}   2.31$^{+117.77}_{-95.12}$&   \textbf{1.01}$^{+0.51}_{-0.36}$\phantom{0}   3.47$^{+128.85}_{-91.56}$&   \textbf{1.02}$^{+0.48}_{-0.38}$\phantom{0}   5.97$^{+122.31}_{-97.51}$&   \textbf{1.04}$^{+0.52}_{-0.38}$\phantom{0}   9.77$^{+128.15}_{-93.91}$&            \textbf{1.03}$^{+0.40}_{-0.21}$\phantom{0}            2.94$^{+40.36}_{-21.06}$&23  &51  \\[3.5pt]
SFG LTS  &   \textbf{1.06}$^{+0.40}_{-0.30}$\phantom{0}   12.82$^{+89.82}_{-66.45}$&   \textbf{1.03}$^{+0.54}_{-0.34}$\phantom{0}   6.71$^{+133.66}_{-83.91}$&   \textbf{1.05}$^{+0.61}_{-0.41}$\phantom{0}   12.24$^{+145.22}_{-98.47}$&   \textbf{1.07}$^{+0.63}_{-0.47}$\phantom{0}   16.62$^{+149.70}_{-112.15}$&   \textbf{1.02}$^{+0.24}_{-0.20}$\phantom{0}   5.76$^{+61.88}_{-51.40}$&                     1.00$^{+0.51}_{-0.20}$\phantom{0}                     0.00$^{+49.02}_{-19.16}$&40  &92  \\[3.5pt]
SFG      &   \textbf{1.05}$^{+0.53}_{-0.36}$\phantom{0}   10.54$^{+118.27}_{-80.34}$&   \textbf{1.05}$^{+0.43}_{-0.31}$\phantom{0}   11.09$^{+101.47}_{-73.17}$&   \textbf{1.06}$^{+0.45}_{-0.36}$\phantom{0}   13.99$^{+105.59}_{-83.30}$&   \textbf{1.06}$^{+0.37}_{-0.29}$\phantom{0}   15.09$^{+87.78}_{-67.96}$&   \textbf{1.01}$^{+0.41}_{-0.37}$\phantom{0}   3.47$^{+104.56}_{-93.38}$&         \underline{0.99}$^{+0.38}_{-0.29}$\phantom{0}         $-$1.20$^{+36.20}_{-27.67}$&52  &124 \\[3.5pt]
all      &   \textbf{1.04}$^{+0.55}_{-0.37}$\phantom{0}   9.27$^{+124.53}_{-84.14}$&   \textbf{1.05}$^{+0.43}_{-0.33}$\phantom{0}   11.39$^{+101.61}_{-76.19}$&   \textbf{1.05}$^{+0.41}_{-0.39}$\phantom{0}   13.03$^{+95.61}_{-88.61}$&   \textbf{1.06}$^{+0.55}_{-0.34}$\phantom{0}   13.59$^{+129.12}_{-79.25}$&   \textbf{1.01}$^{+0.30}_{-0.23}$\phantom{0}   2.77$^{+75.03}_{-56.89}$&            \textbf{1.01}$^{+0.39}_{-0.29}$\phantom{0}            0.57$^{+36.61}_{-26.80}$&62  &160 \\[4pt]
         &\multicolumn{6}{c}{Frequencies of paired spaxels}                                                                                                                                                                         &    &    \\[4pt]
AGN-like &                              1080&                              2155&                              2978&                              3756&                              3396&                                      13365&    &    \\
SFG Red  &                              4468&                              4677&                              5207&                              5074&                              4814&                                      24240&    &    \\
SFG ETS  &                              7018&                              9246&                             10445&                             12288&                              9971&                                      48968&    &    \\
SFG Green&                             16114&                             17057&                             17070&                             16889&                             11998&                                      79128&    &    \\
SFG Blue &                             22815&                             27956&                             30130&                             34247&                             27578&                                     142726&    &    \\
SFG LTS  &                             36169&                             40337&                             41914&                             43883&                             34419&                                     196722&    &    \\
SFG      &                             43380&                             49732&                             52471&                             56207&                             44390&                                     246180&    &    \\
all      &                             44420&                             51866&                             55438&                             59969&                             47786&                                     259479&    &    \\[4pt]
         &\multicolumn{6}{c}{$\sigma_{\mathrm{ion}}$ (see Fig.~\ref{f5})}                                                                                                                                                           &    &    \\[4pt]
AGN-like &   \textbf{1.02}$^{+0.07}_{-0.09}$\phantom{0}   2.47$^{+9.90}_{-13.15}$&   \textbf{1.02}$^{+0.08}_{-0.06}$\phantom{0}   2.54$^{+11.07}_{-8.05}$&   \textbf{1.01}$^{+0.08}_{-0.06}$\phantom{0}   1.03$^{+11.39}_{-8.86}$&            1.00$^{+0.07}_{-0.04}$\phantom{0}            $-$0.29$^{+9.54}_{-6.05}$&\underline{0.99}$^{+0.05}_{-0.05}$\phantom{0}$-$0.98$^{+7.28}_{-6.64}$&            \textbf{1.23}$^{+0.31}_{-0.50}$\phantom{0}            2.02$^{+2.23}_{-3.02}$&9   &34  \\[3.5pt]
SFG Red  &\underline{0.99}$^{+0.04}_{-0.05}$\phantom{0}$-$1.39$^{+6.35}_{-7.65}$&            1.00$^{+0.07}_{-0.06}$\phantom{0}            $-$0.53$^{+9.37}_{-9.01}$&            1.00$^{+0.05}_{-0.05}$\phantom{0}0.42$^{+6.66}_{-7.13}$&   \textbf{1.01}$^{+0.07}_{-0.08}$\phantom{0}1.02$^{+9.81}_{-11.87}$&            1.00$^{+0.06}_{-0.07}$\phantom{0}$-$0.34$^{+9.06}_{-10.00}$&         \underline{0.90}$^{+0.19}_{-0.19}$\phantom{0}         $-$0.87$^{+1.80}_{-1.89}$&10  &17  \\[3.5pt]
SFG ETS  &\underline{0.97}$^{+0.04}_{-0.06}$\phantom{0}$-$3.80$^{+6.01}_{-9.00}$&\underline{0.98}$^{+0.11}_{-0.09}$\phantom{0}$-$3.32$^{+16.40}_{-13.78}$&\underline{0.99}$^{+0.04}_{-0.05}$\phantom{0}$-$1.84$^{+6.35}_{-7.65}$&\underline{0.99}$^{+0.07}_{-0.08}$\phantom{0}$-$0.87$^{+10.19}_{-11.43}$&            1.00$^{+0.05}_{-0.05}$\phantom{0}$-$0.51$^{+7.09}_{-6.47}$&            \textbf{1.08}$^{+0.93}_{-0.35}$\phantom{0}0.63$^{+7.02}_{-2.64}$&11  &31  \\[3.5pt]
SFG Green&\underline{0.99}$^{+0.06}_{-0.05}$\phantom{0}$-$1.78$^{+7.96}_{-7.51}$&\underline{0.98}$^{+0.08}_{-0.09}$\phantom{0}$-$3.20$^{+11.34}_{-13.11}$&            1.00$^{+0.05}_{-0.06}$\phantom{0}$-$0.16$^{+6.61}_{-7.98}$&   \textbf{1.01}$^{+0.04}_{-0.07}$\phantom{0}1.09$^{+6.00}_{-10.37}$&\underline{0.99}$^{+0.04}_{-0.08}$\phantom{0}$-$1.05$^{+6.35}_{-11.44}$&            \textbf{1.05}$^{+0.28}_{-0.38}$\phantom{0}0.45$^{+2.43}_{-3.19}$&15  &55  \\[3.5pt]
SFG Blue &\underline{0.99}$^{+0.07}_{-0.09}$\phantom{0}$-$0.76$^{+9.83}_{-12.15}$&            1.00$^{+0.06}_{-0.07}$\phantom{0}0.61$^{+8.40}_{-9.94}$&            1.00$^{+0.04}_{-0.05}$\phantom{0}0.69$^{+5.06}_{-6.85}$&            1.00$^{+0.05}_{-0.08}$\phantom{0}0.51$^{+7.55}_{-11.36}$&   \textbf{1.01}$^{+0.05}_{-0.06}$\phantom{0}0.70$^{+7.09}_{-8.99}$&            \textbf{1.52}$^{+0.88}_{-0.29}$\phantom{0}3.31$^{+4.20}_{-1.59}$&18  &45  \\[3.5pt]
SFG LTS  &\underline{0.99}$^{+0.05}_{-0.07}$\phantom{0}$-$0.75$^{+7.67}_{-10.49}$&\underline{0.99}$^{+0.04}_{-0.07}$\phantom{0}$-$0.96$^{+6.30}_{-9.49}$&            1.00$^{+0.06}_{-0.07}$\phantom{0}0.64$^{+8.19}_{-10.24}$&   \textbf{1.01}$^{+0.06}_{-0.06}$\phantom{0}0.96$^{+8.85}_{-8.01}$&            1.00$^{+0.06}_{-0.05}$\phantom{0}0.36$^{+7.91}_{-7.65}$&            \textbf{1.08}$^{+0.26}_{-0.47}$\phantom{0}0.71$^{+2.17}_{-3.86}$&34  &86  \\[3.5pt]
SFG      &\underline{0.99}$^{+0.06}_{-0.08}$\phantom{0}$-$1.28$^{+8.28}_{-11.36}$&\underline{0.99}$^{+0.05}_{-0.08}$\phantom{0}$-$1.21$^{+7.34}_{-11.18}$&            1.00$^{+0.05}_{-0.06}$\phantom{0}0.16$^{+6.74}_{-8.44}$&            1.00$^{+0.05}_{-0.07}$\phantom{0}0.49$^{+6.58}_{-9.97}$&            1.00$^{+0.08}_{-0.08}$\phantom{0}$-$0.45$^{+10.72}_{-11.42}$&            \textbf{1.15}$^{+0.38}_{-0.38}$\phantom{0}1.22$^{+2.85}_{-2.78}$&45  &117 \\[3.5pt]
all      &\underline{0.99}$^{+0.06}_{-0.06}$\phantom{0}$-$1.27$^{+8.88}_{-8.47}$&\underline{0.99}$^{+0.06}_{-0.07}$\phantom{0}$-$0.86$^{+8.61}_{-9.63}$&            1.00$^{+0.05}_{-0.08}$\phantom{0}0.38$^{+6.90}_{-11.37}$&            1.00$^{+0.05}_{-0.06}$\phantom{0}0.52$^{+6.64}_{-8.98}$&            1.00$^{+0.07}_{-0.09}$\phantom{0}$-$0.41$^{+9.22}_{-12.14}$&            \textbf{1.07}$^{+0.28}_{-0.41}$\phantom{0}0.61$^{+2.36}_{-3.39}$&54  &151 \\[1ex]
         &\multicolumn{6}{c}{Frequencies of paired spaxels}                                                                                                                                                                         &    &    \\[4pt]
AGN-like &                               380&                               687&                               760&                               740&                               582&                                       3149&    &    \\
SFG Red  &                              1053&                              1099&                               794&                               842&                               926&                                       4714&    &    \\
SFG ETS  &                              2044&                              1731&                              1622&                              1775&                              1454&                                       8626&    &    \\
SFG Green&                              3827&                              2495&                              2010&                              1786&                              1183&                                      11301&    &    \\
SFG Blue &                              4062&                              3099&                              2755&                              3006&                              2454&                                      15376&    &    \\
SFG LTS  &                              6898&                              4962&                              3937&                              3859&                              3109&                                      22765&    &    \\
SFG      &                              8942&                              6693&                              5559&                              5634&                              4563&                                      31391&    &    \\
all      &                              9322&                              7380&                              6319&                              6374&                              5145&                                      34540&    &    \\[4pt]
\hline\\
 \end{tabular}
 \end{scriptsize}
 \end{minipage}
 \end{table*}

\citet{Fac17} investigated the reliability of their $\sigma_{*}$ measurements below their instrumental resolution ($\sigma_{\mathrm{inst.}}$ $\sim$ 72\,km\,s$^{-1}$, 
\textit{i.e.}, the V1200 setup) by performing comparisons with the DiskMass and ATLAS$^{3\mathrm{D}}$ surveys. They determined that their $\sigma_{*}$ values begin 
to systematically deviate, from the one-to-one relation, below $\sim\,$50\,km\,s$^{-1}$. Because they found $\sigma_{*}$ differences below $\sim$\,100\,km\,s$^{-1}$ 
when comparing the two instrumental setups (see Section~\ref{subsec:setups}), it would be appropriate, due to our instrumental resolution, to consider only spaxels with 
measurements of $\sigma_{*}$ clearly above 100\,km\,s$^{-1}$. On this basis, we apply the reliability thresholds (see Section~\ref{subsubsec:Uncer}) as 
follows.

Figure~\ref{f3} shows our velocity dispersion and S/N ratio annular distributions. Starting with $\sigma_{*}$ (top), find at left the distributions of the star-forming 
spaxels as selected for Papers I and II (see Fig.~\ref{f2}, top). Along the annular sequence, note that the medians (midlines of boxes) are closer in value and that the 
spreads (the complete extensions) are different. Therefore, trimming all annular datasets at a common threshold will affect most that of the 100\,\% annulus. We reduce 
this issue per sample as described. In the 100\,\% annulus dataset, we count the number of spaxels with values below the reliability threshold (175\,km\,s$^{-1}$ for 
$\sigma_{*}$) and compare it with the total number of the dataset by computing a fraction. We discard that same fraction of spaxels (that have the lowest values) from 
the remaining annular datasets. We round that fractional number of spaxels to the very next upper integer. This approach slightly reduces the spaxel fraction from 
$\sim$\,0.82 (as it results in Section~\ref{subsubsec:Uncer}) to $\sim$\,0.76 (see the numbers at the top, Fig.~\ref{f3}, middle). In line with this, Fig.~\ref{f3} 
(right) shows the distributions of the same spaxels as in the middle but corresponding to their respective S/N ratios of the stellar continuum. Eventually, we do not 
repeat this approach for the $\sigma_{\mathrm{ion}}$ because the annular medians and spreads are similar in value and extension, respectively (see Fig.~\ref{f3} bottom 
left). For this case, we trim the annular datasets at the common threshold (139\,km\,s$^{-1}$, see Section~\ref{subsubsec:Uncer}). The resulting distributions of 
$\sigma_{\mathrm{ion}}$ and S/N ratio are at the bottom of Fig.~\ref{f3}.

Moreover, for $\sigma_{*}$, the increment in the spread of the datasets overall the annular sequence holds after trimming at the respective threshold (see Fig.~\ref{f3} 
top-middle). This fact agrees with the annular trend of the S/N ratios of the stellar continuum window (top-right). These ratios are computed by \textsc{starlight} 
as the mean divided by the root mean square flux of the synthetic modelled spectra. Notice that the best ratios tend to populate inwards and that 
the worst ones are of the order of $\sim$\,15 (median), which belongs to the 100\,\% annulus. In this regard, \citet{GarL15} and \citet{Fac17} used Voronoi binning 
to reach a certain quality of the stellar continuum (S/N\,$\geq$\,20). In zones where many spectra must be combined, such spatial binning may introduce artificial 
effects, besides the fact that the properties of contiguous SPs must be homogeneous. Most important, spaxel-by-spaxel comparisons like the ones we conduct would 
not be possible if adopting such binning. Our lower limit of the S/N ratios of the stellar continuum is $\geq$\,5 (mainly selected for the estimation of SFRs, see 
Paper I, section 2.3.1). Regarding this, Appendix \ref{sec:app2} treats our uncertainties, as defined in Section~\ref{subsubsec:Uncer}, as a function of the S/N. It shows 
that, in this analysis, the use of low-S/N data, even in the 100\,\% annulus, has no weighty consequences.

For $\sigma_{\mathrm{ion}}$, the similarities in value and extension of the annular medians and spreads increased due to the significant cut-off (see Fig.~\ref{f3}, 
bottom-middle). Despite this, and like the case of $\sigma_{*}$, the associated S/N ratios of the H$\alpha$ emission line decrease outwards (Fig.~\ref{f3}, 
bottom-right). Our line-fitting process computes these ratios (see Section \ref{subsec:SSPstellar}) as the line amplitude divided by the dispersion 
of the continuum near each line. An H$\alpha$ S/N\,$>$\,8 is the threshold of \citet{BaBa14} and \citet{GarL15} for the spectra used to obtain the ionized-gas 
kinematics. Our lower limit of the S/N ratios of the H$\alpha$ line is $\geq$\,3 (also for SFRs, see Paper I, section 2.3.1). Appendix~\ref{sec:app2} 
shows that using this lower limit is much less consequential than using the limit for the stellar component.

\section{Results}
\label{sec:res}

Find here the distributions of velocity dispersion subject to reliability thresholds that result from proper considerations of the nominal spectral resolution. We intend to 
picture whether there is or no influence of tidal perturbations on the kinematics of star-forming regions. 

\subsection{Distributions of the stellar velocity dispersion}
\label{subsec:mssf_1}

Figure \ref{f4} shows the annular distributions of $\sigma_{*}$ per subsample. We merge the star-forming spaxels from the perturbed samples (see 
the end of Section \ref{subsec:samp-prof}) for each one to form a pair with one of the control sample at the minimum difference in stellar velocity shift 
($\mathrm{v_{*}-v_{0}}$). To look for statistical differences, we include results from the two-sample Anderson-Darling (AD), the permutation of equal
densities, and the Mann-Whitney (MW) tests (\textit{i.e.}, the number triplets right below the annular distribution pairs).

Table~\ref{tab:1} (top) lists the fractions and corresponding differences, and their estimated uncertainties, that compare the medians of the annular
distributions (columns 20-100\,\%), and the IQRs of the velocity dispersion distributions per galaxy dataset (column ``Per sample'') in agreement with Fig.~\ref{f4}. For
the medians of the annular distributions we know, that each spaxel with a velocity dispersion value adopts an uncertainty that is the IQR of the velocity dispersion
distribution corresponding to the galaxy dataset the spaxel belongs to (see Section~\ref{subsubsec:Uncer}). We identify the spaxels closest in value to the medians of
each control and perturbed sample annular distribution.\footnote{The median of a distribution with an odd number of spaxels corresponds to the value of a certain single
spaxel. Despite this, taking the spaxel closest in value to that of the median implies minimizing their absolute difference.} We then propagate the uncertainties of
these spaxels closest in value when computing their fractions ($\frac{\mathrm{med_{P}}}{\mathrm{med_{C}}}$, see Table~\ref{tab:1}). To do so, we use the simple approach
for multiplication and division:
\begin{equation} \label{eq:1}
 \frac{\Delta_{\frac{\mathrm{med_{P}}}{\mathrm{med_{C}}}}}{\frac{\mathrm{med_{P}}}{\mathrm{med_{C}}}}\,=\,\frac{\Delta_{\mathrm{med_{C}}}}{\mathrm{med_{C}}}\,+\,\frac{\Delta_{\mathrm{med_{P}}}}{\mathrm{med_{P}}},
\bigskip
\end{equation}

\noindent where $\Delta$ denotes an uncertainty, either upper or lower ($+/-$). For differences ($\mathrm{med_{P}}-\mathrm{med_{C}}$, see Table~\ref{tab:1}), we use the
approach for addition and subtraction
\begin{equation} \label{eq:2}
 \Delta_{\mathrm{med_{P}}-\mathrm{med_{C}}}\,=\,\Delta_{\mathrm{med_{C}}}+\Delta_{\mathrm{med_{P}}}.
\bigskip
\end{equation}

Unfortunately, annular comparisons, like the one above for the medians, can not be done for the IQRs of their annular distributions. Instead, we use the uncertainty of
each spaxel, \textit{i.e.}, each IQR per galaxy dataset, as they are arranged per sample in Fig.~\ref{f4}. In this way, apart from being a variable per galaxy, the IQRs
are weighted on a spaxel basis. We then have two distributions of IQRs per sample for which we compute their IQRs and adopt them as the uncertainties of the medians of
the two distributions. Similarly, we propagate the uncertainties of each median when using them to compute a fraction and a corresponding difference
($\frac{\mathrm{IQR_{P}}}{\mathrm{IQR_{C}}}$, $\mathrm{IQR_{P}}-\mathrm{IQR_{c}}$, see Table~\ref{tab:1}).

\begin{figure*}\centering
   \mbox{\includegraphics[width=.693\columnwidth]{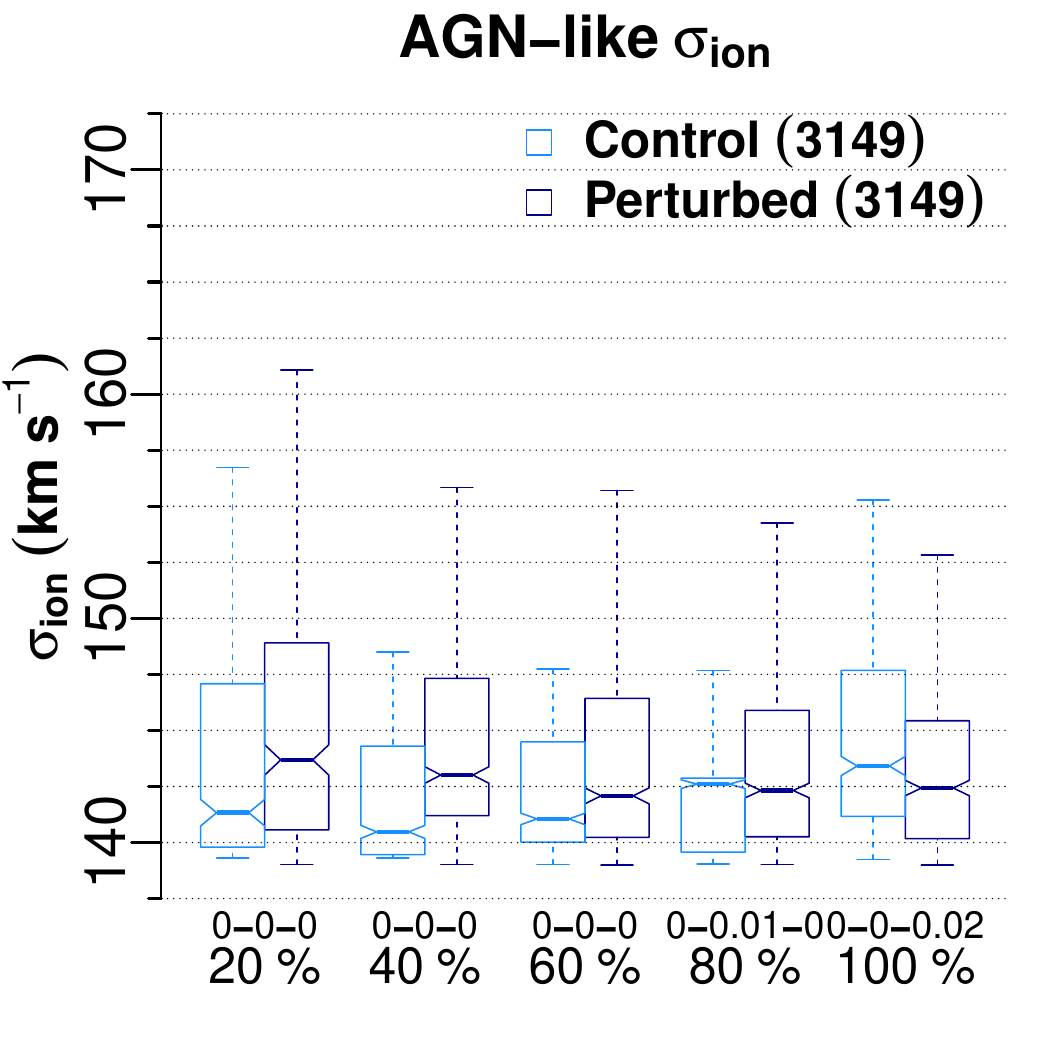}}
   \mbox{\includegraphics[width=.693\columnwidth]{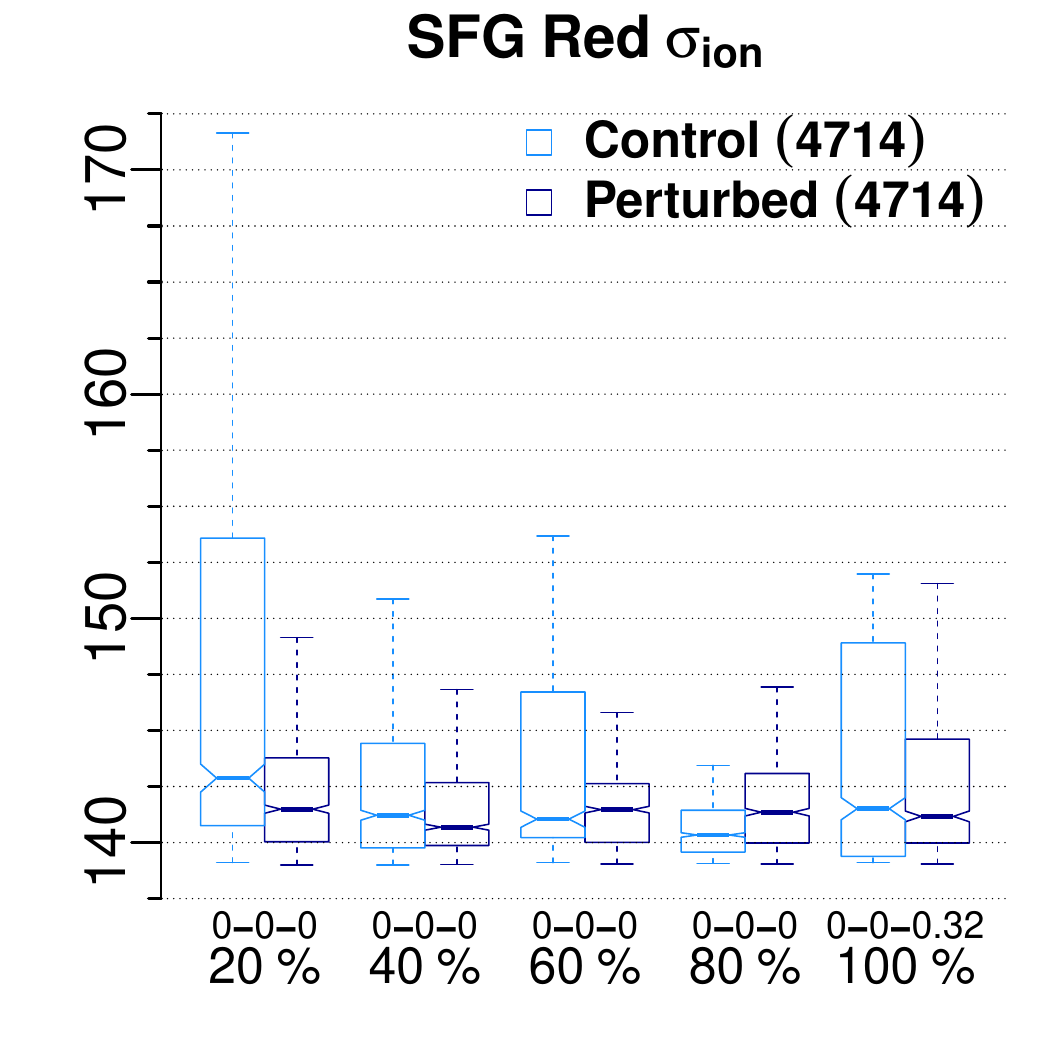}}
   \mbox{\includegraphics[width=.693\columnwidth]{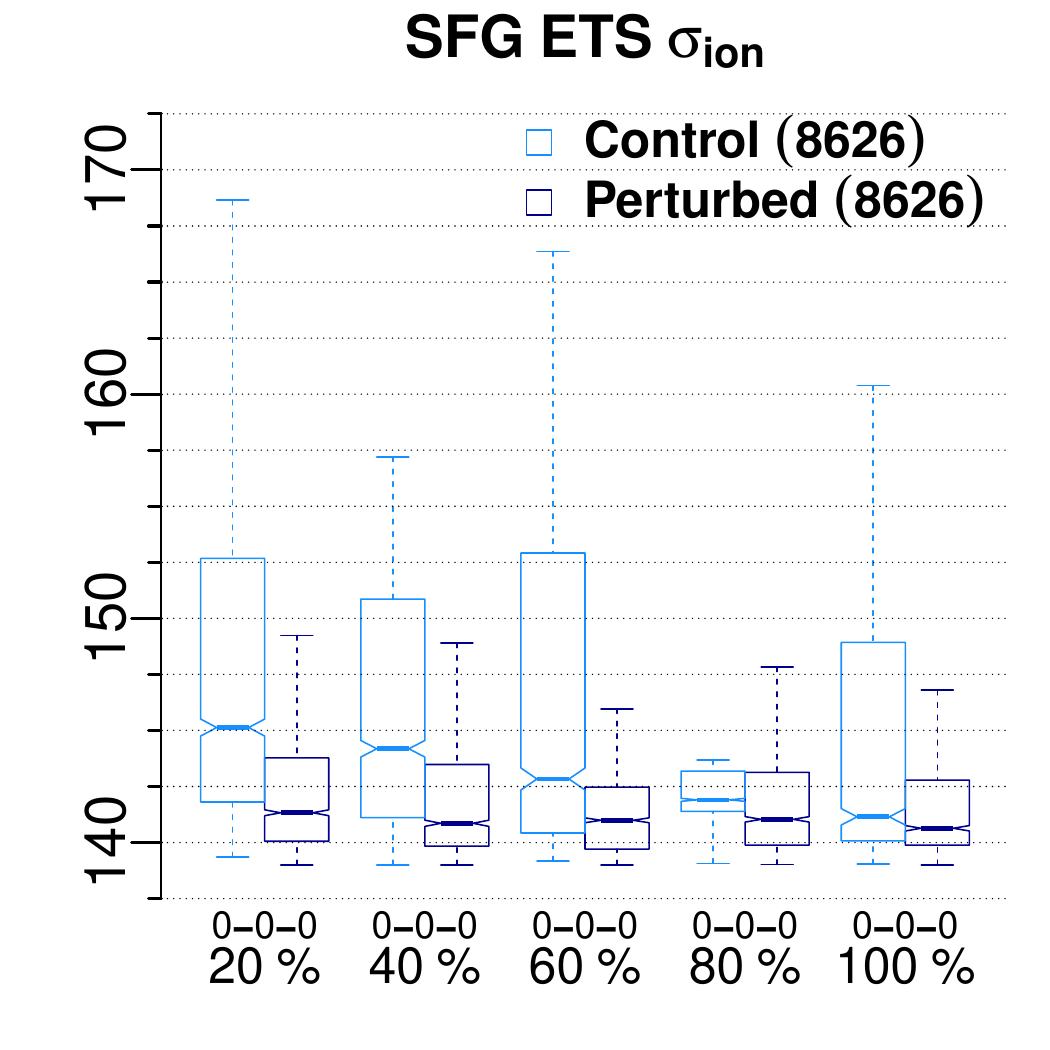}}\\
   \mbox{\includegraphics[width=.693\columnwidth]{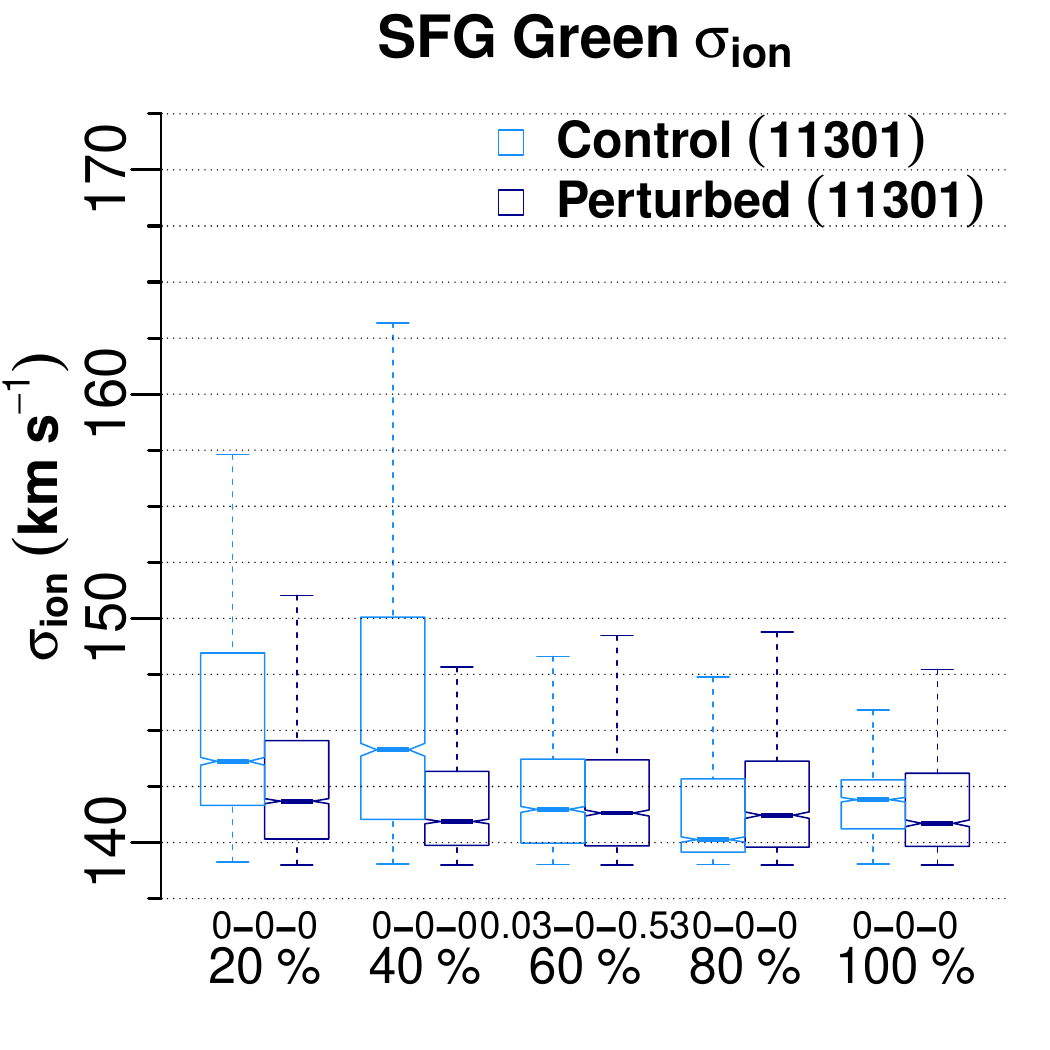}}
   \mbox{\includegraphics[width=.693\columnwidth]{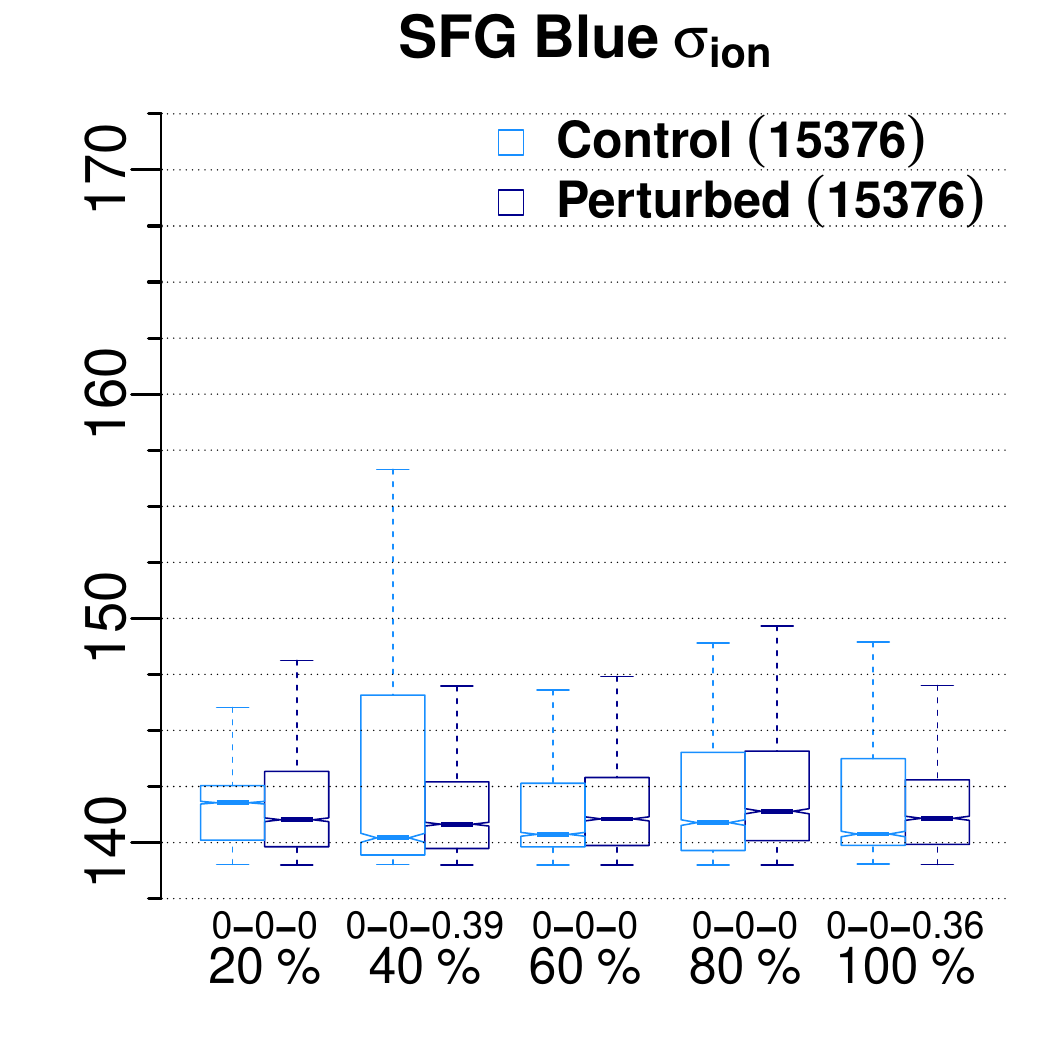}}
   \mbox{\includegraphics[width=.693\columnwidth]{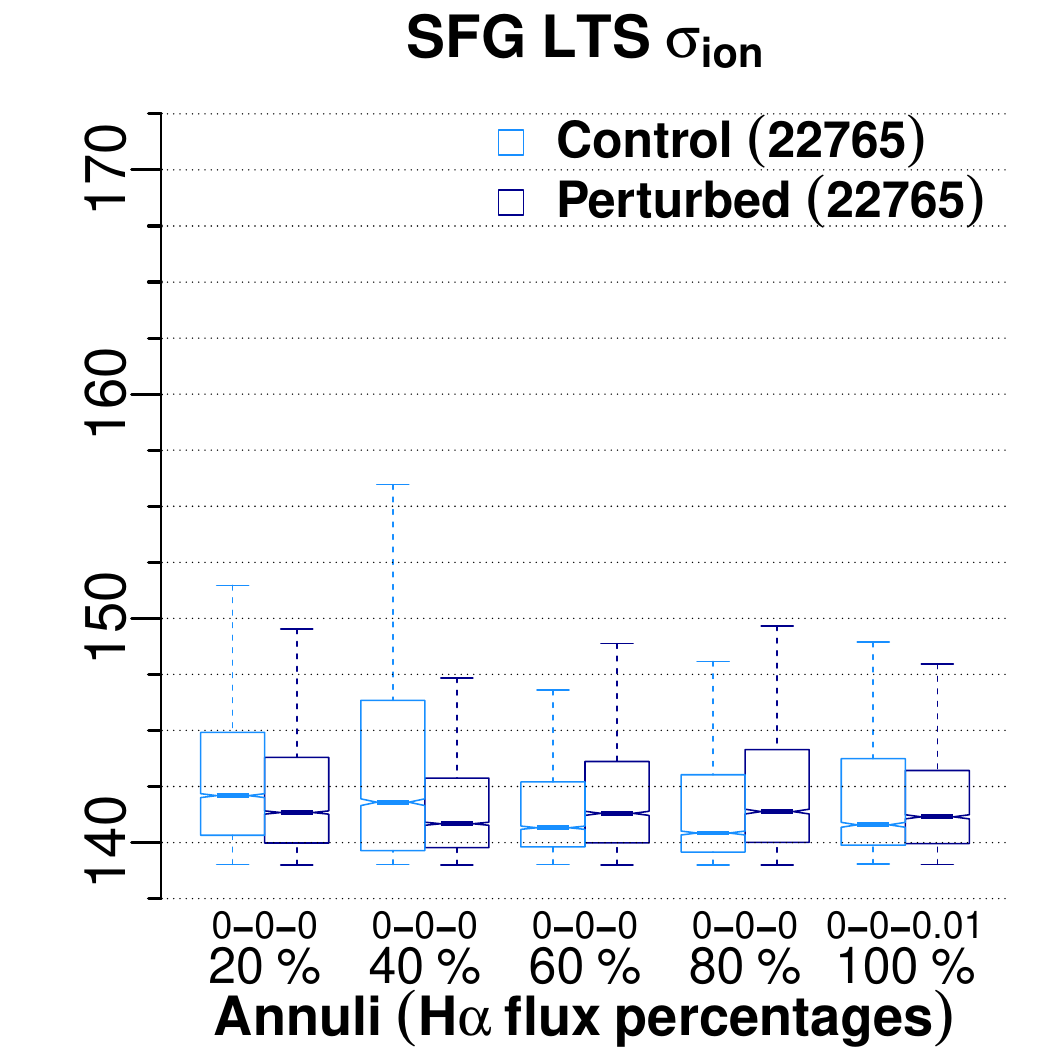}}\\
   \mbox{\includegraphics[width=.693\columnwidth]{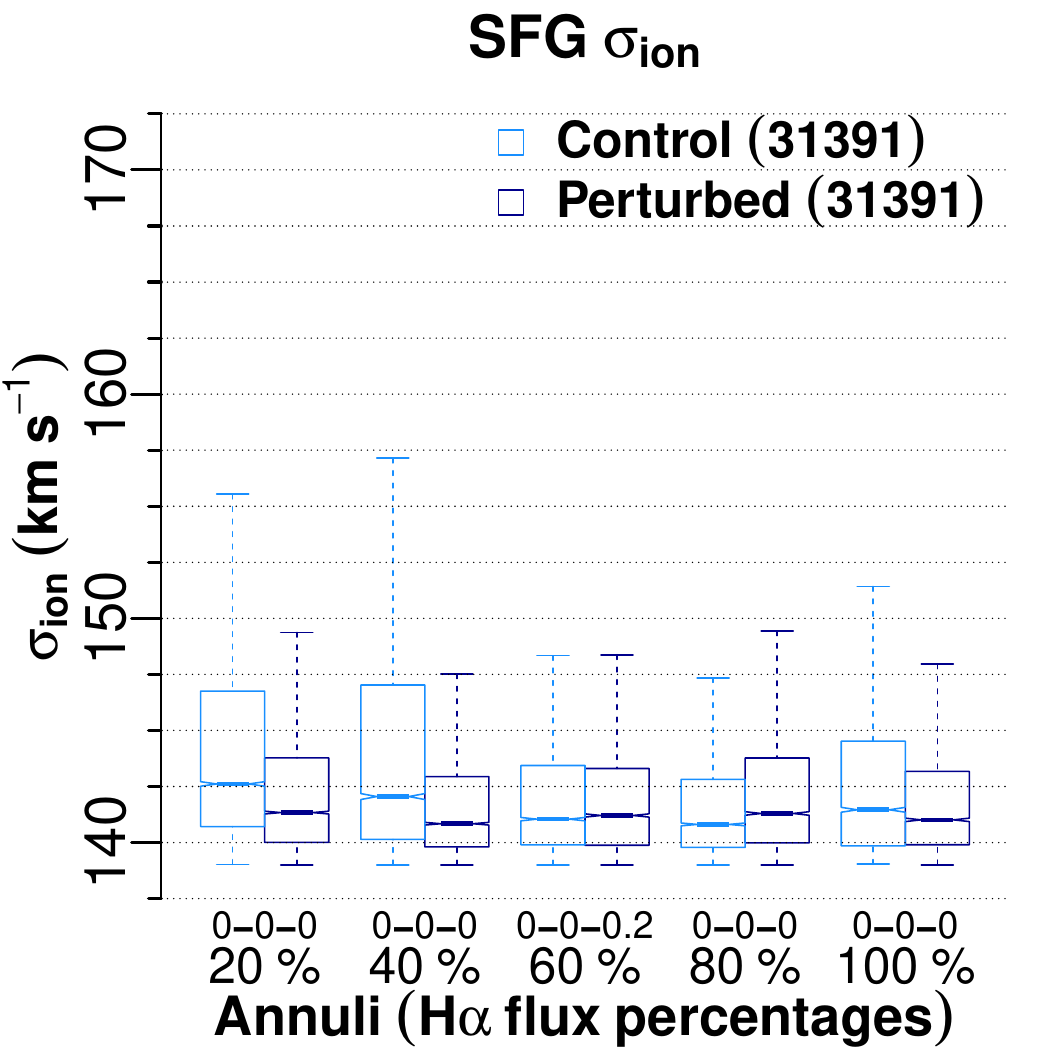}}
   \mbox{\includegraphics[width=.693\columnwidth]{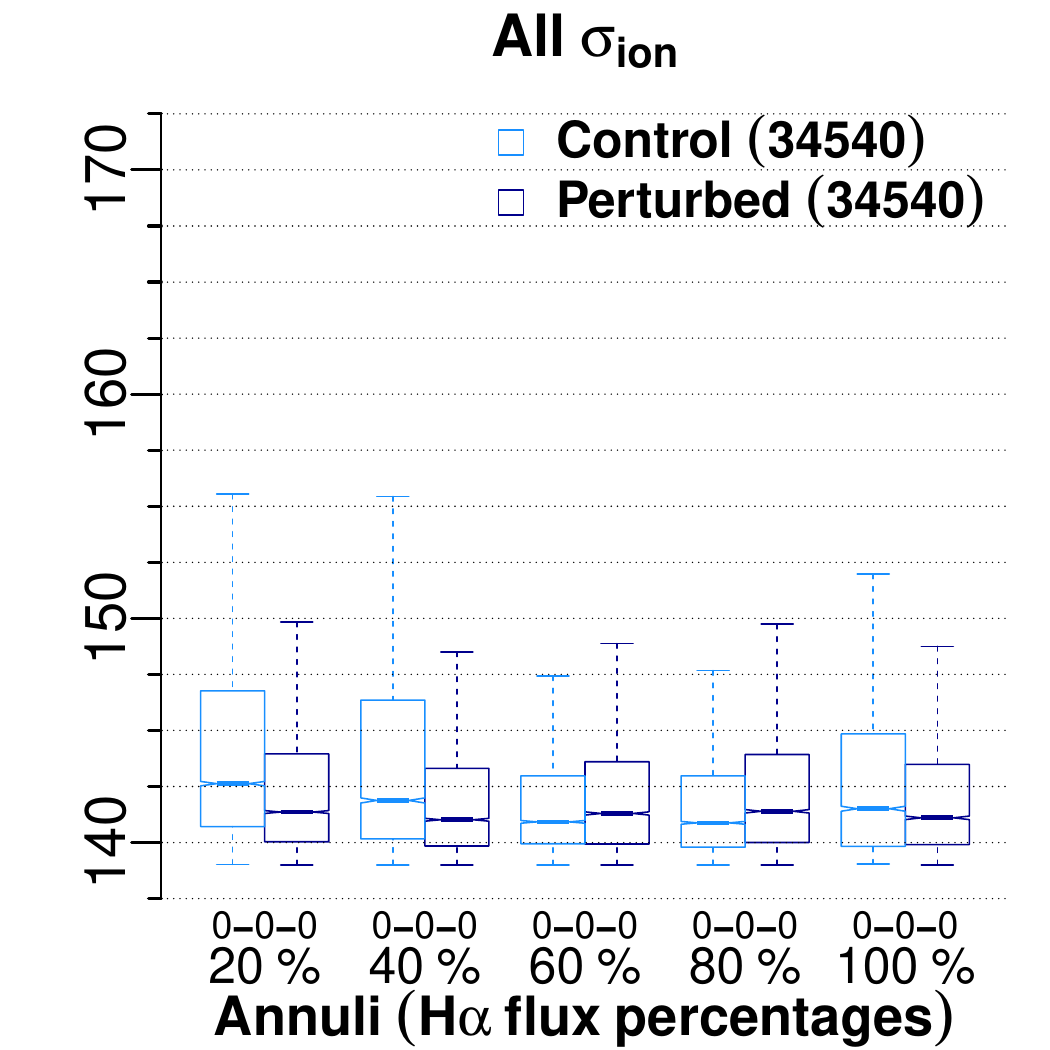}}
   \mbox{\includegraphics[width=.6825\columnwidth]{Nada-eps-converted-to}}
\caption{\scriptsize{Similar to Fig.~\ref{f4} but for the ionized-gas velocity dispersion ($\sigma_{\mathrm{ion}}$). We trim the annular distributions 
according to Section~\ref{subsec:relia}. Star-forming spaxels from the perturbed galaxy samples are merged and paired with their control ones closest in LOS 
rest-framed velocity ($\mathrm{v_{\mathrm{ion}}-v_{0}}$).}}
   \label{f5} 
\end{figure*}

Note in Fig.~\ref{f4}, slightly higher $\sigma_{*}$ median values for perturbed galaxies compared to control ones. In Table~\ref{tab:1} (top), most of the listed fractions are $>$\,1. Considering only those fractions, the following observations can be made:
\begin{itemize}
 \item Among the subsamples, the SFG Red one has the highest fractions, whereas the lowest ones belong to the SFG Blue subsample (with respective medians of 1.07$^{+0.40}_{-0.26}$ in the 60\,\% annulus and 1.02$^{+0.48}_{-0.38}$ in the 80\,\% one). The SFG LTS, SFG and all subsamples together have similar intermediate fractions (with medians of 1.05$^{+0.61}_{-0.41}$ in the 60\,\% annulus, 1.05$^{+0.53}_{-0.36}$ in the 20\,\%, and 1.05$^{+0.41}_{-0.39}$ in the 60\,\% one).
 \item Among the annuli, the highest fractions characterize the 20\,\% and 60\,\% ones, whereas the lowest fractions are in the 100\,\% annulus (with medians of 1.05$^{+0.53}_{-0.36}$ and 1.05$^{+0.61}_{-0.41}$ in the SFG and SFG LTS, and 1.02$^{+0.24}_{-0.20}$ in the SFG LTS subsample, respectively). Intermediate fractions characterize the 80\,\% annulus (median of 1.03$^{+0.43}_{-0.31}$ in the SFG Green subsample).
\end{itemize}

Therefore, the central trend is that star-forming regions in perturbed galaxies show slightly higher fractions, $\sim$(2 to 5-7)\,\%, corresponding to a $\sigma_{*}$ difference of $\sim$(6 to 10-15) km\,s$^{-1}$. Moreover, note that when considering whether the fractions are $>$\,1 or not, the upper uncertainties are all larger than the lower ones. However, the \emph{highest}, \emph{intermediate} and \emph{lowest} fractions are very close to each other, within $\sim$(3 to 5)\,\%, and clearly consistent with 1 within their larger uncertainties ($\sim$(10 to 15-20)\,\%, $\sim$(25 to 35-40)\,km\,s$^{-1}$).

Regarding the IQRs per sample, see in Table~\ref{tab:1} that the spread of the $\sigma_{*}$ also tends to be higher for perturbed galaxies. Five out of eight subsamples 
have fractions $>$\,1. The SFG Red subsample has the highest fraction, whereas the lowest one characterizes all subsamples together (1.17$^{+0.76}_{-0.38}$ and 
1.01$^{+0.39}_{-0.29}$, respectively). Within these ends lies the SFG ETS subsample (1.05$^{+0.19}_{-0.50}$). This subsample is the only exclusion for larger
upper uncertainties, discarding the fact of the fractions as whether or not $>$\,1. Therefore, a slightly higher spread in the $\sigma_{*}$ values
characterizes perturbed galaxies.

Lastly, in Fig.~\ref{f4}, the results from the statistical tests suggest that the pairs of annular distributions are not statistically similar.\footnote{The AD test 
shows whether or not two distributions come from a population with a common unspecified distribution function. The permutation test shuffles the data and compares 
their density estimates by looking for tie patterns. The resulting likelihoods for both tests are fractions from 0 to 1. Finally, the distributions differ mainly 
by their medians if the MW test results are lower than the statistical level.} Notice that the MW test occasionally indicates alikeness in the distributions, 
which often occurs if the box notches overlap, like in the AGN-like and SFG Green subsamples (80 and 100\,\% annuli, respectively).

\subsection{Distributions of the ionized-gas velocity dispersion}
\label{subsec:mssf_2}

Figure \ref{f5} shows the annular distributions of $\sigma_{\mathrm{ion}}$ per subsample. Similarly, star-forming spaxels of the perturbed samples merge and form pairs with those spaxels of the control sample at the closest values of $\mathrm{v_{\mathrm{ion}}-v_{0}}$. We include results from the AD, permutation and MW tests too.

In the same manner, Table~\ref{tab:1} (bottom) lists the fractions, corresponding differences, and uncertainties that compare the annular medians and the galaxy-set IQRs in agreement with Fig.~\ref{f5}.
The fractions and differences come from the spaxels closest in value to the medians of each pair of control and perturbed sample annular distributions. We propagate the respective
adopted uncertainties. We compute the medians of the distributions of the galaxy-set IQRs as arranged in Fig.~\ref{f5}. We adopt the IQRs of such distributions as 
the respective uncertainties, which we propagate and list alongside the fractions and differences.

In Table~\ref{tab:1} (bottom), the majority of frequencies for fractions $>$\,1 are in the AGN-like subsample (median of 1.02$^{+0.07}_{-0.09}$ in the 20\,\% annulus). In contrast, the SFG and all subsamples together show fractions fluctuating from 0.99 to 1. The lowest fractions belong to the SFG ETS subsample (with a minimum of 0.97$^{+0.04}_{-0.06}$ in the 20\,\% annulus).

Regarding the annuli, fractions $<$\,1 are generally observed in both the 20\,\% and 40\,\% annuli. In the remaining ones, fractions of 1 prevail. The 80\,\% annulus shows 3 out of 8 subsamples where fractions are minimally $>$\,1 (a difference of $\sim$1\,km\,s$^{-1}$). Moreover, lower uncertainties, which are just slightly larger than the upper ones by $\sim$(1-4)\,\%, dominate overall.

In summary, for star-forming regions in perturbed galaxies, the central trend suggests that the dispersion of the gaseous velocity does not vary.

The spread in $\sigma_{\mathrm{ion}}$ apparently disagrees with the results of the medians. In Table~\ref{tab:1} (bottom-right), the fractions that compare the IQRs per sample are
$>$\,1 for almost all subsamples. The SFG Blue subsample has the highest fraction, whereas the lowest one characterizes the SFG Green one (1.52$^{+0.88}_{-0.29}$ and 
1.05$^{+0.28}_{-0.38}$, respectively). Within these ends lies the SFG ETS subsample (1.08$^{+0.93}_{-0.35}$). However, when expressed as differences, the contrast in
these spreads is not significant. Larger, lower uncertainties are still present alongside the fractions (4 out of 8 subsamples). Despite this, a minimally higher
spread in the values of $\sigma_{\mathrm{ion}}$ characterizes perturbed galaxies.

Like the case of $\sigma_{*}$, apart from a few annuli where the MW test suggests statistical similitudes (see Fig.~\ref{f5}, the SFG Red, SFG Green, SFG Blue and the SFG 
subsamples), we may say again that the annular distributions of control and perturbed samples are not statistically alike.

\subsection{Distributions from the velocity shift\,-\,velocity dispersion space}
\label{subsec:conf}

\begin{figure*}\centering
   \mbox{\includegraphics[width=.51665\columnwidth]{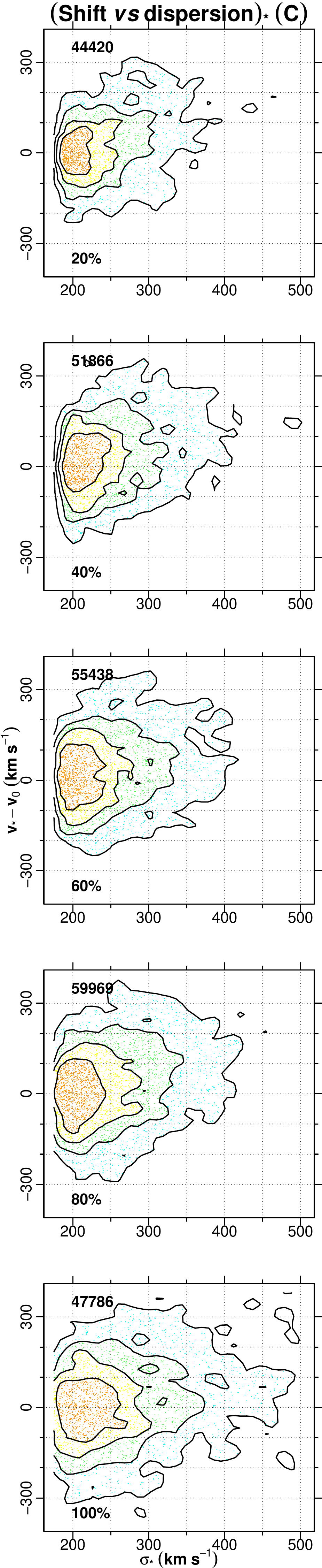}}
   \mbox{\includegraphics[width=.51665\columnwidth]{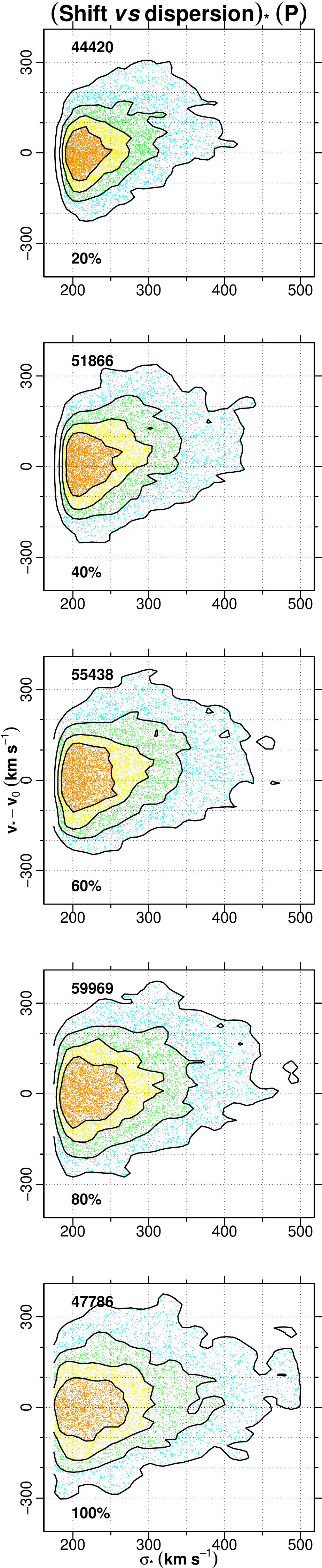}}
   \mbox{\includegraphics[width=.51665\columnwidth]{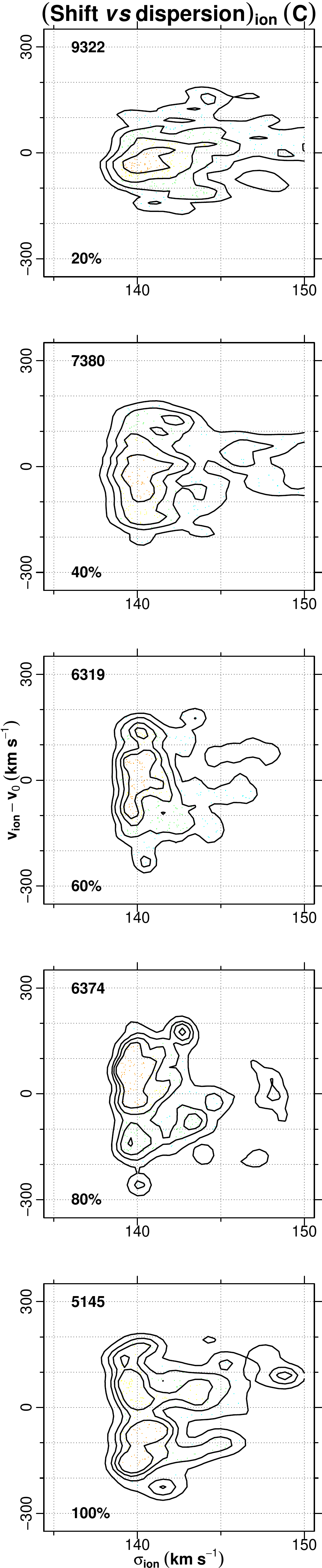}}
   \mbox{\includegraphics[width=.51665\columnwidth]{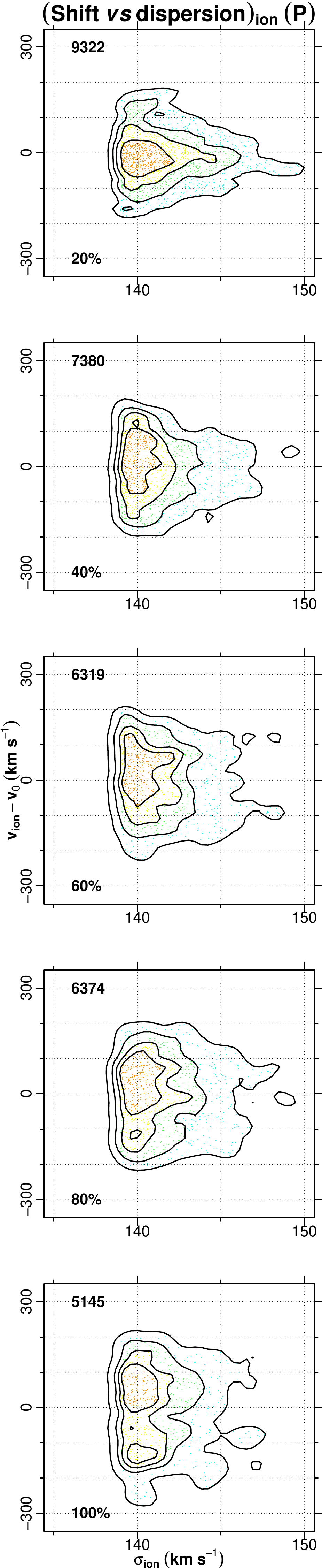}}
\caption{\scriptsize{Annular velocity shift\,-\,velocity dispersion space for star-forming spaxels of all subsamples. We pair star-forming spaxels of the perturbed samples 
(P) with those of the control (C) one at the closest velocity shifts, either $\mathrm{v_{*}-v_{0}}$ (left) or $\mathrm{v_{\mathrm{ion}}-v_{0}}$ (right). Numbers at the 
top-left are the totals of star-forming spaxels paired. Density contours/segments (colours) encircle 90 (aqua), 70 (green), 50 (yellow) and 30\,\% (gold) of star-forming 
spaxels from outside in. We perform 2D K-S/Peacock two-sample tests \citep{Yua17} between the values of the totals of control and perturbed spaxels paired per annulus. The 
maximum differences ($\mathrm{D_{2DKS}}$) are 20\,\% (0.11), 40\,\% (0.08), 60\,\% (0.08), 80\,\% (0.10) and 100\,\% (0.03) for the stellar component; and 20\,\% (0.18), 
40\,\% (0.19), 60\,\% (0.11), 80\,\% (0.12) and 100\,\% (0.16) for the ionized gas.}}
   \label{f6} 
\end{figure*}

To confirm our results, we plot the velocity shift, with respect to zero (\textit{i.e.}, the nuclear value) for the stars and with respect to the rest-frame wavelength 
for the gas, against the velocity dispersion in Fig.~\ref{f6}. Star-forming spaxels from perturbed galaxies form pairs with those of control ones at the closest 
velocity shifts. Annularly, we show density contours that encircle different percentages of spaxels to the totals paired. We take the velocity dispersion distributions 
of the encircled spaxels to select their medians and adopted uncertainties. We take over these uncertainties, the IQRs of the velocity dispersion distributions per 
galaxy dataset, to define one distribution per sample, compute IQRs, and adopt these as the uncertainties of the medians of the two defined distributions. As in Section 
\ref{subsec:mssf_1}, we compute $\frac{\mathrm{med_{P}}}{\mathrm{med_{C}}}$ and $\frac{\mathrm{IQR_{P}}}{\mathrm{IQR_{C}}}$ and propagated uncertainties (see Table 
\ref{tab:2}).

Table~\ref{tab:2} shows $\sigma_{*}$ median values of perturbed galaxies just slightly higher than those of control ones. All contours show fractions $>$\,1 (with the exclusion of the 30\,\% one at the 100\,\% annulus). Among the annuli, the 80\,\% and 100\,\% ones have the highest and lowest fractions, respectively. Additionally, all upper uncertainties tend to be larger than the lower ones. As for the $\sigma_{\mathrm{ion}}$ median values, the fractions are clearly 1. Therefore, there is a slight/there is no distinction in stellar/gaseous velocity dispersion between star-forming regions in perturbed and control galaxies. Despite the presence of larger upper/lower uncertainties, the differences are minimal.

Table~\ref{tab:2} (bottom) shows the spread in velocity dispersion, the IQRs. These are still slightly higher for stars in perturbed galaxies. The $\frac{\mathrm{IQR_{P}}}{\mathrm{IQR_{C}}}$
fractions mostly resemble the $\frac{\mathrm{med_{P}}}{\mathrm{med_{C}}}$ ones along the 20 to 60\,\% annuli. However, there is a dominant presence of slightly
larger lower uncertainties (3 out of 4 contours). On the other hand, the spread in velocity dispersion of the gas still disagrees with the median values (clearly 
unmatching $\frac{\mathrm{IQR_{P}}}{\mathrm{IQR_{C}}}$ and $\frac{\mathrm{med_{P}}}{\mathrm{med_{C}}}$ fractions). The presence of perceptively larger lower
uncertainties alongside the fractions repeats.

Finally, we perform the 2D K-S/Peacock test\footnote{Which implements the original definition of the K-S test extended to multi-dimensional space \citep{Pea83}. The test 
partitions the points and computes the maximum absolute difference between cumulative distribution functions in all partitions. That is very demanding, computationally. 
\citet{Yua17} alleviates this problem with a fast and more efficient algorithm.} on the velocity shift\,-\,velocity dispersion space. In all annular comparisons between 
control and perturbed sample distribution functions at the statistical level, the maximum differences computed by the test (see the caption of Fig.~\ref{f6}) reject the 
null hypothesis of a parent distribution as the origin. 

\begin{table}
   \setlength{\tabcolsep}{0.30\tabcolsep}
 \begin{minipage}{\columnwidth}
\caption{\scriptsize{Fractions and uncertainties computed as described in Section~\ref{subsec:mssf_1}. \textit{Top}: comparison of the medians of the velocity 
dispersion annular distributions. \textit{Bottom}: comparison of the IQRs of the velocity dispersion distributions per galaxy dataset. These comparisons are conducted 
based on the velocity shift\,-\,velocity dispersion space by considering contours/segments that encircle four different percentages (one per colour, see Fig.~\ref{f6}) 
of spaxels from the totals paired per annulus ($\frac{\mathrm{med_{P}}}{\mathrm{med_{C}}}$) and per sample ($\frac{\mathrm{IQR_{P}}}{\mathrm{IQR_{C}}}$). Fractions 
$>$\,1 are in bold font and those $<$\,1 are underlined. Middle rows list the frequencies of encircled paired spaxels (Control\,-\,Perturbed).}
 \label{tab:2}}
  \centering
 \begin{scriptsize}
 \begin{tabular}{@{\hspace{0.5\tabcolsep}}lccccc}
 \hline
contour&\multicolumn{5}{c}{Annuli (H$\alpha$ flux percentages)}                                                                                                              \\
      /&20\%                              &40\%                           &60\%                           &80\%                           &100\%                             \\
segment&\multicolumn{5}{c}{$\frac{\mathrm{med_{P}}}{\mathrm{med_{C}}}$}                                                                                                      \\
\cline{1-6}                                                                                                                                                                  \\
       &\multicolumn{5}{c}{$\sigma_{*}$}                                                                                                                                     \\[4pt]
 90\,\%&   \textbf{1.04}$^{+0.52}_{-0.44}$&\textbf{1.04}$^{+0.39}_{-0.28}$&\textbf{1.05}$^{+0.54}_{-0.43}$&\textbf{1.06}$^{+0.38}_{-0.32}$&   \textbf{1.01}$^{+0.48}_{-0.35}$\\[3.5pt]
 70\,\%&   \textbf{1.05}$^{+0.63}_{-0.44}$&\textbf{1.05}$^{+0.39}_{-0.31}$&\textbf{1.05}$^{+0.35}_{-0.23}$&\textbf{1.07}$^{+0.71}_{-0.49}$&   \textbf{1.01}$^{+0.51}_{-0.43}$\\[3.5pt]
 50\,\%&   \textbf{1.04}$^{+0.67}_{-0.44}$&\textbf{1.04}$^{+0.36}_{-0.23}$&\textbf{1.04}$^{+0.66}_{-0.47}$&\textbf{1.07}$^{+0.61}_{-0.42}$&   \textbf{1.02}$^{+0.44}_{-0.31}$\\[3.5pt]
 30\,\%&   \textbf{1.05}$^{+0.62}_{-0.52}$&\textbf{1.02}$^{+0.13}_{-0.13}$&\textbf{1.04}$^{+0.43}_{-0.34}$&\textbf{1.06}$^{+0.46}_{-0.42}$&\underline{0.99}$^{+0.61}_{-0.43}$\\[4pt]
       &\multicolumn{5}{c}{Frequencies of spaxels (Control\,-\,Perturbed)}                                                                                                   \\[4pt]
 90\,\%&                       40354-40071&                    46680-46722&                    50106-49552&                    53610-53346&                       42125-42431\\
 70\,\%&                       31545-31542&                    36532-36754&                    39060-39070&                    41749-41947&                       32684-32871\\
 50\,\%&                       23103-23206&                    26425-26441&                    28258-27868&                    30148-30072&                       22809-23295\\
 30\,\%&                       14138-13241&                    16406-15691&                    17060-17329&                    18013-17913&                       13557-13515\\[4pt]
       &\multicolumn{5}{c}{$\sigma_{\mathrm{ion}}$}                                                                                                                          \\[4pt]
 90\,\%&\underline{0.99}$^{+0.06}_{-0.09}$&         1.00$^{+0.07}_{-0.07}$&         1.00$^{+0.05}_{-0.09}$&         1.00$^{+0.06}_{-0.06}$&            1.00$^{+0.06}_{-0.08}$\\[3.5pt]
 70\,\%&            1.00$^{+0.08}_{-0.08}$&         1.00$^{+0.06}_{-0.08}$&         1.00$^{+0.05}_{-0.05}$&         1.00$^{+0.05}_{-0.05}$&            1.00$^{+0.05}_{-0.07}$\\[3.5pt]
 50\,\%&            1.00$^{+0.07}_{-0.07}$&         1.00$^{+0.04}_{-0.06}$&         1.00$^{+0.06}_{-0.06}$&         1.00$^{+0.07}_{-0.08}$&            1.00$^{+0.09}_{-0.09}$\\[3.5pt]
 30\,\%&            1.00$^{+0.07}_{-0.08}$&         1.00$^{+0.07}_{-0.07}$&         1.00$^{+0.06}_{-0.08}$&         1.00$^{+0.07}_{-0.08}$&            1.00$^{+0.04}_{-0.06}$\\[4pt]
       &\multicolumn{5}{c}{Frequencies of spaxels (Control\,-\,Perturbed)}                                                                                                   \\[4pt]
 90\,\%&                         7148-8006&                      5839-6370&                      5299-5362&                      5288-5321&                         4167-4334\\
 70\,\%&                         5266-6142&                      4262-5071&                      4245-4295&                      4503-4180&                         3469-3485\\
 50\,\%&                         3581-4334&                      3057-3716&                      3132-3106&                      2590-3041&                         2607-2437\\
 30\,\%&                         1503-2461&                      1102-2025&                      1557-1669&                      1889-1614&                          876-1018\\[4pt]
\hline
                       &\multicolumn{4}{c}{contour/segment}                                                                                            &                     \\
                       &90\,\%                         &70\,\%                         &50\,\%                         &30\,\%                         &                     \\       
                       &\multicolumn{4}{c}{$\frac{\mathrm{IQR_{P}}}{\mathrm{IQR_{C}}}$}                                                                &                     \\
\cline{1-6}                                                                                                                                                                  \\
           $\sigma_{*}$&\textbf{1.01}$^{+0.35}_{-0.33}$&\textbf{1.04}$^{+0.40}_{-0.46}$&\textbf{1.05}$^{+0.41}_{-0.44}$&\textbf{1.05}$^{+0.41}_{-0.50}$&                     \\[4pt]
$\sigma_{\mathrm{ion}}$&\textbf{1.07}$^{+0.26}_{-0.41}$&\textbf{1.07}$^{+0.24}_{-0.41}$&\textbf{1.07}$^{+0.24}_{-0.41}$&\textbf{1.06}$^{+0.25}_{-0.40}$&                     \\[4pt]
\hline\\
 \end{tabular}
 \end{scriptsize}
 \end{minipage}
 \end{table}

\section{Discussion}
\label{sec:dis}

We analyse galaxies with close companions or forming pairs, like the classification of ``pre-merger'' stage galaxies by \citet{BaBa15a}. In the 10\,$<$\,log$_{10}$\,M$_{*}$ (M$_{\odot}$) $<$\,10.5 range, they find an increment of misalignments (\textit{i.e.}, photometric against kinematic position angles) in their interacting galaxies compared with their control ones. In such a range, most of their sampled galaxies are pairs.\footnote{From simulations of \citet{Mor13}, a similar M$_{*}$ range is the peak of the mass distribution function of galaxy pairs.}
When comparing photometric against kinematic orientations, \citet{BaBa15a} report that the median misalignments are the largest for merging galaxies. Similarly, the 
median misalignments for pre-merger-stage galaxies are the smallest. In summary, the misalignments in pre-merging galaxies are the closest to those in their control 
galaxies. They also report only a few pre-merging objects with kinematic stellar against ionized-gas misalignments larger than the median value for their control 
galaxies. They conclude that pre-merging and secular processes appear to have similar impacts on the kinematics of both stars and gas.

Moreover, \citet{Cas22} conclude that the abrupt and simultaneous decrease in stellar and star-forming gas mass is more associated with misalignments than mergers. 
Galaxies that become misaligned seem to preferentially suffer a reduction of their star-forming gas mass rather than an increase. \citet{Cas22} study the tidal field 
around galaxies and find that they are more affected by interaction with nearby galaxies. They suggest this likely reduces their gas content and leads to misalignments.
In our case, the star-forming regions in perturbed galaxies show lower gas fractions compared with those for regions in control galaxies (see Paper II, figure 10).

These analyses suggest that the perturbed galaxies here, perhaps not officially declared as forming pairs but with close companions, are likely subject to kinematic 
misalignments. Though we do not prove that so because that is beyond the scope of this work, it is worth mentioning that misalignments appear to be usual in galaxies with 
close companions.

Returning to our analysis, the SFG Blue, SFG LTS, SFG and all subsamples together own the highest frequencies of star-forming spaxels paired and the richest presence of galaxies (see Fig.~\ref{f4} and Table~\ref{tab:1}). For $\sigma_{*}$, these subsamples show fractions in the 1.02\,$\leq\,\frac{\mathrm{med_{P}}}{\mathrm{med_{C}}}\,\leq$\,1.05 range (differences of $\sim$(6 to 12) km\,s$^{-1}$) and uncertainties of $\sim$(10 to 20)\,\% ($\sim$(25 to 50)\,km\,s$^{-1}$) along the annuli. However, these low fractions and their associated large uncertainties can not be decisive in setting a dominant presence of higher values of stellar velocity dispersion for star-forming regions in perturbed galaxies.

The SFG Red and SFG Blue subsamples have the highest and lowest fractions (1.07$^{+0.40}_{-0.26}$ and 1.02$^{+0.48}_{-0.38}$, respectively). Unfavourably, the former possesses lower and poorer frequencies of paired spaxels and present galaxies compared to the latter. Among the annuli, the lowest fractions (median of 1.02$^{+0.24}_{-0.20}$) characterize the 100\,\% one. This fact might suggest that slightly less disturbed kinematics define the outermost star-forming regions of perturbed galaxies. The AGN-like subsample is the only exclusion (see Table~\ref{tab:1}, top). In sum, these fractions could suggest that less disturbed stellar kinematics prevail in the outskirts of galaxy types where star-forming regions are typical, in contrast with the case of AGN-like galaxies.\footnote{Regularly, in galaxies, SF is concentrated towards the centres \citep[\textit{e.g.}][]{Hop13,Mor15,ArFe16}. In this series, centrally-concentrated SF has been pictured for SFGs in general (see Paper I, figure 8). Apparently, the higher degree of SF activity has no effect or is somehow unrelated to more disordered kinematics.} Additionally, all the upper uncertainties of the fractions are larger than the lower ones. However, their differences are commonly $\sim$10\,\% ($\sim$25\,km\,s$^{-1}$). Hence, the uncertainties are weak to confirm a clear dominant trend of higher $\sigma_{*}$ median values for star-forming regions in perturbed galaxies. This fact appears to be supported by the spread of $\sigma_{*}$; notice that the $\frac{\mathrm{IQR_{P}}}{\mathrm{IQR_{C}}}$ fractions diminish towards galaxy types where we expect star-forming regions to be more typical (see Table~\ref{tab:1}).

On the other hand, for $\sigma_{\mathrm{ion}}$, the SFG subsample and all together, the ones with the highest frequencies of star-forming spaxels paired and the richest 
presence of galaxies (see Fig.~\ref{f5} and Table~\ref{tab:1}, bottom), lack of $\frac{\mathrm{med_{P}}}{\mathrm{med_{C}}}$ fractions $>$\,1 (maxima of 1$^{+0.08}_{-0.08}$ and
1$^{+0.07}_{-0.09}$, respectively). The AGN-like subsample is the only one where fractions $>$\,1 dominate (median of 1.02$^{+0.07}_{-0.09}$). Per annulus, the 20 and 
40\,\% ones show fractions $<$\,1 (respective medians of 0.99$^{+0.06}_{-0.05}$ and 0.99$^{+0.04}_{-0.07}$). The 80\,\% annulus is the only case of several fractions $>$\,1
(median of 1.01$^{+0.06}_{-0.06}$). In sum, the fractions suggest that the ionized-gas kinematics is not disturbed by close companions since the $\sigma_{\mathrm{ion}}$ median
values of perturbed galaxies tend to be similar to those of control ones. Despite the spread in $\sigma_{\mathrm{ion}}$ shown by the subsamples
disagrees with the results of the medians (the fractions represent minimal differences). Respectively, the SFG Blue and the AGN-like subsamples have the highest and second-highest
fractions (1.52$^{+0.88}_{-0.29}$ and 1.23$^{+0.31}_{-0.50}$). However, as we emphasize, their respective differences are minimal (3.31$^{+4.20}_{-1.59}$ and 2.02$^{+2.23}_{-3.02}$ km\,s$^{-1}$).

The results above, all coming from statistically-unalike distributions as indicated by the AD, permutation and MW tests, are confirmed in the velocity shift\,-\,velocity 
dispersion space for all subsamples. The fractions behave roughly the same, \textit{i.e.}, $\frac{\mathrm{IQR_{P}}}{\mathrm{IQR_{C}}}$ trying to resemble 
$\frac{\mathrm{med_{P}}}{\mathrm{med_{C}}}$ for the stellar component, and both not matching for the ionized gas. Not least in importance, the 2D K-S/Peacock two-sample
test on the velocity shift\,-\,velocity dispersion space indicates, at the statistical level, that the control and perturbed sample distribution functions do not share
a common origin from a parent distribution function. The maximum differences, higher for the ionized gas, agree with the more irregular shapes of the contours for that
component (see Fig.~\ref{f6}).

Finding just slightly disturbed and undisturbed motions for the bulks of stars and ionized gas under the presence of close companions is a direct consequence of the statistical
variability of the respective datasets. The variability in velocity dispersion for the stars is greater than that for the gas (\textit{e.g.} Fig.~\ref{f2}), making it easier to depict differences for the
former compared to the latter.

Moreover, the uncertainties of the fractions that compare the medians of the annular distributions (see Tables~\ref{tab:1} and \ref{tab:2})
assist in determining how likely, and by how much, each fraction is higher or lower than 1. The reason for these larger upper and larger lower uncertainties for $\sigma_{*}$ and $\sigma_{\mathrm{ion}}$,
respectively, is the difference between the 3rd and 1st quartiles with the median in each velocity dispersion distribution per galaxy. For $\sigma_{*}$, the differences of the 3rd quartile are
higher than those of the 1st. Summaries of these differences (1st quartile, median and 3rd quartile, respectively) are 31.2, 51.2 and 62.1 (3rd quartile) against 24.2, 36.0 and 44.6 (1st quartile)
km\,s\,$^{-1}$. Contrarily, for $\sigma_{\mathrm{ion}}$, the differences of the 3rd quartile are lower than those of the 1st. Similarly, summaries of these differences are 2.9, 3.8 and 5.1 against
2.9, 4.3 and 5.4 km\,s\,$^{-1}$. Note finally, for these $\sigma_{*}$ and $\sigma_{\mathrm{ion}}$ summaries respectively, differences (1st-3rd quartiles) of $\sim$(7-17.5) and $\sim$(0-0.3) km\,s\,$^{-1}$.
Hence, for all measurements, the larger upper/lower uncertainties confirm a minimal/not at all proof of higher velocity dispersions for star-forming regions in perturbed galaxies.

\section{Summary and conclusions}
\label{sec:conc}

Using IFS and spectral synthesis of SPs, we obtain along the line of sight, spatially-resolved kinematic distributions for star-forming regions that inhabit CALIFA 
survey tidally perturbed (perturbed) and non-tidally perturbed (control) galaxies. Considering our limitations in resolution, we compare the velocity dispersions of the 
stars (synthetic model line absorption) and ionized gas (H$\alpha$ line emission) in such regions at their closest values in velocity shift (rest frame-like). For 
deep and fair comparisons, we further distinguish our samples according to the dominant source of ionization of the gas, photometry, and morphological group of 
their constituent galaxies. 

Our median uncertainties, derived from the distributions of velocity dispersion themselves, reach up to 50\,\% and are less than 8\,\% for the stars and ionized gas,
respectively (Fig.~\ref{f2}). We set thresholds of reliability based on the nominal resolution of the spectral setup ($\geq$\,175 and $\geq$\,139\,km\,s$^{-1}$, for 
the stars and gas, respectively). 

Our results and conclusions are:
\begin{enumerate}
 \item From the annular distributions of $\sigma_{*}$, the medians for perturbed galaxies are higher than those for control galaxies by typically $\sim$5\,\%/$\sim$12
 km\,s$^{-1}$, with uncertainties of $\sim$20\,\%/$\sim$50 km\,s$^{-1}$ (Fig.~\ref{f4} and Table~\ref{tab:1}). Though the upper uncertainties of the fractions that compare  these medians
 are all larger than the lower ones, their differences are typically $\sim$10\,\% ($\sim$25 km\,s$^{-1}$). These low fractions and their associated larger uncertainties indicate a
 minimal difference in velocity dispersion. The spread of $\sigma_{*}$, per galaxy dataset and weighted by each star-forming spaxel, confirms  this, as the fractions diminish towards
 galaxy types where star-forming regions are more typical. Slightly less disturbed stellar kinematics characterize the outskirts  of these types of galaxies. Even though that, in general,
 the $\sigma_{*}$ annular distributions of the star-forming spaxels in control and perturbed galaxies are statistically unalike, we find stellar motions minimally affected  by close
 companions.
 \item From the annular distributions of $\sigma_{\mathrm{ion}}$, the medians for perturbed galaxies are practically equal to those for control galaxies (Fig.~\ref{f5}
 and Table~\ref{tab:1}). Fractions of 1 (0.99-1) characterize those subsamples with the highest frequencies of star-forming spaxels paired and the richest presence of
 galaxies. Fractions of 1 prevail, excluding the two innermost annuli. Expressed as a difference, the spread of $\sigma_{\mathrm{ion}}$, weighted by each star-forming
 spaxel, suggests no difference between the random motions of the ionized gas for regions in perturbed and control galaxies. Despite the narrow range of
 $\sigma_{\mathrm{ion}}$, there is disagreement regarding a statistically common origin between the annular distribution pairs.
 \item The velocity dispersion median values and median galaxy-set IQRs from the velocity shift\,-\,velocity dispersion space confirm our results per annulus (Fig.
 \ref{f6} and Table~\ref{tab:2}). In line with the other statistical tests, the 2D K-S/Peacock test suggests that the control and perturbed sample 2D distribution functions
 lack a common parent distribution function.
\end{enumerate}

In this analysis, the appearance of the stellar kinematics as being more easily affected by tidal perturbations is a consequence of its richer data
variability. However, within the uncertainties, the velocity dispersion fractions and differences are minimal. In contrast, our data for the gas component shows
much less variability, resulting in a similarity between control and perturbed galaxies. As we do not treat stronger interactions such as pre-mergers or mergers,
$\sigma_{\mathrm{ion}}$ might be the only unexpected case. Regarding this, we remark that we exclusively treat regions that are characterized by high line intensities
(H$\alpha$). Narrow line widths are typical (characteristic of \ion{H}{ii} regions), which result in  low velocity dispersions, an intrinsic characteristic of such sharp
intensities. In short, studies treating a higher statistical variability of star-forming regions only should be referred to test our results.

Throughout this series, we have found evidence that confirms that tidal interactions can contribute to the modulation of SF. Through the histories of SF, we have demonstrated
that star-forming regions in tidally-perturbed galaxies were more active when compared with star-forming regions in isolated galaxies. We have shown that these regions are characterized by
more prominent gas inflows, which explains their differences in SF and consequent metal content. Lastly, while only their stellar velocity dispersions are higher, all values are consistent
(within the uncertainties) with those in regions from isolated galaxies, indicating that the kinematics of both components are not affected by the influence of close companions.


\section*{Acknowledgements}
\footnotesize{
Authors wish to thank an anonymous Referee for her/his comments and suggestions that improved this work. A. Morales-Vargas thanks Assistant Editor Bella Lock for her kindness.

All figures for this paper were possible by the use of \textit{R: A language and environment for statistical computing}\footnote{https://www.R-project.org/}. 

The \textsc{starlight}\footnote{http://www.starlight.ufsc.br/} project is supported by the Brazilian agencies CNPq, CAPES and FAPESP and by the 
France-Brazil CAPES/Cofecub program.

The SDSS\footnote{http://www.sdss.org/} is managed by the Astrophysical Research Consortium for the Participating 
Institutions: the Brazilian Participation Group, the Carnegie Institution for Science, Carnegie Mellon University, the Chilean Participation
Group, the French Participation Group, Harvard-Smithsonian centre for Astrophysics, Instituto de Astrof\'{i}sica de Canarias, The Johns Hopkins 
University, Kavli Institute for the Physics and Mathematics of the Universe (IPMU)/University of Tokyo, Lawrence Berkeley National Laboratory, 
Leibniz Institut f\"{u}r Astrophysik Potsdam (AIP), Max-Planck-Institut f\"{u}r Astronomie (MPIA Heidelberg), Max-Planck-Institut f\"{u}r 
Astrophysik (MPA Garching), MaxPlanck-Institut f\"{u}r Extraterrestrische Physik (MPE), National Astronomical Observatories of China, New Mexico 
State University, New York University, Notre Dame University, Observat\'{o}rio Nacional/MCTI, Ohio State University, Pennsylvania State 
University, Shanghai Astronomical Observatory, United Kingdom Participation Group, Universidad Nacional Aut\'{o}noma de M\'{e}xico, University 
of Arizona, University of Colorado Boulder, University of Oxford, University of Portsmouth, University of Utah, University of Virginia, University 
of Washington, University of Wisconsin, Vanderbilt University, and Yale University.

This study uses data provided by the Calar Alto Legacy Integral Field Area (CALIFA) survey.

The Calar Alto Legacy Integral Field Area survey is the first legacy survey being performed at Calar Alto and is managed by the CALIFA survey Collaboration. All members 
would like to thank the IAA-CSIC and MPIA-MPG as major partners of the observatory, and CAHA itself, for the unique access to telescope time and support in manpower and 
infrastructures. The CALIFA survey Collaboration also thanks the CAHA staff for the dedication to the project.

Esperemos pacientes por mi d\'{i}a Mateo. A partir de ah\'{i} ser\'{a} para siempre.


\section*{DATA AVAILABILITY}
The data underlying this article will be shared on reasonable request to the corresponding author. 

The CALIFA survey data can be found here (https://califaserv22.caha.es/CALIFA\_WEB/public\_html/?q=content/).





\bsp	


\appendix
\normalsize{
\section{The galaxy subsamples}
\label{sec:app1}

In this series, we distinguish the SFG galaxies based on their morphological group and colour due to their frequencies. Among control galaxies, SFGs constitute
$\sim$\,84\,\% (52/62), whereas among perturbed galaxies, they account for $\sim$\,77\,\% (125/162) (refer to Paper I, table A1).

While splitting into morphological groups and colours represent different methods, there exists a widespread belief that the SFG red and SFG blue galaxies predominantly
belong to the SFG ETS and SFG LTS groups, respectively. However, this is not always the case. As outlined in Paper I, Table~\ref{tab:A1} lists the relationship between
morphological group and colour. The upper and middle rows present percentages of galaxies, while the bottom ones pertain to percentages of star-forming spaxels. The
middle and bottom rows exclusively concern SFGs. Notice, in the middle and bottom rows that:
\begin{enumerate}
 \item in the case of the ETS group, the green colour makes a considerable contribution of $\sim$\,40\,\% for control galaxies. For perturbed and all galaxies together,
 the percentages of Green galaxies clearly surpass those of Red ones. As for the spaxels (bottom row), the percentages related to Green galaxies are even more significant.
 \item For the Red colour, the percentages of LTS galaxies are $\sim$\,40\,\% (which is non-negligible). When it comes to the spaxels, the contribution of LTS galaxies
 exceeds that of the ETS galaxies.
\end{enumerate}
Hence, in our series, SFG Red galaxies do not strongly dominate SFG ETS galaxies, and the contribution of spaxels within SFG ETS galaxies that are red does not dominate
either. Consequently, we can not discard either the SFG Red or the SFG ETS subsample.

Moreover, for the middle and bottom rows of Table~\ref{tab:A1}:
\begin{enumerate}
 \item for the LTS group, the contribution of non-blue galaxies is $>$\,40\,\%. When it comes to the spaxels, their percentage in non-blue galaxies is $\sim$\,30\,\%
 (which is not negligible).
 \item Regarding the Blue colour, all SFG Blue galaxies are practically LTS, meaning that all spaxels in SFG Blue galaxies are most contained in the SFG LTS subsample.
 However, not all SFG LTS galaxies are blue, and not all of their spaxels belong to the SFG Blue subsample.
\end{enumerate}
Likewise, we can not discard either the SFG Blue or the SFG LTS subsample. The former represents pure star-forming content, whereas the latter exhibits a significant
``quenched'' or ``retired'' nature. Table~\ref{tab:A1} demonstrates that the division into subsamples for SFGs is not redundant in percentages of either galaxies or
star-forming spaxels contained.

The degree of overlap among our subsamples, both in terms of galaxies and star-forming spaxels, can be deduced from the percentages in Table~\ref{tab:A1}. As noticed
before, the only unbalanced case is that of the SFG Blue subsample, which is practically contained in the SFG LTS subsample (the presence of Blue galaxies and hence
their contribution of spaxels in the SFG ETS subsample is minimal). Hence, all SFG Blue galaxies belong to the SFG LTS category, and their associated star-forming
spaxels are found exclusively in SFG LTS galaxies. Additionally, note that both the SFG Green and Red subsamples are importantly conformed by the SFG ETS and SFG LTS
subsamples.

Finally, the individual presence of the SFG subsample is important. It brings together the approximations of ages (colours) and level of SF (either ``active'' or
``quenched'') of the star-forming regions.

\scriptsize{
\begin{table*}
 \setlength{\tabcolsep}{0.5\tabcolsep}
\begin{minipage}{\textwidth}
\centering
 \caption{\scriptsize{Morphological group-galaxy colour relation in percentages according to Paper I (see table A1, \textit{i.e.}, 62 control and 162 perturbed 
 galaxies). The first three columns correspond to each morphological group. The remaining columns correspond to colours per morphological group and vice versa. The 
 top  and middle rows give the percentages of galaxies for both dominant gas excitations, AGN-like \& SFG (all galaxies) and the SFG one, respectively. The bottom 
 rows list the percentages of star-forming spaxels for SFG excitation.}
  \label{tab:A1}}
 \begin{scriptsize}
 \begin{tabular}{lccccccccccccccccccccc}
 \hline\\
         &                                                           &                                                                    &                                                                       &\multicolumn{3}{c}{ET}&\multicolumn{3}{c}{ETS}&\multicolumn{3}{c}{LTS}&\multicolumn{3}{c}{Red}&\multicolumn{3}{c}{Green}&\multicolumn{3}{c}{Blue}\\
         &ET\footnote{\scriptsize{E and S0, early type (ET) galaxy.}}&ETS\footnote{\scriptsize{Sa to Sb, early type spiral (ETS) galaxy.}}&LTS\footnote{\scriptsize{Sbc to later, late type spiral (LTS) galaxy.}}&Red  &Green   &Blue   &Red   &Green   &Blue   &Red   &Green   &Blue   &ET    &ETS     &LTS    &ET     &ETS     &LTS     &ET     &ETS    &LTS     \\[1ex]
\hline\\
         &\multicolumn{21}{c}{all galaxies}                                                                                                                                                                                                                                                                                                                         \\
Control  &2                                                          &32                                                                  &66                                                                     &0    &100     &0      &55    &45      &0      &12    &32      &56     &0     &69      &31     &4      &39      &57      &0      &0      &100     \\
Perturbed&2                                                          &39                                                                  &59                                                                     &67   &33      &0      &50    &48      &2      &8     &39      &53     &5     &76      &19     &1      &45      &54      &0      &2      &98      \\
all      &2                                                          &37                                                                  &61                                                                     &50   &50      &0      &51    &48      &1      &9     &37      &54     &4     &74      &22     &2      &44      &54      &0      &1      &99      \\[1ex]  
         &\multicolumn{21}{c}{SFGs}                                                                                                                                                                                                                                                                                                                                 \\
Control  &0                                                          &23                                                                  &77                                                                     &0    &0       &0      &58    &42      &0      &12    &30      &58     &0     &58      &42     &0      &29      &71      &0      &0      &100     \\
Perturbed&0                                                          &26                                                                  &74                                                                     &0    &0       &0      &36    &61      &3      &8     &38      &54     &0     &63      &37     &0      &36      &64      &0      &2      &98      \\
all      &0                                                          &25                                                                  &75                                                                     &0    &0       &0      &42    &56      &2      &9     &36      &55     &0     &61      &39     &0      &35      &65      &0      &1      &99      \\[1ex]
         &\multicolumn{21}{c}{star-forming regions (SFGs)}                                                                                                                                                                                                                                                                                                          \\
Control  &0                                                          &12                                                                  &88                                                                     &0    &0       &0      &49    &51      &0      &8     &23      &69     &0     &45      &55     &0      &24      &76      &0      &0      &100     \\
Perturbed&0                                                          &16                                                                  &84                                                                     &0    &0       &0      &20    &69      &11     &5     &30      &65     &0     &44      &56     &0      &30      &70      &0      &3      &97      \\
all      &0                                                          &15                                                                  &85                                                                     &0    &0       &0      &28    &64      &8      &6     &28      &66     &0     &44      &56     &0      &28      &72      &0      &2      &98      \\[1ex]
\hline\\
 \end{tabular}
 \end{scriptsize}
 \end{minipage}
\end{table*}

\section{Uncertainties and S/N ratios}
\label{sec:app2}
\normalsize{
A low signal-to-noise (S/N) value is an inherent characteristic of the outskirts of galaxies. Low S/N is due to reduced flux intensity, whether by surface or volume. Our
intention to using low S/N is, among other things, to mantain the spatial resolution of the data. This choice allows for spaxel-to-spaxel comparisons. In sum, we want to
leverage the IFS information by preserving the spatial resolution in our analysis.

Figure~\ref{B1} shows our adopted uncertainties as a function of the S/N ratio. With only one exception (see caption), all slopes ($s$) are statistically significant,
aswell as the correlation coefficient results (top, left and right, respectively). Note that both the linear regression and the coefficients agree to suggesting an
anticorrelation between our uncertainties and the S/N values. However, though statistically valid, this level of anticorrelation is rather low. For example, the
coefficient results are generally $\leq$\,20\,\%, \textit{i.e.}, only the 40, 60 and 80\,\% annuli for control galaxies (stellar component), and the 20\,\% annulus 
for control galaxies (ionized gas), are the ones that exceed a value of 0.2.

Concerning the linear regression, see that the slopes are rather flat. For the stars, theslopes are steeper than $-$0.4 only in four annuli (40, 60 and 80\,\%, and
40\,\%, for control and perturbed galaxies, respectively). The steepest slope is observed in control galaxies in the 60\,\% annulus ($s\,=\,-$0.6), which implies a
difference in the uncertainty of $\sim$\,50 km\,s$^{-1}$ for a difference in S/N of $\sim$\,80.

When it comes to the ionized gas, all the slopes are practically flat. Due to the reduced scale in the uncertainties and the much wider range in S/N, it is practically unimportant
to consider any variation.

Regarding the uncertainty values within the regularly spaced S/N bins, note that, for the stars, they indicate non-linear trends that resemble the line regression only
for ranges of S/N where the spaxels saturate (20 to 40, and 10 to 30, for the three inner and the two outer annuli, respectively). When considering all values at all
paces of S/N, they are most inclined to be flat rather than indicating any anticorrelation.

Of particular importance is the 100\,\% annulus of perturbed galaxies (regarding Fig.~\ref{f3}, top-right). For S/N values of $\sim$\,5 and $\sim$\,20 (the first three
open circles from left to right, Fig.~\ref{B1}), the difference in uncertainty is $\sim$\,10 km\,s$^{-1}$. The most remarkable differences in uncertainty ($\sim$\,40
km\,s$^{-1}$) for S/N values of $\sim$\,5 and $\sim$\,20 are in the 80\,\% and 40\,\% annuli of control and perturbed galaxies, respectively.

Concerning the ionized gas, the only clear annular cases of agreement between the uncertainties within the bins and the linear regression are in perturbed galaxies (60\,\%,
80\,\% and 100\,\% annuli).

In summary, Fig.~\ref{B1} suggests that using low-S/N data in our analysis, even in the 100\,\% annulus, does not have significant consequences.

\begin{figure*}\centering
   \mbox{\includegraphics[width=.51665\columnwidth]{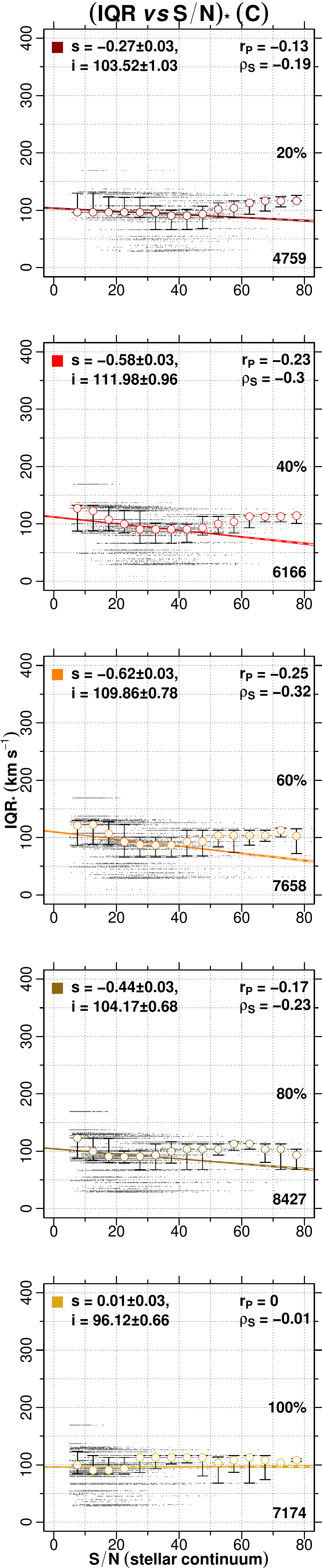}}
   \mbox{\includegraphics[width=.51665\columnwidth]{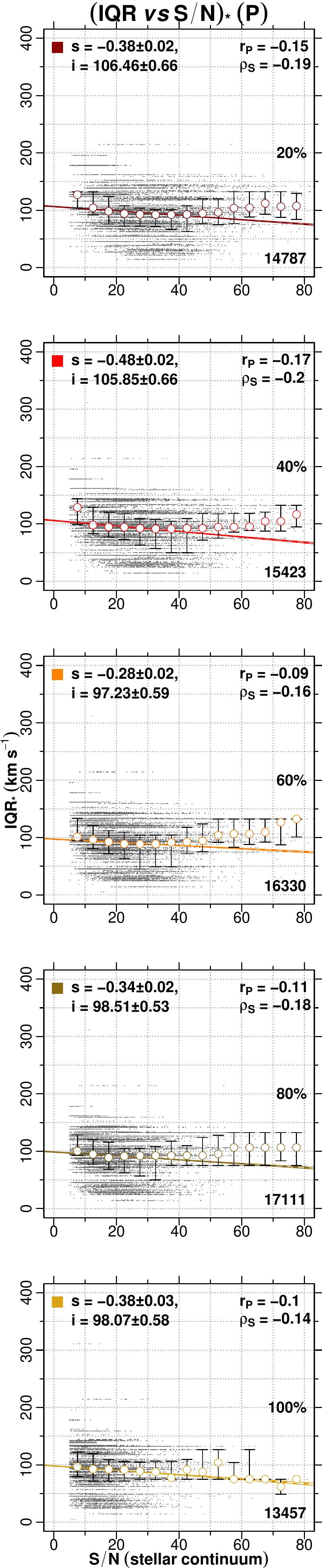}}
   \mbox{\includegraphics[width=.51665\columnwidth]{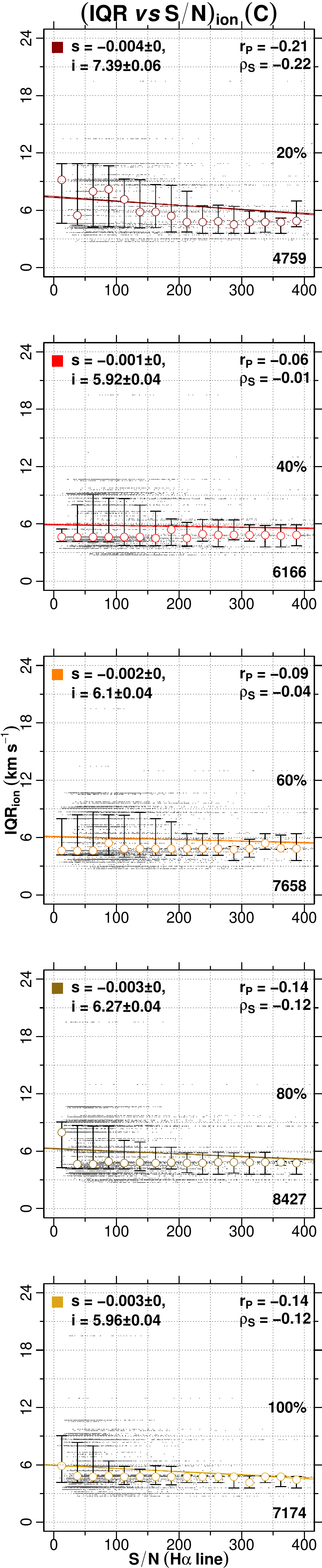}}
   \mbox{\includegraphics[width=.51665\columnwidth]{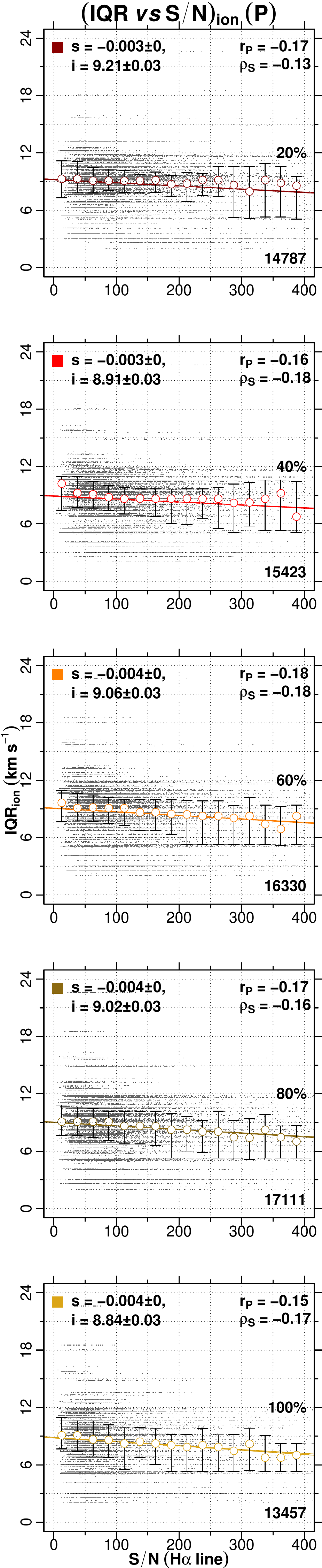}}
\caption{\scriptsize{Interquartile Range (IQR) of the velocity dispersion distribution per galaxy, which represents the adopted uncertainty for a star-forming spaxel within a galaxy dataset
(see Section~\ref{subsubsec:Uncer}) \textit{vs} the corresponding S/N ratio, either that of the continuum (left), or that of the line (right), per star-forming spaxel (see also Fig.~\ref{f3},
panels at left). (C) and (P) stand for control and perturbed galaxies, respectively. Linear model regression results, all statistically significant at the statistical level (except that of
the stellar component for the 100\,\% annulus in control galaxies), are shown at the top-left. Pearson and Spearman correlation results, all statistically significant at the statistical level
(except for the same case as mentioned for the linear regression), are included too (top-right). We also bin the data at regular paces of S/N and computed medians (open circles) and error bars (arrows, \textit{i.e.}, IQRs) for the IQR values corresponding to the binned S/N values. The numbers at the bottom-right are the frequencies of spaxels per annulus.}}
   \label{B1} 
\end{figure*}

\label{lastpage}
\end{document}